\documentclass[aps,twocolumn,superscriptaddress,longbibliography,showkeys]{revtex4-2}

\usepackage{graphicx}
\usepackage{amsmath,amsfonts,amssymb}
\usepackage{booktabs}
\usepackage[colorlinks=true,allcolors=blue]{hyperref}

\begin{document}

\title{Low energy excitations in a long prism geometry:\\
testing the lower critical dimension of the Ising spin glass}

\author{M.~Bernaschi}
\affiliation{Istituto per le Applicazioni del Calcolo, CNR - Via dei Taurini 19, 00185 Rome, Italy}

\author{L.A.~Fern\'{a}ndez}
\affiliation{Departamento de F\'isica T\'eorica, Universidad Complutense de Madrid, 28040 Madrid, Spain}

\author{I.~Gonz\'{a}lez-Adalid Pemart\'{i}n}
\email{i.gonzalez@iac.cnr.it}
\affiliation{Istituto per le Applicazioni del Calcolo, CNR - Via dei Taurini 19, 00185 Rome, Italy}

\author{V.~Mart\'{i}n-Mayor}
\affiliation{Departamento de F\'isica T\'eorica, Universidad Complutense de Madrid, 28040 Madrid, Spain}

\author{G.~Parisi}
\affiliation{International Research Center of Complexity Sciences, Hangzhou International Innovation Institute, Beihang University, Hangzhou 311115, China}
\affiliation{Dipartimento di Fisica, Sapienza Universit\`a di Roma, and CNR-Nanotec, Rome Unit, and INFN, Sezione di Roma, 00185 Rome, Italy}

\author{F.~Ricci-Tersenghi}
\affiliation{Dipartimento di Fisica, Sapienza Universit\`a di Roma, and CNR-Nanotec, Rome Unit, and INFN, Sezione di Roma, 00185 Rome, Italy}

\begin{abstract}
We propose a general method for studying systems that display low-energy excitations in their low-temperature phase. We argue that in a rectangular right-prism geometry (the 3-dimensional generalization of a strip), with one longitudinal size much larger than the transverse size, correlations decay exponentially (at all temperatures) along the longitudinal direction. Still, the scaling of the correlation length with the transverse size carries crucial information from which the lower critical dimension can be inferred. The method is applied in the particularly demanding context of Ising spin glasses at zero magnetic field. The lower critical dimension and the multifractal spectrum for the correlation functions are computed from large-scale numerical simulations. Several technical novelties (such as the unexpectedly crucial performance of Houdayer's cluster method or the convenience of using open---rather than periodic---boundary conditions) allow us to study three-dimensional prisms with transverse dimensions up to $L=24$ and effectively infinite longitudinal dimensions, down to low temperatures. The value that we find for the exponent controlling the behavior of the correlation length agrees with predictions from the Replica Symmetry Breaking (RSB) theory. We argue that our novel setting holds promise in clarifying which of the two competing theories, RSB or the Droplet Model, more accurately describes three-dimensional spin glasses.
\end{abstract}


\maketitle

Excitations of arbitrarily low energy due to some form of spontaneous symmetry breaking that generates massless Goldstone modes are a fairly standard feature in many-body physics and in quantum and/or statistical field theory (crucially, the broken symmetry must be \emph{continuous}; see, e.g., Refs.~\cite{parisi:88, zinn-justin:05}). When present, these low-energy excitations have profound physical implications (think, for instance,  of Debye's specific heat for solids, or of the Higgs mechanism responsible for the Meissner effect in superconductors~\cite{anderson:63} and the generation of masses in particle physics~\cite{higgs:64}) and help us to find far-reaching theoretical statements of broad applicability (the  Mermin-Wagner-Hohenberg-Coleman theorem provides a celebrated example). However, the physical origin of the low-energy excitations is not always clear, and the magnetically disordered alloys known as spin glasses~\cite{mydosh:93,charbonneau:23,dahlberg:25} provide a particularly controversial example.

In Ising spin glasses, the only evident symmetry is \emph{discrete}, namely the global spin-reversal symmetry [see \eqref{eq:H-3D} below], which raises the question of the origin of the soft excitations in these systems. In fact, these Goldstone-like modes were suggested long ago within the mean-field approach by Thouless, Anderson, and Palmer~\cite{thouless:77, anderson:84}. Their nature was later clarified by de Dominicis and Kondor using perturbative field theory~\cite{dedominicis:84}. Physically, these soft excitations arise from nearly degenerate free-energy minima that are structurally very similar and exchange dominance under tiny perturbations (a manifestation of the marginal stability of the equilibrium state~\cite{parisi:23}).
Unfortunately, it has not yet been possible to extend the analysis of de Dominicis and Kondor beyond perturbation theory. The goal of the present work is to study correlations in a different regime:  we found quantities that can be computed both numerically and theoretically.

Rather than the standard cubic geometry, we choose a rectangular right prism in $D$ spatial dimensions of size $L^{D-1}M$, where $M$ is the prism length along the longitudinal $D$-th dimension, and $L$ is the size along the other $D-1$ transverse dimensions~\footnote{This is one of the geometries employed by Brezin in his investigation of Finite Size Scaling at the critical temperature $T_\text{c}$~\cite{brezin:82}.} ($M$ will eventually be sent to $\infty$ at fixed $L$, see Fig.~\ref{fig:clusters}). The prism geometry was used in careful studies of spin glasses at zero temperature in $D=2$~\cite{cheung:83,carter:02,fisch:08}, but the lack of the spin glass phase in $D=2$ prevented the study of soft-excitations in the low-temperature phase.

Consider now a non-zero temperature below the critical one, $0<T<T_\text{c}$, and impose mutually incongruent boundary conditions at the end planes of the prism along the longitudinal direction, $x_D=0$ and $x_D=M-1$. In the case of a ferromagnetic model, for instance, we could impose that the spins at $x_D=0$ and those at $x_D=M-1$ align in opposite directions. The twist causes an excess of free energy $\Delta F$ that scales with both system sizes as
\begin{equation}\label{eq:scaling-1}
\Delta F \sim L^{D-1}/M^b\,,
\end{equation}
where the exponent $b$ is problem-specific but does not depend on $D$ or $T$. In particular, provided $D>b+1$, one has $b=0$ for the Ising ferromagnet and $b=1$ for the Heisenberg ferromagnet~\footnote{In a Heisenberg ferromagnet, the boundary twist $\Delta \varphi$ can be equally distributed among the planes, giving an excess free energy of order $O([\Delta\varphi/(M-1)]^2 L^{D-1})$ per consecutive plane pair. The sum of $M-1$ such terms gives the scaling in \eqref{eq:scaling-1} (writing $M$ or $M-1$ is irrelevant).}.

As explained in Materials and Methods, \eqref{eq:scaling-1} determines: (i) the lower critical dimension, $D_{\text{lc}}$ (the low-temperature phase disappears for $D\leq D_{\text{lc}}$, since the critical temperature approaches $T_\text{c}=0$ as $D\to D_{\text{lc}}^+$), and (ii) the correlation length $\xi(L)$ governing the exponential decay of correlations along the principal axis of the prism:
\begin{equation}\label{eq:scaling-2}
\xi(L) \propto L^{a_D}\,,\quad a_D= \frac{D-1}{b} \,\,,\quad D_\text{lc}=1+\frac{D-1}{a_D}.
\end{equation}
From $b^{\text{Heis.}}\!=\!1$ for Heisenberg ferromagnets we get $a_D^{\text{Heis.}}\!=\!D-1$ and $D_{\text{lc}}^{\text{Heis.}}\!=\!2$. In the Ising case, $b\!=\!0$, soft excitations are absent and $\log\xi(L)\propto L^{D-1}$, see the Supporting Information ({\bf SI}).

The physical reason why~\eqref{eq:scaling-1} applies to Ising spin glasses is marginal stability: the existence of many states allows the system to deform gradually from one boundary condition to the other incongruent one, spreading the free-energy cost over the entire length of the prism. This argument can be made fully quantitative using replicas and mean-field theory~\cite{franz:94}, which yields $b=3/2$ and $D_\text{lc}=5/2$ via \eqref{eq:scaling-2}. Unfortunately, simulations at $D=5/2$ are not possible, so alternative ways of testing the theory directly in dimensions $D\geq 3$ are needed.

Direct measurements of $\Delta F$ have been made in a cubic geometry~\cite{maiorano:18}. Yet, the limit $M\gg L$ is difficult to study because $\Delta F\to 0$ in the limit $M\to\infty$. Here, we circumvent this difficulty by computing the correlation length in prism geometry. A clear roadmap opens up. We shall specialize to $D=3$ and obtain $\xi(L)$ in prisms of increasing transverse sizes $L$. The scaling of $\xi(L)$ in \eqref{eq:scaling-2} gives the exponent $a_\text{3}$.  The predictions of the Mean Field Theory (MFT) are $b^\text{\tiny MFT}=3/2$, $a^\text{\tiny MFT}_\text{3}=4/3$, and $D^\text{\tiny MFT}_\text{lc}=5/2$~\cite{franz:94}, which are in nice numerical agreement with the results of our Monte Carlo simulations.

Finally, let us recall the prediction for $\xi(L)$ of the competing Droplet Model (DM), which is very similar (even numerically) to that of MFT. One simply needs to substitute $a_D=1+y_D$~\cite{fisher:88b,carter:02} into \eqref{eq:scaling-2}, where the stiffness exponent in $D=3$ is $y_3=0.24(1)$~\cite{palassini:99b,hartmann:99c,boettcher:04}. Within the DM, this scaling does not rely on marginal stability. Instead, the prism is mapped onto a one-dimensional spin chain with effective random couplings that have a small probability of taking very small values (see {\bf SI}) . Surprisingly, a remarkable feature of this one-dimensional spin chain appears to have gone unnoticed: multifractal scaling~\cite{benzi:84,frisch:85,castellani:86,halsey:86,halsey:86b,harte:01} (our derivation is presented in {\bf SI}, while a quantitative assessment is provided in Materials and Methods).

\section*{The Ising spin glass in a long prism geometry.}
Ising spins $S_{\boldsymbol{x}}=\pm 1$ occupy the nodes $\boldsymbol{x}$ of a cubic lattice, which form a rectangular right prism of size $L$ along the $X$ and $Y$ directions and size $M$ ($\gg L$) along the $Z$ direction. The spins interact with their lattice nearest neighbors through the Hamiltonian~\cite{edwards:75,edwards:76}
\begin{equation}\label{eq:H-3D}
H^{\text{3D}}=-\sum_{\langle\boldsymbol{x},\boldsymbol{y}\rangle}\, J_{\boldsymbol{x},\boldsymbol{y}} S_{\boldsymbol{x}}S_{\boldsymbol{y}}\,,
\end{equation}
We choose the usual periodic boundary conditions in the $X$ and $Y$ directions (the $Z$ direction warrants a more careful discussion; see below).
The couplings, $J_{\boldsymbol{x},\boldsymbol{y}}=\pm 1$ with equal probability, are independent and identically distributed random variables. The disorder is quenched: for any observable we first compute the thermal average, $\langle\ldots\rangle $, over different real replicas (copies of the system with the same couplings evolving independently) and only afterward we compute the average over the $J$s, $\overline{(\ldots)}$. As usual, we compute correlation functions using different replicas. For every plane $x_3=z$, we define a local overlap $Q(z)=\sum_{x_1,x_2} S_{x_1,x_2,z}^{(a)} S_{x_1,x_2,z}^{(b)}/L^2$ where $a\neq b$ are replica indices. Our basic equilibrium correlation functions are
\begin{equation}\label{eq:eq-C3D}
C^{(n)} (z_1,z_2)= \overline{\langle Q(z_1) Q(z_2)\rangle^n}\,.
\end{equation}
Considering different powers $n$ in $C^{(n)}(z_1,z_2)$ is motivated by the recent finding of multifractal scaling in out-of-equilibrium spin glass dynamics~\cite{janus:24} \footnote{However, the multifractal spectrum that we compute here cannot be directly compared with the results of~\cite{janus:24}, because of the different geometry. The importance of geometry was emphasized in~\cite{marinari:24}.} (see also~\cite{fisch:08} in $D=2$).
$C^{(n)}$ is a function of $|z_2-z_1|$ for $M\to\infty$, but for finite $M$ and open boundary conditions the dependence on $z_1$ and $z_2$ should be dealt with care, see {\bf SI}.

\section*{Two peculiarities of our numerical simulations.}
We aim to effectively reach the $M\to\infty$ limit on prisms with increasingly large transverse size $L$ to extract the correlation length $\xi_{n=1}$ from the correlation functions $C^{(1)}(z_1,z_2)$, see \eqref{eq:eq-C3D}. In the following, we will use $\xi$ as a shorthand for $\xi_{n=1}$ whenever no ambiguity arises, and $\xi(L)$ will be used in the scaling analysis of \eqref{eq:scaling-2}. The limit $M,L\gg1$ is hard to achieve because thermalizing a large spin glass model at $T<T_{\text{c}}$ is computationally very demanding~\cite{billoire:18}, even with the help of dedicated hardware~\cite{janus:10,janus:14} and optimized algorithms~\cite{hukushima:96}. Two ingredients have been invaluable in this task.

\begin{figure}[t]
    \centering\includegraphics[width=\linewidth]{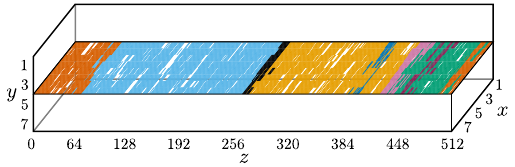}
    \caption{The eight largest clusters for a typical configuration on a lattice $8\times8\times512$ with periodic boundary conditions at $T=0.7$ are depicted at $y=4$. There is no percolation through the lattice along the Z direction. The Z scale has been reduced by a factor of 10 in order to improve visibility.}
    \label{fig:clusters}
\end{figure}
First, Houdayer's cluster move~\cite{houdayer:01}, very successful in 2D~\cite{fernandez:16b} but not so much for cubic systems~\cite{zhu:15,chilin:26b}, turns out to be very helpful in a prism geometry because, at low $T$, the clusters do not percolate along the $Z$ direction, see Fig.~\ref{fig:clusters}, which has resulted in a three-order-of-magnitude speed-up for large $M$, see {\bf SI}.

Second, we use open boundary conditions (OBC) along the Z direction. Indeed, we were surprised to find that, when using periodic BC (PBC), a size ratio as large as $M^\text{\tiny PBC}/\xi\approx 11$ was needed to reach the limit $M\to\infty$ (for the Heisenberg ferromagnet, $M^\text{\tiny PBC}/\xi=6$ suffices).  We show in the Materials and Methods that OBC simulations are free from this pathology (see {\bf SI} for a deeper discussion).

In this way, and using highly tuned GPU codes, we have been able to equilibrate 2000 samples (or more) for prisms of transverse dimensions $L=4,6,8,12,16$ and $24$, effectively reaching the large-$M$ limit at a temperature $T=0.7\approx 0.63 T_{\text{c}}$~\cite{janus:13}. For more details on our simulations and data analysis, see {\bf SI}.

\begin{figure}[t]
    \centering\includegraphics[width=\linewidth,height=0.6\linewidth]{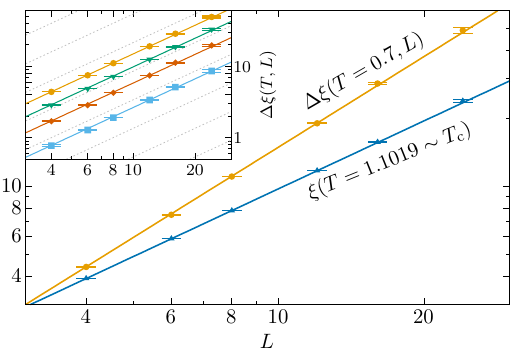}
    \caption{Correlation length at the critical point, $\xi(T_\text{c},L)$, and the difference $\Delta\xi(T,L)=\xi(T,L)-\xi(T_\text{c},L)$ at $T=0.7\simeq 0.635\, T_\text{c}$ vs.~the prism transverse size $L$.
    \textbf{Inset:} $\Delta\xi(T,L)$ vs.~$L$ for $T=0.7, 0.8, 0.9$ and 1 (from top to bottom) in double logarithmic scale. Solid lines are fits, see Eqs.~\eqref{eq:Rxi} and~\eqref{eq:U(T)-fit}. Dotted lines are guide-to-eyes parallel to $L^{4/3}$.}
    \label{fig:xi_vs_L}
\end{figure}

\begin{figure*}[t]
    \centering\includegraphics[width=\linewidth]{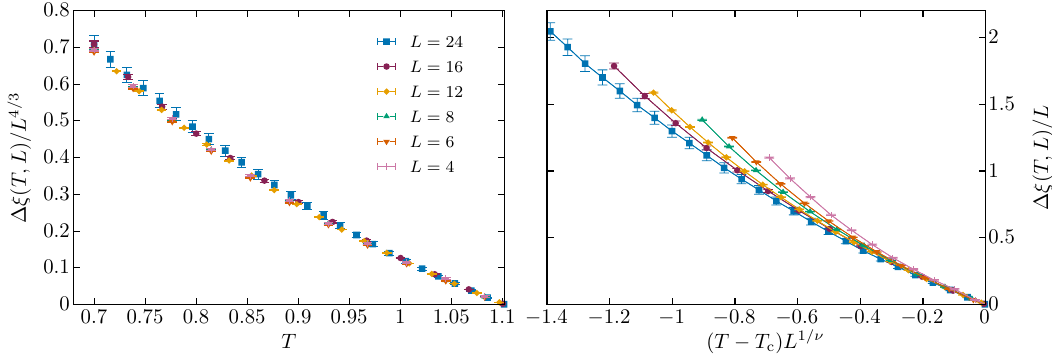}
    \caption{\textbf{Left:} the good collapse of $\Delta\xi(T,L)/L^{4/3}$ vs.~$T$ for all the transverse sizes $L$ supports the mean-field theory prediction. \textbf{Right}: standard critical finite size scaling, $\Delta\xi(T, L)/L$ vs.~$(T-T_\text{c})L^{1/\nu}$ (with $T_\text{c}$ and $\nu$ values from~\cite{janus:13}) fails for $T<T_\text{c}$. Lines joining data are just guides for the eyes.}
    \label{fig:temperature_crossover}
\end{figure*}
\section*{Temperature crossover: from \boldmath$T_\text{c}$ to low \boldmath$T$.}
Figure~\ref{fig:xi_vs_L} shows that, at the critical point $T_\text{c}=1.1019(29)$~\cite{janus:13}, $\xi$ grows linearly with $L$, as expected~\cite{brezin:82}. However, as predicted by \eqref{eq:scaling-2}, at $T=0.7$ the growth of $\xi$ with $L$ is superlinear. To highlight this behavior, we focus on
$\Delta\xi(T,L)\equiv \xi(T,L)-\xi(T_\text{c},L)$ \footnote{The subtraction $\xi(T,L)-\xi(T_\text{c},L)$ highlights the differences from critical scaling. Since $a_3>1$, see \eqref{eq:scaling-2}, $\xi(T_\text{c},L)$ is subdominant compared with $\xi(T,L)$ at any fixed $T<T_\text{c}$. An additional benefit is that the subtraction largely suppresses scaling corrections.}, which turns out to be perfectly compatible with the MFT prediction $\Delta\xi(T,L) \propto L^{4/3}$ for any $T<T_\text{c}$ (see inset in Fig.~\ref{fig:xi_vs_L}). How are these two scaling behaviors connected?

On the one hand, MFT predicts that, for $T<T_\text{c}$ and $L\to\infty$, $\xi(T,L)/L=L^{a_3-1}u(T),$ with $u(T_\text{c}^-)=0$. Let us assume that $u(T)\propto (T_\text{c}-T)^{p_D}$ as $T\to T_\text{c}^-$. On the other hand, if $|T-T_\text{c}|$ is sufficiently small, standard finite-size scaling (FSS) predicts~\cite{barber:83} $\Delta\xi(T,L)/L=f(x)$, where $f$ is an analytic function of $x=L^{1/\nu}(T-T_\text{c})$. The two asymptotic expansions differ in the quantity that is held fixed as $L$ grows: either $x$ (FSS) or $T$ (the low-temperature regime). Following Ref.~\cite{barber:83}, we note that the two expansions match if $f(x\to-\infty)\sim (-x)^{p_D}$ and
\begin{equation}\label{eq:matching}
p_D=(a_D-1)\nu\,.
\end{equation}
The two scalings therefore coincide whenever $p_D=1$, and disentangling FSS from genuine low-temperature behavior may be difficult when $p_D\approx 1$.
This is not uncommon:
(i) for both the Ising~\cite{wallace:79} and Heisenberg~\cite{brezin:76} ferromagnets, $1/\nu_D=a_D-1+O(\epsilon^2)$, where $\epsilon=D-D_\text{lc}$ (in the limit of an infinite number of spin components, $1/\nu_D=a_D-1$ exactly; see, e.g.,~\cite{zinn-justin:05});
(ii) for Ising spin glasses, we obtain $p_3=0.854(14)$, $p_4=1.068(7)$, and $p_5=1.215(18)$ by inserting into \eqref{eq:matching} the values of $a_D$ obtained from \eqref{eq:scaling-2} with $b=3/2$, together with the estimates of $\nu_D$ from Refs.~\cite{janus:13,banos:12,aguilar-janita:25}.

\section*{Numerical results.}
The value $p_3\neq 1$ prompts us to disentangle FSS from genuine low-temperature scaling behavior, which we achieve in Fig.~\ref{fig:temperature_crossover}.
The right panel shows that FSS works fine only close to $T_\text{c}$, but fails to describe the data for $T<T_\text{c}$.
On the other hand, the good data collapse in the left panel supports the scaling predicted by the MFT, $\Delta \xi(T,L)\propto L^{4/3}$, at any fixed $T<T_\text{c}$.
Note that Fig.~\ref{fig:temperature_crossover} contains no adjustable parameters, as the values for $T_\text{c}$ and $\nu$ are taken from~\cite{janus:13}.

Let us make the analysis more quantitative.
At $T_\text{c}$, the ratio $\xi(T_\text{c},L)/L$, see Fig.~\ref{fig:xi_vs_L} and Table~\ref{tab:xi}, converges to a (universal) $L$-independent value. Fitting data with $L\geq L_\text{min}=4$ we get ($\chi^2/\text{dof}=3.21/5$, p-value=0.67):
\begin{equation}\label{eq:Rxi}
    R_\xi\equiv \lim_{L\to\infty} \left[\frac{\xi(T_\text{c},L)}{L}\right]=0.9776(10)\,,
\end{equation}
As expected~\cite{cardy:84b,burkhardt:85}, $R_\xi$ is geometry-dependent. For instance, in a cubic lattice $R_\xi=0.6513(32)$~\cite{janus:13}.

We explore two possible analyses for $T<T_\text{c}$: first, we assess the statistical compatibility of data with the MFT; afterward, we compute the exponent $a_3$.
To test the MFT predictions, data at a fixed $T$ in Fig.~\ref{fig:xi_vs_L} are fit to a $L$-independent $u(T)=\Delta\xi(T,L)/L^{4/3}$.
To obtain fits of good quality, we have to take $L_\text{min}=12$:
\footnote{The p-values from the $\chi^2$ tests with two degrees of freedom are $0.19\,(T\!=\!0.7)$, $0.18\,(T\!=\!0.8),$ $0.05\, (T\!=\!0.9),$ and $0.11\,(T\!=\!1)$. These four $\chi^2$ tests are not statistically independent, because data at different $T$ come from the same simulation.
Although we have not computed $u(T)$ close enough to $T_\text{c}$ to perform a stringent test of \eqref{eq:matching}, these $u(T\geq 0.8)$ can be fitted to
$u(T)=(T_\text{c}-T)^{p_3}[h_1+h_2(T_\text{c}-T)]$ with $p_3=0.854$ ($\chi^2/\text{dof}=0.95/1$, p-value=0.33, the fit parameters being $h_1$ and $h_2$).}
\begin{equation}\label{eq:U(T)-fit}
\begin{array}{ll}
 u(0.7)=0.694(2),&u(0.8)=0.4573(15),\\
 u(0.9)=0.2717(10),&u(1.0)=0.1236(7).
\end{array}
\end{equation}

Alternatively, we estimate the exponent $a_3$ from the fit $\Delta \xi(0.7,L) =B L^{a_3}$ with fit parameters $B$ and $a_3$. We use the data at the lowest temperature to minimize contamination from $T_ \text{c}$.
As the reader may check from Table~\ref{tab:xi}, the fit with $L_\text{min}=6$ is good, $a_3=1.349(6)$ ($\chi^2/\text{dof}=2.34/3$, p-value=0.50), and compatible with the MFT prediction (only 2.6 standard deviations away).
For larger $L_\text{min}$ the error grows [e.g., for the largest possible $L_\text{min}=12: \ a_3=1.39(3)$, $\chi^2/\text{dof}=0.49/1$, p-value=0.49]. Hence, our final estimate
takes the central value from the fit with $L_\text{min}=6$ and, to account for systematic errors, the error from the fit with $L_\text{min}=12$:
\begin{equation}\label{eq:Dlc}
a_3=1.35(3),\quad D_\text{lc}=2.48(3)\,.
\end{equation}
For comparison, we recall the predictions by the MFT, $a^\text{\tiny MFT}_3=4/3\simeq 1.33$, and by the DM, $a^\text{\tiny DM}_3=1.24(1)$.

\begin{figure}[t]
\centering\includegraphics[width=\linewidth]{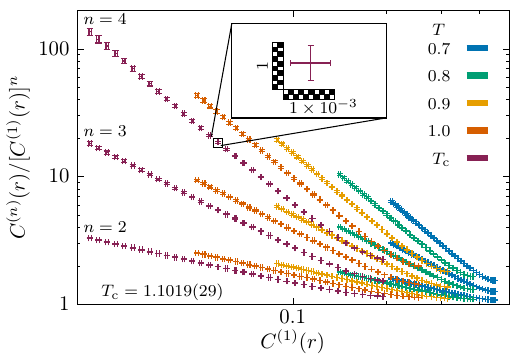}
    \caption{Ratio of the correlation functions $C^{(n)}/[C^{(1)}]^n$ versus $C^{(1)}$, measured for $L=16$ at several temperatures $T\le T_\text{c}$, and plotted parametrically increasing $r$ from right to left. For $C^{(1)}\ll 1$ (large $r$), the slope equals $\tau_n-n$ where $\tau_n=\xi/\xi_n$ is the so-called multifractal spectrum (whose values are in Table~\ref{tab:multifractal-spectrum}). The negative slopes indicate that $\tau_n<n$, hence $C^{(n)}\gg [C^{(1)}]^n$ at long distances (multifractal scaling).}
    \label{fig:Multifractal}
\end{figure}

Fig.~\ref{fig:Multifractal} reports the evidence for multifractal scaling in the correlation functions. Assuming $C^{(n)}(r)\sim e^{-r/\xi_n}$ for $r\to\infty$, the negative slopes in the data in Fig.~\ref{fig:Multifractal} imply $1/\xi_n<n/\xi$, that is $C^{(n)}\gg [C^{(1)}]^n$ at long distances. This is clear evidence for multifractal scaling at equilibrium in 3D spin glasses; previous results were obtained either in the out-of-equilibrium regime~\cite{janus:24} or in the 2D case where the spin glass phase is absent~\cite{fisch:08}.

\section*{Discussion.}
 We have shown how to exploit the elongated prism geometry to study soft excitations in the low-temperature phase of disordered systems.
In close analogy to path-integral Quantum Monte Carlo, tiny free energies that are difficult to compute in cubic geometry~\cite{maiorano:18} translate in the prism geometry into large correlation lengths that allow for a reliable scaling analysis: $\xi(L) \propto L^{a_3}$.
Our best estimate $a_3=1.35(3)$ is in perfect agreement with the MFT prediction $a^\text{\tiny MFT}_3=4/3$ and implies a lower critical dimension $D_\text{lc}=2.48(3)$, which is compatible with the known results in the literature.
The existence of a continuous phase transition at $T_\text{c}>0$ in $D=3$ \cite{gunnarsson:91,palassini:99,ballesteros:00} and the power-law scaling in $T$ of the correlation length in $D=2$ \cite{fernandez:16b} prove that the lower critical dimension is in the range $2\!<\!D_\text{lc}\!<\!3$ (confirmed by studies in a film geometry \cite{guchhait:14,fernandez:19b}).
$D_\text{lc}$ was previously calculated by varying $D$ at $T=0$, finding $D_\text{lc}\approx 2.4986$ \cite{boettcher:05}, or by varying $D$ at $T_\text{c}$, finding $D_\text{lc}\approx 2.43(3)$ \cite{aguilar-janita:25}.
Both determinations, particularly the one at $T=0$, are in good agreement with our estimate, which is the first one obtained at $D=3$ and $0<T<T_\text{c}$.
It is worth noticing that the scaling $\Delta\xi(L) \propto L^{4/3}$ holds for any $T<T_\text{c}$, making our analysis and results very robust.

We have computed dimensionless quantities, such as $R_\xi$ and the ratios of correlation lengths $\tau_n$ that provide the multifractal spectrum at and below $T_\text{c}$ (see Materials and Methods and {\bf SI} for more details).
This is clear evidence for multifractality in equilibrium correlation functions measured in a spin glass phase.
No $T$ dependence has been identified when $T<T_\text{c}$, as in the out-of-equilibrium case \cite{janus:24}.
Computing this multifractal spectrum analytically is a new challenge to the theory.
The simple $D=1$ model discussed in {\bf SI} leads us to conjecture that multifractality is generic in disordered systems supporting Goldstone excitations in a prism geometry. It also suggests the DM needs to be improved to reproduce the observed spectrum.

We stress the success of Houdayer's cluster method~\cite{houdayer:01} in a prism geometry. The dynamic speed-up of a factor $\sim 10^3$ is the first real success of a cluster method in a $D=3$ spin glass. It allowed to equilibrate a system containing $16\times16\times 512=2^{17}$ spins down to $T \approx 0.63 T_{\text{c}}$. The previous world record at this temperature was $2^{15}$ spins, achieved only with dedicated hardware~\cite{janus:10b}.

Extending these computations to higher dimensions may give crucial information on the proper spin glass theory: for example, RSB predicts that the exponent $b$ does not depend on the space dimension $D$, while the opposite is true for the DM.

We also envisage several future applications for the prism geometry in $D=3$. For instance, it could help us to clarify the controversy regarding the existence of chiral-decoupling in Heisenberg spin-glasses~\cite{kawamura:92,hukushima:00,kawamura:15,ogawa:20}, or lack thereof~\cite{lee:03,fernandez:09b,nakamura:19}. Other applications envisaged beyond the spin-glass realm include the Random Field Ising Model~\cite{nattermann:98} and the KPZ equation~\cite{kardar:86}.

\section*{Materials and Methods}

\subsection*{From Eq.~(\ref{eq:scaling-1}) to Eq.~(\ref{eq:scaling-2})} To compute  $D_{\text{lc}}$  let us set $M=L$. A low-temperature ordered phase occurs only if $\Delta F\sim L^{D-1-b}$ diverges for large $L$, which requires $D>D_{\text{lc}}=1+b$.
If we take the longitudinal size $M\to\infty$ instead, Wilson's Renormalization Group (RG)~\cite{wilson:75,parisi:88} suggests that the long-distance behavior is ruled by an effective one-dimensional Hamiltonian (the precise form of the effective Hamiltonian is a delicate issue). One-dimensionality implies that the correlations decay as $\text{e}^{-x_D/\xi(L)}$.
To estimate $\xi(L)$ as a function of $L$, we simply put $M=\xi(L)$ in \eqref{eq:scaling-1} and require $\Delta F\sim L^0$ [recall that the probability that the order parameter differs significantly in the two planes $x_D=n$ and  $x_D=n+d$ is $\text{e}^{-\Delta F(L, M=d)}$].

\subsection*{On the use of Open Boundary conditions} 
\begin{figure}[h]
\centering\includegraphics[width=\linewidth]{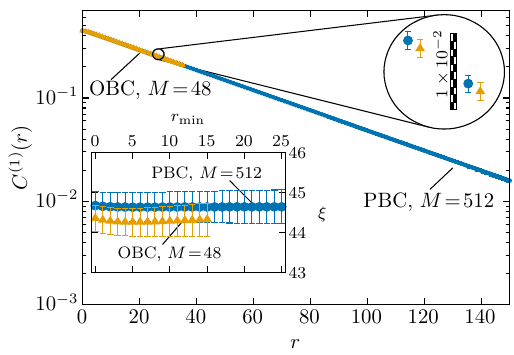}
    \caption{A comparison of the correlation functions $C^{(1)}(r)$ obtained on a lattice $16\times16\times512$ with PBC and on a lattice $16\times16\times48$ with OBC shows full compatibility within the errors. {\bf Inset:} $\xi$ as a function of the minimal distance considered $r_\text{min}$ in the analysis, see {\bf SI}, comes out compatible with both boundary conditions.}
    \label{fig:C_PBC_vs_OBC}
\end{figure}

As Fig.~\ref{fig:C_PBC_vs_OBC} shows, our OBC simulations at a modest size ratio $M^\text{\tiny OBC}/\xi\approx 1.1$ have turned out to provide $\xi$ with a similar accuracy to the much more expensive PBC simulations with $M^{\text{\tiny PBC}}/\xi\approx 11$ (unfortunately, for modest $M/\xi$ the Houdayer's clusters percolate and the cluster move is not quite as effective).

\subsection*{On sub-leading correlation lengths} We have estimated correlation lengths with high accuracy (about 2\% for our largest system; see Table~\ref{tab:xi}). This is quite a serious numerical challenge, because one should be confident that subdominant contributions to the correlation functions are under reasonable control. Here, we provide a first discussion of this problem, which will be continued in the {\bf SI}.

\begin{table}[htb]
    \centering
    \caption{\boldmath$\xi(T,L)$ and $\Delta\xi(T,L)$ for the different system sizes, $L^2\times M$, boundary conditions along the Z-axis (BC-Z), and temperatures. Data are plotted in Fig.~\ref{fig:xi_vs_L}.}
    \label{tab:xi}
    \begin{tabular}{cccccc}
        $L$ & $M$ & BC-Z & $\xi(T=0.7)$ & $\Delta\xi(T=0.7)$ & $\xi(T\simeq T_\mathrm{c})$ \\
        \midrule
         $4$ & $192$ & PBC & 8.316(15) & 4.402(18) & 3.914(16)\\
         $6$ & $192$ & PBC & 13.363(24) & 7.492(22) & 5.871(12)\\
         $8$ & $192$ & PBC & 18.87(3) & 11.06(3) & 7.807(13)\\
         $12$ & $320$ & PBC & 30.79(6) & 19.04(6) & 11.742(23)\\
         $16$ & $512$ & PBC & 44.6(4) & 28.7(4) & 15.96(14)\\
         $16$ & $48$ & OBC & 44.3(4) & 28.6(4) & 15.70(10)\\
         $24$ & $88$ & OBC & 73.0(16) & 49.1(16) & 23.9(4)\\
         \bottomrule
   \end{tabular}
\end{table}
Up to this point, our analysis has focused on the behavior of the correlation functions $C^{(n)}(z_1,z_2)$  when the plane-to-plane separation $r=|z_2-z_1|$ is very large. However, the behavior at small $r$ is also interesting, given that in this region subdominant terms provide a non-negligible contribution and can thus be reliably estimated.

To model our expectations, we go back to the replica trick~\cite{edwards:75} that restores translational invariance at the price of considering a number $\hat{n}$ of replicated systems that are mutually interacting. At the end of the computation, the limit $\hat{n}\to 0$ should be taken, which is not only counterintuitive, but is also far from trivial~\cite{mezard:87,charbonneau:23}. The advantage is that the recovery of translation invariance makes it natural to use standard tools such as the transfer matrix $\mathcal{T}$~\cite{parisi:88,kogut:79} (see Ref.~\cite{lucibello:14} for applications of the replicated transfer matrix to disordered systems).

For simplicity, let us focus on the correlation function $C^{(1)}$. The replicated system has a transfer matrix $\mathcal{T}$ with eigenvectors $|k\rangle$ and eigenvalues $\text{e}^{-E_k}$, with $E_0=0<E_1^\text{odd}<E_2^{\text{odd}}<\ldots$. The eigenvalue corresponding to the ground state, $E_0=0$, is non-degenerate due to the one-dimensional geometry of the prism, which precludes any spontaneous symmetry breaking. The superscript ``odd" emphasizes that---apart from the ground state, which is of even parity---only eigenvectors of odd parity with respect to global spin-reversal must be considered.
Hence, in the limit $M\to\infty$, the correlation function at $r=|z_2-z_1|$ can be written as~\cite{parisi:88}
\begin{equation}\label{EMeq:transfer}
C^{(1)}(r)=\sum_{k=1}^{\infty}\, |\langle 0| \hat Q|k\rangle|^2\,\text{e}^{-E_k^\text{odd}\, r}\,,
\end{equation}
where $\hat Q$ is the operator that represents the plane-overlap $Q(z)$. We immediately identify the correlation length $\xi=1/E_1^{\text{odd}}$ from the leading term in the above expansion. However, the asymptotic exponential decay is accompanied by sub-leading exponential terms $\text{e}^{-r/\xi_{n=1;k}}$, with $\xi_{n=1;k}=1/E^{\text{odd}}_k < \xi$. Of particular interest are those contributions whose ratio $\xi_{n=1;k}/\xi$ remains non-zero in the large $L$ limit---the so called scaling limit where $\xi$ diverges, recall \eqref{eq:scaling-2}---because they would indicate that $\hat Q$ produces further soft excitations in the system.

\begin{figure}[t]
    \centering
    \includegraphics[width=\linewidth]{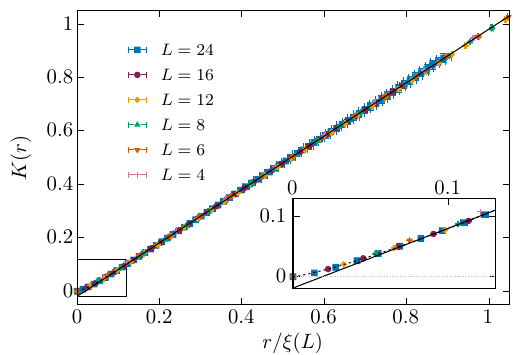}
    \caption{Scaling of the logarithm of the (normalized) correlation function $C^{(1)}(r)$, see \eqref{EMeq:K-def}. Data for systems of size $L=4,6,8,12,16$, and $24$ at temperature $T=0.7$. The black line is the best fit to $K(r)=(r/\xi)+d$ for the $L=16$ data in the range $0.05 < r/\xi < 1.1$. {\bf Inset:} zoom in near the origin. The dotted line is the best fit to $K(r)=A(r/\xi)^a$ for the $L=16$ data in the range $r/\xi < 0.15$ (with exponent $a\approx 1.23$) and interpolates well data for all $L$ values.}
    \label{fig:scaling_one_length}
\end{figure}

To assess whether or not this scenario is realized for Ising spin glass models in the elongated prism geometry, we represent in Fig.~\ref{fig:scaling_one_length} the following quantity
\begin{equation}\label{EMeq:K-def}
    K(r)=-\log \frac{C^{(1)}(r)}{C^{(1)}(r=0)}\,.
\end{equation}
as a function of $r/\xi$ (the values of $\xi$ are in Table~\ref{tab:xi}). In order to reduce any residual problem with translation invariance in OBC prisms, see {\bf SI}, we actually use the normalized ratio $C^{(1)}(z_1,z_2)/\sqrt{C^{(1)}(z_1,z_1)C^{(1)}(z_2,z_2)}$ in Eq.~\ref{EMeq:K-def} averaged over pairs of planes with $z_2-z_1=r$. If the leading term in Eq.~\ref{EMeq:transfer} were the only contribution surviving in the large $L$ limit, one would have exactly $K(r)=r/\xi$ in that limit.

The first important observation about the data in Fig.~\ref{fig:scaling_one_length} is their good scaling.
Interestingly enough, when the $K(r;L)$ functions computed for different transverse sizes $L$ are plotted versus $r/\xi(L)$, they all fall onto a single master curve, $K(r;L)=\widehat{K}[r/\xi(L)]$.
The scaling is excellent, even close to the origin (see the inset in Fig.~\ref{fig:scaling_one_length}) down to our resolution of order $1/\xi(L=24)$.
Indeed, the dotted line in the inset of Fig.~\ref{fig:scaling_one_length}, obtained via a fit to the $L=16$ data, perfectly interpolates the $L=24$ data points too.
This dependence on a single length scale $\xi$ strongly suggests that all the sub-leading correlation lengths $\xi_{n=1;k>1}$ (or at least all those that we can resolve within the limit of our statistical accuracy) scale proportionally to $\xi$, and the corresponding ratios $\xi_{n=1;k>1}/\xi$ reach a non-zero limit for large $L$.
Let us stress that the fit in Fig.~\ref{fig:scaling_one_length} should be regarded as an illustration of the \emph{single-length scaling} property of the data. A more quantitative analysis is presented in {\bf SI}.

The second observation is that the linear behavior of the scaling function $\widehat{K}[r/\xi(L)]$, represented by the solid black line in Fig.~\ref{fig:scaling_one_length}, breaks down for $r/\xi(L)\lesssim 0.07$. This is the length scale where subdominant contributions to the leading behavior become relevant. And it is just a few percent of the leading correlation length, $\xi$, indicating a large length-scale separation.

As a concluding remark, let us note that the scaling of the sub-dominant length scales $\xi_{n=1;k>1}$ tells us that the 1D toy model is not the right effective model for the prism, because those length scales are missing in such a 1D toy model (see {\bf SI}).

\begin{table}[hb]
    \centering
    \caption{\boldmath$\tau_n=\xi_{n=1}/\xi_n$ for the two OBC systems, $L^2\times M$ with $M=48$ ($L=16$) and $M=88$ ($L=24$), at several temperatures and at the critical point, read off at $r_\mathrm{min}=8$ as in Table~\ref{tab:xi}. The last row is the one-dimensional Droplet Model prediction, $\sum_{k=1}^{n}1/(2k-1)$.}
    \label{tab:multifractal-spectrum}
    \begin{tabular}{cclll}
        $T$ & $L$ & \multicolumn{1}{c}{$\tau_{2}\quad$} & \multicolumn{1}{c}{$\tau_{3}\quad$} & \multicolumn{1}{c}{$\tau_{4}\quad$}\\
        \midrule
         $0.7$ & $16$ & 1.445(5) & 1.730(11) & 1.941(16)\\
         $0.7$ & $24$ & 1.435(12) & 1.72(3) & 1.93(4)\\
         $0.8$ & $16$ & 1.448(5) & 1.732(11) & 1.943(17)\\
         $0.8$ & $24$ & 1.456(12) & 1.75(3) & 1.97(4)\\
         $0.9$ & $16$ & 1.450(5) & 1.736(11) & 1.941(19)\\
         $0.9$ & $24$ & 1.448(12) & 1.730(24) & 1.95(4)\\
         $1.0$ & $16$ & 1.463(6) & 1.759(13) & 1.968(22)\\
         $1.0$ & $24$ & 1.445(14) & 1.72(3) & 1.91(4)\\
         $T_\mathrm{c}$ & $16$ & 1.478(7) & 1.782(16) & 2.00(3)\\
         $T_\mathrm{c}$ & $24$ & 1.442(17) & 1.71(4) & 1.88(6)\\
         \midrule
         \multicolumn{2}{c}{1D Droplet:} & $\frac43 \approx 1.3333$ & $\frac{23}{15} \approx 1.5333$ & $\frac{176}{105} \approx  1.6762$\\
         \bottomrule
   \end{tabular}
\end{table}

\subsection*{Numerical results: the singularity spectrum}
The evidence for multifractal scaling is presented in Fig.~\ref{fig:Multifractal}. We have studied the multifractal spectrum \cite{benzi:84,frisch:85,castellani:86,halsey:86,halsey:86b,harte:01} through $\tau_n=\xi/\xi_n$ and found that $\tau_n<n$. Our results
for $\tau_n$ at temperatures $T=0.7, 0,.8, 0.9, 1.0$ and $1.1019\approx T_\text{c}$ are in Table~\ref{tab:multifractal-spectrum}.
The results are remarkably stable with varying $T$ and $L$, provided that $T<T_\text{c}$.
The reader may check that the ansatz $\tau_n=1+ A\log n$ fits our results well for $T<T_\text{c}$. Interestingly enough, a similar fit $\tau_n\sim \log n$ works for the multifractal spectrum of out-of-equilibrium correlations~\cite{janus:24}.

Notice as well in Table~\ref{tab:multifractal-spectrum} that the multifractal spectrum measured in the elongated prism differs from the one predicted by the DM at $\lambda=0$ (see {\bf SI} for the detailed derivation and the meaning of the $\lambda$ parameter).
Interestingly enough, varying the $\lambda$ parameter one finds that $\{\tau_2,\tau_3,\tau_4\}_{\text{1D},\lambda=0.36} \approx \{1.43,1.71,1.92\}$ agrees with the prism singularity spectrum, suggesting that the DM prediction for the 1D effective Hamiltonian needs to be improved.

\begin{acknowledgments}
We are indebted to Mike Moore and David Huse for their very useful correspondence regarding the DM predictions.
L.~A.~F. and V.~M.-M. have received financial support from Ministerio de Ciencia, Innovación y Universidades (MICIU, Spain), Agencia Estatal de Investigación (AEI, Spain, MCIN/AEI/10.13039/501100011033), European Regional Development Fund (ERDF, A way of making Europe) through Grant no.~PID2022-136374NB-C21. I.~G.-A.~P and F.~R.-T. were supported by the “National Centre for HPC, Big Data and Quantum Computing - HPC”, Project CN\_00000013, CUP B83C22002940006, NRP Mission 4 Component 2 Investment 1.4, Funded by the European Union - NextGenerationEU.
\end{acknowledgments}

\paragraph*{Competing interests.}
The authors declare no competing interests.

\section*{Data Availability}
All study data are available in the GitHub repository \url{https://github.com/IsidoroGlez/EA-LCD-MC}. The repository contains the numerical data used to generate the figures in the main text, along with the simulation codes used to obtain the results.

\bibliographystyle{apsrev4-2}
\bibliography{biblio}

\begin{thebibliography}{17}%
\makeatletter
\providecommand \@ifxundefined [1]{%
 \@ifx{#1\undefined}
}%
\providecommand \@ifnum [1]{%
 \ifnum #1\expandafter \@firstoftwo
 \else \expandafter \@secondoftwo
 \fi
}%
\providecommand \@ifx [1]{%
 \ifx #1\expandafter \@firstoftwo
 \else \expandafter \@secondoftwo
 \fi
}%
\providecommand \natexlab [1]{#1}%
\providecommand \enquote  [1]{``#1''}%
\providecommand \bibnamefont  [1]{#1}%
\providecommand \bibfnamefont [1]{#1}%
\providecommand \citenamefont [1]{#1}%
\providecommand \href@noop [0]{\@secondoftwo}%
\providecommand \href [0]{\begingroup \@sanitize@url \@href}%
\providecommand \@href[1]{\@@startlink{#1}\@@href}%
\providecommand \@@href[1]{\endgroup#1\@@endlink}%
\providecommand \@sanitize@url [0]{\catcode `\\12\catcode `\$12\catcode
  `\&12\catcode `\#12\catcode `\^12\catcode `\_12\catcode `\%12\relax}%
\providecommand \@@startlink[1]{}%
\providecommand \@@endlink[0]{}%
\providecommand \url  [0]{\begingroup\@sanitize@url \@url }%
\providecommand \@url [1]{\endgroup\@href {#1}{\urlprefix }}%
\providecommand \urlprefix  [0]{URL }%
\providecommand \Eprint [0]{\href }%
\providecommand \doibase [0]{https://doi.org/}%
\providecommand \selectlanguage [0]{\@gobble}%
\providecommand \bibinfo  [0]{\@secondoftwo}%
\providecommand \bibfield  [0]{\@secondoftwo}%
\providecommand \translation [1]{[#1]}%
\providecommand \BibitemOpen [0]{}%
\providecommand \bibitemStop [0]{}%
\providecommand \bibitemNoStop [0]{.\EOS\space}%
\providecommand \EOS [0]{\spacefactor3000\relax}%
\providecommand \BibitemShut  [1]{\csname bibitem#1\endcsname}%
\let\auto@bib@innerbib\@empty
%</preamble>
\bibitem [{\citenamefont {Barma}\ and\ \citenamefont {{Sriram
  Shastry}}(1977)}]{barma:77}%
  \BibitemOpen
  \bibfield  {author} {\bibinfo {author} {\bibfnamefont {M.}~\bibnamefont
  {Barma}}\ and\ \bibinfo {author} {\bibfnamefont {B.}~\bibnamefont {{Sriram
  Shastry}}},\ }\href {https://doi.org/10.1016/0375-9601(77)90248-1} {\bibfield
   {journal} {\bibinfo  {journal} {Physics Letters A}\ }\textbf {\bibinfo
  {volume} {61}},\ \bibinfo {pages} {15–18} (\bibinfo {year}
  {1977})}\BibitemShut {NoStop}%
\bibitem [{\citenamefont {Bernaschi}\ \emph
  {et~al.}(2024{\natexlab{a}})\citenamefont {Bernaschi}, \citenamefont
  {{González-Adalid Pemartín}}, \citenamefont {Martín-Mayor},\ and\
  \citenamefont {Parisi}}]{bernaschi:24}%
  \BibitemOpen
  \bibfield  {author} {\bibinfo {author} {\bibfnamefont {M.}~\bibnamefont
  {Bernaschi}}, \bibinfo {author} {\bibfnamefont {I.}~\bibnamefont
  {{González-Adalid Pemartín}}}, \bibinfo {author} {\bibfnamefont
  {V.}~\bibnamefont {Martín-Mayor}},\ and\ \bibinfo {author} {\bibfnamefont
  {G.}~\bibnamefont {Parisi}},\ }\href
  {https://doi.org/10.1016/j.cpc.2024.109101} {\bibfield  {journal} {\bibinfo
  {journal} {Comp. Phys. Comm.}\ }\textbf {\bibinfo {volume} {298}},\ \bibinfo
  {pages} {109101} (\bibinfo {year} {2024}{\natexlab{a}})}\BibitemShut
  {NoStop}%
\bibitem [{\citenamefont {Swendsen}\ and\ \citenamefont
  {Wang}(1987)}]{swendsen:87}%
  \BibitemOpen
  \bibfield  {author} {\bibinfo {author} {\bibfnamefont {R.~H.}\ \bibnamefont
  {Swendsen}}\ and\ \bibinfo {author} {\bibfnamefont {J.-S.}\ \bibnamefont
  {Wang}},\ }\href {https://doi.org/10.1103/PhysRevLett.58.86} {\bibfield
  {journal} {\bibinfo  {journal} {Phys. Rev. Lett.}\ }\textbf {\bibinfo
  {volume} {58}},\ \bibinfo {pages} {86} (\bibinfo {year} {1987})}\BibitemShut
  {NoStop}%
\bibitem [{\citenamefont {Wolff}(1989)}]{wolff:89}%
  \BibitemOpen
  \bibfield  {author} {\bibinfo {author} {\bibfnamefont {U.}~\bibnamefont
  {Wolff}},\ }\href {https://doi.org/10.1103/PhysRevLett.62.361} {\bibfield
  {journal} {\bibinfo  {journal} {Phys. Rev. Lett.}\ }\textbf {\bibinfo
  {volume} {62}},\ \bibinfo {pages} {361} (\bibinfo {year} {1989})}\BibitemShut
  {NoStop}%
\bibitem [{\citenamefont {Houdayer}(2001)}]{houdayer:01}%
  \BibitemOpen
  \bibfield  {author} {\bibinfo {author} {\bibfnamefont {J.}~\bibnamefont
  {Houdayer}},\ }\href {https://doi.org/10.1007/PL00011151} {\bibfield
  {journal} {\bibinfo  {journal} {Eur. Phys. J. B}\ }\textbf {\bibinfo {volume}
  {22}},\ \bibinfo {pages} {479} (\bibinfo {year} {2001})}\BibitemShut
  {NoStop}%
\bibitem [{\citenamefont {Münster}\ and\ \citenamefont
  {Weigel}(2024)}]{munster:24}%
  \BibitemOpen
  \bibfield  {author} {\bibinfo {author} {\bibfnamefont {L.}~\bibnamefont
  {Münster}}\ and\ \bibinfo {author} {\bibfnamefont {M.}~\bibnamefont
  {Weigel}},\ }\bibfield  {journal} {\bibinfo  {journal} {Frontiers in
  Physics}\ }\textbf {\bibinfo {volume} {12}},\ \href
  {https://doi.org/10.3389/fphy.2024.1448175} {10.3389/fphy.2024.1448175}
  (\bibinfo {year} {2024})\BibitemShut {NoStop}%
\bibitem [{\citenamefont {Komura}(2015)}]{komura:15}%
  \BibitemOpen
  \bibfield  {author} {\bibinfo {author} {\bibfnamefont {Y.}~\bibnamefont
  {Komura}},\ }\href
  {https://doi.org/https://doi.org/10.1016/j.cpc.2015.04.015} {\bibfield
  {journal} {\bibinfo  {journal} {Comp. Phys. Comm.}\ }\textbf {\bibinfo
  {volume} {194}},\ \bibinfo {pages} {54} (\bibinfo {year} {2015})}\BibitemShut
  {NoStop}%
\bibitem [{\citenamefont {Zhu}\ \emph {et~al.}(2015)\citenamefont {Zhu},
  \citenamefont {Ochoa},\ and\ \citenamefont {Katzgraber}}]{zhu:15}%
  \BibitemOpen
  \bibfield  {author} {\bibinfo {author} {\bibfnamefont {Z.}~\bibnamefont
  {Zhu}}, \bibinfo {author} {\bibfnamefont {A.~J.}\ \bibnamefont {Ochoa}},\
  and\ \bibinfo {author} {\bibfnamefont {H.~G.}\ \bibnamefont {Katzgraber}},\
  }\href {https://doi.org/10.1103/PhysRevLett.115.077201} {\bibfield  {journal}
  {\bibinfo  {journal} {Phys. Rev. Lett.}\ }\textbf {\bibinfo {volume} {115}},\
  \bibinfo {pages} {077201} (\bibinfo {year} {2015})}\BibitemShut {NoStop}%
\bibitem [{git(2025)}]{github}%
  \BibitemOpen
  \href@noop {} {\bibinfo {title} {Data and code available at}},\ \bibinfo
  {howpublished} {\url{https://github.com/IsidoroGlez/EA-LCD-MC}} (\bibinfo
  {year} {2025}),\ \bibinfo {note} {accessed: December 2025}\BibitemShut
  {NoStop}%
\bibitem [{\citenamefont {Billoire}\ \emph {et~al.}(2011)\citenamefont
  {Billoire}, \citenamefont {Fernandez}, \citenamefont {Maiorano},
  \citenamefont {Marinari}, \citenamefont {Mart{\'i}n-Mayor},\ and\
  \citenamefont {Yllanes}}]{billoire:11}%
  \BibitemOpen
  \bibfield  {author} {\bibinfo {author} {\bibfnamefont {A.}~\bibnamefont
  {Billoire}}, \bibinfo {author} {\bibfnamefont {L.~A.}\ \bibnamefont
  {Fernandez}}, \bibinfo {author} {\bibfnamefont {A.}~\bibnamefont {Maiorano}},
  \bibinfo {author} {\bibfnamefont {E.}~\bibnamefont {Marinari}}, \bibinfo
  {author} {\bibfnamefont {V.}~\bibnamefont {Mart{\'i}n-Mayor}},\ and\ \bibinfo
  {author} {\bibfnamefont {D.}~\bibnamefont {Yllanes}},\ }\href
  {https://doi.org/10.1088/1742-5468/2011/10/P10019} {\bibfield  {journal}
  {\bibinfo  {journal} {J. Stat. Mech.}\ }\textbf {\bibinfo {volume} {2011}},\
  \bibinfo {pages} {P10019} (\bibinfo {year} {2011})},\ \Eprint
  {https://arxiv.org/abs/arXiv:1108.1336} {arXiv:1108.1336} \BibitemShut
  {NoStop}%
\bibitem [{\citenamefont {Cooper}\ \emph {et~al.}(1982)\citenamefont {Cooper},
  \citenamefont {Freedman},\ and\ \citenamefont {Preston}}]{cooper:82}%
  \BibitemOpen
  \bibfield  {author} {\bibinfo {author} {\bibfnamefont {F.}~\bibnamefont
  {Cooper}}, \bibinfo {author} {\bibfnamefont {B.}~\bibnamefont {Freedman}},\
  and\ \bibinfo {author} {\bibfnamefont {D.}~\bibnamefont {Preston}},\ }\href
  {https://doi.org/10.1016/0550-3213(82)90240-1} {\bibfield  {journal}
  {\bibinfo  {journal} {Nucl. Phys. B}\ }\textbf {\bibinfo {volume} {210}},\
  \bibinfo {pages} {210} (\bibinfo {year} {1982})}\BibitemShut {NoStop}%
\bibitem [{\citenamefont {Edwards}\ and\ \citenamefont
  {Sokal}(1989)}]{edwards:89}%
  \BibitemOpen
  \bibfield  {author} {\bibinfo {author} {\bibfnamefont {R.~G.}\ \bibnamefont
  {Edwards}}\ and\ \bibinfo {author} {\bibfnamefont {A.~D.}\ \bibnamefont
  {Sokal}},\ }\href {https://doi.org/10.1103/PhysRevD.40.1374} {\bibfield
  {journal} {\bibinfo  {journal} {Phys. Rev. D}\ }\textbf {\bibinfo {volume}
  {40}},\ \bibinfo {pages} {1374} (\bibinfo {year} {1989})}\BibitemShut
  {NoStop}%
\bibitem [{\citenamefont {Parisi}(1994)}]{parisi:94}%
  \BibitemOpen
  \bibfield  {author} {\bibinfo {author} {\bibfnamefont {G.}~\bibnamefont
  {Parisi}},\ }\href@noop {} {\emph {\bibinfo {title} {Field Theory, Disorder
  and Simulations}}}\ (\bibinfo  {publisher} {World Scientific},\ \bibinfo
  {year} {1994})\BibitemShut {NoStop}%
\bibitem [{\citenamefont {Fisher}\ and\ \citenamefont
  {Huse}(1988)}]{fisher:88b}%
  \BibitemOpen
  \bibfield  {author} {\bibinfo {author} {\bibfnamefont {D.~S.}\ \bibnamefont
  {Fisher}}\ and\ \bibinfo {author} {\bibfnamefont {D.~A.}\ \bibnamefont
  {Huse}},\ }\href {https://doi.org/10.1103/PhysRevB.38.386} {\bibfield
  {journal} {\bibinfo  {journal} {Phys. Rev. B}\ }\textbf {\bibinfo {volume}
  {38}},\ \bibinfo {pages} {386} (\bibinfo {year} {1988})}\BibitemShut
  {NoStop}%
\bibitem [{\citenamefont {Carter}\ \emph {et~al.}(2002)\citenamefont {Carter},
  \citenamefont {Bray},\ and\ \citenamefont {Moore}}]{carter:02}%
  \BibitemOpen
  \bibfield  {author} {\bibinfo {author} {\bibfnamefont {A.~C.}\ \bibnamefont
  {Carter}}, \bibinfo {author} {\bibfnamefont {A.~J.}\ \bibnamefont {Bray}},\
  and\ \bibinfo {author} {\bibfnamefont {M.~A.}\ \bibnamefont {Moore}},\ }\href
  {https://doi.org/10.1103/PhysRevLett.88.077201} {\bibfield  {journal}
  {\bibinfo  {journal} {Phys. Rev. Lett.}\ }\textbf {\bibinfo {volume} {88}},\
  \bibinfo {pages} {077201} (\bibinfo {year} {2002})}\BibitemShut {NoStop}%
\bibitem [{\citenamefont {Bernaschi}\ \emph
  {et~al.}(2024{\natexlab{b}})\citenamefont {Bernaschi}, \citenamefont
  {{González-Adalid Pemartín}}, \citenamefont {Martín-Mayor},\ and\
  \citenamefont {Parisi}}]{bernaschi:24b}%
  \BibitemOpen
  \bibfield  {author} {\bibinfo {author} {\bibfnamefont {M.}~\bibnamefont
  {Bernaschi}}, \bibinfo {author} {\bibfnamefont {I.}~\bibnamefont
  {{González-Adalid Pemartín}}}, \bibinfo {author} {\bibfnamefont
  {V.}~\bibnamefont {Martín-Mayor}},\ and\ \bibinfo {author} {\bibfnamefont
  {G.}~\bibnamefont {Parisi}},\ }\href
  {https://doi.org/10.1038/s41586-024-07647-y} {\bibfield  {journal} {\bibinfo
  {journal} {Nature}\ }\textbf {\bibinfo {volume} {631}},\ \bibinfo {pages}
  {749–754} (\bibinfo {year} {2024}{\natexlab{b}})}\BibitemShut {NoStop}%
\bibitem [{\citenamefont {Widom}(1965)}]{widom:65}%
  \BibitemOpen
  \bibfield  {author} {\bibinfo {author} {\bibfnamefont {B.}~\bibnamefont
  {Widom}},\ }\href {https://doi.org/10.1063/1.1696617} {\bibfield  {journal}
  {\bibinfo  {journal} {J. Chem. Phys.}\ }\textbf {\bibinfo {volume} {43}},\
  \bibinfo {pages} {3892} (\bibinfo {year} {1965})}\BibitemShut {NoStop}%
\end{thebibliography}%


\begin{thebibliography}{72}%
\makeatletter
\providecommand \@ifxundefined [1]{%
 \@ifx{#1\undefined}
}%
\providecommand \@ifnum [1]{%
 \ifnum #1\expandafter \@firstoftwo
 \else \expandafter \@secondoftwo
 \fi
}%
\providecommand \@ifx [1]{%
 \ifx #1\expandafter \@firstoftwo
 \else \expandafter \@secondoftwo
 \fi
}%
\providecommand \natexlab [1]{#1}%
\providecommand \enquote  [1]{``#1''}%
\providecommand \bibnamefont  [1]{#1}%
\providecommand \bibfnamefont [1]{#1}%
\providecommand \citenamefont [1]{#1}%
\providecommand \href@noop [0]{\@secondoftwo}%
\providecommand \href [0]{\begingroup \@sanitize@url \@href}%
\providecommand \@href[1]{\@@startlink{#1}\@@href}%
\providecommand \@@href[1]{\endgroup#1\@@endlink}%
\providecommand \@sanitize@url [0]{\catcode `\\12\catcode `\$12\catcode
  `\&12\catcode `\#12\catcode `\^12\catcode `\_12\catcode `\%12\relax}%
\providecommand \@@startlink[1]{}%
\providecommand \@@endlink[0]{}%
\providecommand \url  [0]{\begingroup\@sanitize@url \@url }%
\providecommand \@url [1]{\endgroup\@href {#1}{\urlprefix }}%
\providecommand \urlprefix  [0]{URL }%
\providecommand \Eprint [0]{\href }%
\providecommand \doibase [0]{https://doi.org/}%
\providecommand \selectlanguage [0]{\@gobble}%
\providecommand \bibinfo  [0]{\@secondoftwo}%
\providecommand \bibfield  [0]{\@secondoftwo}%
\providecommand \translation [1]{[#1]}%
\providecommand \BibitemOpen [0]{}%
\providecommand \bibitemStop [0]{}%
\providecommand \bibitemNoStop [0]{.\EOS\space}%
\providecommand \EOS [0]{\spacefactor3000\relax}%
\providecommand \BibitemShut  [1]{\csname bibitem#1\endcsname}%
\let\auto@bib@innerbib\@empty
%</preamble>
\bibitem [{\citenamefont {Parisi}(1988)}]{parisi:88}%
  \BibitemOpen
  \bibfield  {author} {\bibinfo {author} {\bibfnamefont {G.}~\bibnamefont
  {Parisi}},\ }\href@noop {} {\emph {\bibinfo {title} {Statistical Field
  Theory}}}\ (\bibinfo  {publisher} {Addison-Wesley},\ \bibinfo {year}
  {1988})\BibitemShut {NoStop}%
\bibitem [{\citenamefont {Zinn-Justin}(2005)}]{zinn-justin:05}%
  \BibitemOpen
  \bibfield  {author} {\bibinfo {author} {\bibfnamefont {J.}~\bibnamefont
  {Zinn-Justin}},\ }\href@noop {} {\emph {\bibinfo {title} {Quantum Field
  Theory and Critical Phenomena}}},\ \bibinfo {edition} {4th}\ ed.\ (\bibinfo
  {publisher} {Clarendon Press},\ \bibinfo {address} {Oxford},\ \bibinfo {year}
  {2005})\BibitemShut {NoStop}%
\bibitem [{\citenamefont {Anderson}(1963)}]{anderson:63}%
  \BibitemOpen
  \bibfield  {author} {\bibinfo {author} {\bibfnamefont {P.~W.}\ \bibnamefont
  {Anderson}},\ }\href {https://doi.org/10.1103/PhysRev.130.439} {\bibfield
  {journal} {\bibinfo  {journal} {Phys. Rev.}\ }\textbf {\bibinfo {volume}
  {130}},\ \bibinfo {pages} {439} (\bibinfo {year} {1963})}\BibitemShut
  {NoStop}%
\bibitem [{\citenamefont {Higgs}(1964)}]{higgs:64}%
  \BibitemOpen
  \bibfield  {author} {\bibinfo {author} {\bibfnamefont {P.~W.}\ \bibnamefont
  {Higgs}},\ }\href {https://doi.org/10.1103/PhysRevLett.13.508} {\bibfield
  {journal} {\bibinfo  {journal} {Phys. Rev. Lett.}\ }\textbf {\bibinfo
  {volume} {13}},\ \bibinfo {pages} {508} (\bibinfo {year} {1964})}\BibitemShut
  {NoStop}%
\bibitem [{\citenamefont {Mydosh}(1993)}]{mydosh:93}%
  \BibitemOpen
  \bibfield  {author} {\bibinfo {author} {\bibfnamefont {J.~A.}\ \bibnamefont
  {Mydosh}},\ }\href@noop {} {\emph {\bibinfo {title} {Spin Glasses: an
  Experimental Introduction}}}\ (\bibinfo  {publisher} {Taylor and Francis},\
  \bibinfo {address} {London},\ \bibinfo {year} {1993})\BibitemShut {NoStop}%
\bibitem [{\citenamefont {Charbonneau}\ \emph {et~al.}(2023)\citenamefont
  {Charbonneau}, \citenamefont {Marinari}, \citenamefont {Mézard},
  \citenamefont {Parisi}, \citenamefont {Ricci-Tersenghi}, \citenamefont
  {Sicuro},\ and\ \citenamefont {Zamponi}}]{charbonneau:23}%
  \BibitemOpen
  \bibfield  {author} {\bibinfo {author} {\bibfnamefont {P.}~\bibnamefont
  {Charbonneau}}, \bibinfo {author} {\bibfnamefont {E.}~\bibnamefont
  {Marinari}}, \bibinfo {author} {\bibfnamefont {M.}~\bibnamefont {Mézard}},
  \bibinfo {author} {\bibfnamefont {G.}~\bibnamefont {Parisi}}, \bibinfo
  {author} {\bibfnamefont {F.}~\bibnamefont {Ricci-Tersenghi}}, \bibinfo
  {author} {\bibfnamefont {G.}~\bibnamefont {Sicuro}},\ and\ \bibinfo {author}
  {\bibfnamefont {F.}~\bibnamefont {Zamponi}},\ }\href
  {https://doi.org/10.1142/13341} {\emph {\bibinfo {title} {Spin Glass Theory
  and Far Beyond}}}\ (\bibinfo  {publisher} {WORLD SCIENTIFIC},\ \bibinfo
  {year} {2023})\ \Eprint
  {https://arxiv.org/abs/https://www.worldscientific.com/doi/pdf/10.1142/13341}
  {https://www.worldscientific.com/doi/pdf/10.1142/13341} \BibitemShut
  {NoStop}%
\bibitem [{\citenamefont {Dahlberg}\ \emph {et~al.}(2025)\citenamefont
  {Dahlberg}, \citenamefont {Pemart{\'\i}n}, \citenamefont {Marinari},
  \citenamefont {Parisi}, \citenamefont {Ricci-Tersenghi}, \citenamefont
  {Martin-Mayor}, \citenamefont {Moreno-Gordo}, \citenamefont {Orbach},
  \citenamefont {Paga}, \citenamefont {Ruiz-Lorenzo} \emph
  {et~al.}}]{dahlberg:25}%
  \BibitemOpen
  \bibfield  {author} {\bibinfo {author} {\bibfnamefont {E.}~\bibnamefont
  {Dahlberg}}, \bibinfo {author} {\bibfnamefont {I.~G.-A.}\ \bibnamefont
  {Pemart{\'\i}n}}, \bibinfo {author} {\bibfnamefont {E.}~\bibnamefont
  {Marinari}}, \bibinfo {author} {\bibfnamefont {G.}~\bibnamefont {Parisi}},
  \bibinfo {author} {\bibfnamefont {F.}~\bibnamefont {Ricci-Tersenghi}},
  \bibinfo {author} {\bibfnamefont {V.}~\bibnamefont {Martin-Mayor}}, \bibinfo
  {author} {\bibfnamefont {J.}~\bibnamefont {Moreno-Gordo}}, \bibinfo {author}
  {\bibfnamefont {R.}~\bibnamefont {Orbach}}, \bibinfo {author} {\bibfnamefont
  {I.}~\bibnamefont {Paga}}, \bibinfo {author} {\bibfnamefont {J.}~\bibnamefont
  {Ruiz-Lorenzo}}, \emph {et~al.},\ }\href {https://doi.org/10.1103/ctp2-zwyr}
  {\bibfield  {journal} {\bibinfo  {journal} {Rev. Mod. Phys.}\ }\textbf
  {\bibinfo {volume} {97}},\ \bibinfo {pages} {045005} (\bibinfo {year}
  {2025})}\BibitemShut {NoStop}%
\bibitem [{\citenamefont {Thouless}\ \emph {et~al.}(1977)\citenamefont
  {Thouless}, \citenamefont {Anderson},\ and\ \citenamefont
  {Palmer}}]{thouless:77}%
  \BibitemOpen
  \bibfield  {author} {\bibinfo {author} {\bibfnamefont {D.~J.}\ \bibnamefont
  {Thouless}}, \bibinfo {author} {\bibfnamefont {P.~W.}\ \bibnamefont
  {Anderson}},\ and\ \bibinfo {author} {\bibfnamefont {R.~G.}\ \bibnamefont
  {Palmer}},\ }\href {https://doi.org/10.1080/14786437708235992} {\bibfield
  {journal} {\bibinfo  {journal} {Phil. Mag.}\ }\textbf {\bibinfo {volume}
  {35}},\ \bibinfo {pages} {593} (\bibinfo {year} {1977})},\ \Eprint
  {https://arxiv.org/abs/http://dx.doi.org/10.1080/14786437708235992}
  {http://dx.doi.org/10.1080/14786437708235992} \BibitemShut {NoStop}%
\bibitem [{\citenamefont {Anderson}(1984)}]{anderson:84}%
  \BibitemOpen
  \bibfield  {author} {\bibinfo {author} {\bibfnamefont {P.~W.}\ \bibnamefont
  {Anderson}},\ }in\ \href {https://doi.org/10.1142/0031} {\emph {\bibinfo
  {booktitle} {Ill-Condensed Matter, Les Houches Session XXXI}}},\ \bibinfo
  {editor} {edited by\ \bibinfo {editor} {\bibfnamefont {R.}~\bibnamefont
  {Balian}}, \bibinfo {editor} {\bibfnamefont {R.}~\bibnamefont {Maynard}},\
  and\ \bibinfo {editor} {\bibfnamefont {G.}~\bibnamefont {Toulouse}}}\
  (\bibinfo  {publisher} {World Scientific},\ \bibinfo {address} {Singapore},\
  \bibinfo {year} {1984})\BibitemShut {NoStop}%
\bibitem [{\citenamefont {De~Dominicis}\ and\ \citenamefont
  {Kondor}(1984)}]{dedominicis:84}%
  \BibitemOpen
  \bibfield  {author} {\bibinfo {author} {\bibfnamefont {C.}~\bibnamefont
  {De~Dominicis}}\ and\ \bibinfo {author} {\bibfnamefont {I.}~\bibnamefont
  {Kondor}},\ }\href {https://doi.org/10.1051/jphyslet:01984004505020500}
  {\bibfield  {journal} {\bibinfo  {journal} {J. Physique Lett.}\ }\textbf
  {\bibinfo {volume} {45}},\ \bibinfo {pages} {205 } (\bibinfo {year}
  {1984})}\BibitemShut {NoStop}%
\bibitem [{\citenamefont {Parisi}(2023)}]{parisi:23}%
  \BibitemOpen
  \bibfield  {author} {\bibinfo {author} {\bibfnamefont {G.}~\bibnamefont
  {Parisi}},\ }\href {https://doi.org/10.1103/RevModPhys.95.030501} {\bibfield
  {journal} {\bibinfo  {journal} {Rev. Mod. Phys.}\ }\textbf {\bibinfo {volume}
  {95}},\ \bibinfo {pages} {030501} (\bibinfo {year} {2023})}\BibitemShut
  {NoStop}%
\bibitem [{Note1()}]{Note1}%
  \BibitemOpen
  \bibinfo {note} {This is one of the geometries employed by Brezin in his
  investigation of Finite Size Scaling at the critical temperature $T_\protect
  \text {c}$~\cite {brezin:82}.}\BibitemShut {Stop}%
\bibitem [{\citenamefont {Cheung}\ and\ \citenamefont
  {McMillan}(1983)}]{cheung:83}%
  \BibitemOpen
  \bibfield  {author} {\bibinfo {author} {\bibfnamefont {H.~F.}\ \bibnamefont
  {Cheung}}\ and\ \bibinfo {author} {\bibfnamefont {W.~L.}\ \bibnamefont
  {McMillan}},\ }\href {https://doi.org/10.1088/0022-3719/16/36/017} {\bibfield
   {journal} {\bibinfo  {journal} {Journal of Physics C: Solid State Physics}\
  }\textbf {\bibinfo {volume} {16}},\ \bibinfo {pages} {7027} (\bibinfo {year}
  {1983})}\BibitemShut {NoStop}%
\bibitem [{\citenamefont {Carter}\ \emph {et~al.}(2002)\citenamefont {Carter},
  \citenamefont {Bray},\ and\ \citenamefont {Moore}}]{carter:02}%
  \BibitemOpen
  \bibfield  {author} {\bibinfo {author} {\bibfnamefont {A.~C.}\ \bibnamefont
  {Carter}}, \bibinfo {author} {\bibfnamefont {A.~J.}\ \bibnamefont {Bray}},\
  and\ \bibinfo {author} {\bibfnamefont {M.~A.}\ \bibnamefont {Moore}},\ }\href
  {https://doi.org/10.1103/PhysRevLett.88.077201} {\bibfield  {journal}
  {\bibinfo  {journal} {Phys. Rev. Lett.}\ }\textbf {\bibinfo {volume} {88}},\
  \bibinfo {pages} {077201} (\bibinfo {year} {2002})}\BibitemShut {NoStop}%
\bibitem [{\citenamefont {Fisch}(2008)}]{fisch:08}%
  \BibitemOpen
  \bibfield  {author} {\bibinfo {author} {\bibfnamefont {R.}~\bibnamefont
  {Fisch}},\ }\href {https://doi.org/10.1007/s10955-007-9436-4} {\bibfield
  {journal} {\bibinfo  {journal} {J. Stat. Phys}\ }\textbf {\bibinfo {volume}
  {130}},\ \bibinfo {pages} {561} (\bibinfo {year} {2008})}\BibitemShut
  {NoStop}%
\bibitem [{Note2()}]{Note2}%
  \BibitemOpen
  \bibinfo {note} {In a Heisenberg ferromagnet, the boundary twist $\Delta
  \varphi $ can be equally distributed among the planes, giving an excess free
  energy of order $O([\Delta \varphi /(M-1)]^2 L^{D-1})$ per consecutive plane
  pair. The sum of $M-1$ such terms gives the scaling in \protect \eqref
  {eq:scaling-1} (writing $M$ or $M-1$ is irrelevant).}\BibitemShut {Stop}%
\bibitem [{\citenamefont {Franz}\ \emph {et~al.}(1994)\citenamefont {Franz},
  \citenamefont {Parisi},\ and\ \citenamefont {Virasoro}}]{franz:94}%
  \BibitemOpen
  \bibfield  {author} {\bibinfo {author} {\bibfnamefont {S.}~\bibnamefont
  {Franz}}, \bibinfo {author} {\bibfnamefont {G.}~\bibnamefont {Parisi}},\ and\
  \bibinfo {author} {\bibfnamefont {M.}~\bibnamefont {Virasoro}},\ }\href
  {https://doi.org/10.1051/jp1:1994213} {\bibfield  {journal} {\bibinfo
  {journal} {J. Phys. (France)}\ }\textbf {\bibinfo {volume} {4}},\ \bibinfo
  {pages} {1657} (\bibinfo {year} {1994})}\BibitemShut {NoStop}%
\bibitem [{\citenamefont {Maiorano}\ and\ \citenamefont
  {Parisi}(2018)}]{maiorano:18}%
  \BibitemOpen
  \bibfield  {author} {\bibinfo {author} {\bibfnamefont {A.}~\bibnamefont
  {Maiorano}}\ and\ \bibinfo {author} {\bibfnamefont {G.}~\bibnamefont
  {Parisi}},\ }\href {https://doi.org/10.1073/pnas.1720832115} {\bibfield
  {journal} {\bibinfo  {journal} {Proc. Natl. Acad. Sci. USA}\ }\textbf
  {\bibinfo {volume} {115}},\ \bibinfo {pages} {5129} (\bibinfo {year}
  {2018})}\BibitemShut {NoStop}%
\bibitem [{\citenamefont {Fisher}\ and\ \citenamefont
  {Huse}(1988)}]{fisher:88b}%
  \BibitemOpen
  \bibfield  {author} {\bibinfo {author} {\bibfnamefont {D.~S.}\ \bibnamefont
  {Fisher}}\ and\ \bibinfo {author} {\bibfnamefont {D.~A.}\ \bibnamefont
  {Huse}},\ }\href {https://doi.org/10.1103/PhysRevB.38.386} {\bibfield
  {journal} {\bibinfo  {journal} {Phys. Rev. B}\ }\textbf {\bibinfo {volume}
  {38}},\ \bibinfo {pages} {386} (\bibinfo {year} {1988})}\BibitemShut
  {NoStop}%
\bibitem [{\citenamefont {Palassini}\ and\ \citenamefont
  {Young}(1999)}]{palassini:99b}%
  \BibitemOpen
  \bibfield  {author} {\bibinfo {author} {\bibfnamefont {M.}~\bibnamefont
  {Palassini}}\ and\ \bibinfo {author} {\bibfnamefont {A.~P.}\ \bibnamefont
  {Young}},\ }\href {https://doi.org/10.1103/PhysRevLett.83.5126} {\bibfield
  {journal} {\bibinfo  {journal} {Phys. Rev. Lett.}\ }\textbf {\bibinfo
  {volume} {83}},\ \bibinfo {pages} {5126} (\bibinfo {year}
  {1999})}\BibitemShut {NoStop}%
\bibitem [{\citenamefont {Hartmann}(1999)}]{hartmann:99c}%
  \BibitemOpen
  \bibfield  {author} {\bibinfo {author} {\bibfnamefont {A.~K.}\ \bibnamefont
  {Hartmann}},\ }\href@noop {} {\bibfield  {journal} {\bibinfo  {journal}
  {Phys. Rev. E.}\ }\textbf {\bibinfo {volume} {59}},\ \bibinfo {pages} {84}
  (\bibinfo {year} {1999})},\ \Eprint
  {https://arxiv.org/abs/arXiv:cond-mat/9806114} {arXiv:cond-mat/9806114}
  \BibitemShut {NoStop}%
\bibitem [{\citenamefont {Boettcher}(2004)}]{boettcher:04}%
  \BibitemOpen
  \bibfield  {author} {\bibinfo {author} {\bibfnamefont {S.}~\bibnamefont
  {Boettcher}},\ }\href {https://doi.org/10.1140/epjb/e2004-00102-5} {\bibfield
   {journal} {\bibinfo  {journal} {Eur. Phys. J. B}\ }\textbf {\bibinfo
  {volume} {38}},\ \bibinfo {pages} {83} (\bibinfo {year} {2004})},\ \Eprint
  {https://arxiv.org/abs/arXiv:cond-mat/0310698} {arXiv:cond-mat/0310698}
  \BibitemShut {NoStop}%
\bibitem [{\citenamefont {Benzi}\ \emph {et~al.}(1984)\citenamefont {Benzi},
  \citenamefont {Paladin}, \citenamefont {Parisi},\ and\ \citenamefont
  {Vulpiani}}]{benzi:84}%
  \BibitemOpen
  \bibfield  {author} {\bibinfo {author} {\bibfnamefont {R.}~\bibnamefont
  {Benzi}}, \bibinfo {author} {\bibfnamefont {G.}~\bibnamefont {Paladin}},
  \bibinfo {author} {\bibfnamefont {G.}~\bibnamefont {Parisi}},\ and\ \bibinfo
  {author} {\bibfnamefont {A.}~\bibnamefont {Vulpiani}},\ }\href
  {https://doi.org/10.1088/0305-4470/17/18/021} {\bibfield  {journal} {\bibinfo
   {journal} {Journal of Physics A: Mathematical and General}\ }\textbf
  {\bibinfo {volume} {17}},\ \bibinfo {pages} {3521} (\bibinfo {year}
  {1984})}\BibitemShut {NoStop}%
\bibitem [{\citenamefont {Frisch}\ and\ \citenamefont
  {Parisi}(1985)}]{frisch:85}%
  \BibitemOpen
  \bibfield  {author} {\bibinfo {author} {\bibfnamefont {U.}~\bibnamefont
  {Frisch}}\ and\ \bibinfo {author} {\bibfnamefont {G.}~\bibnamefont
  {Parisi}},\ }in\ \href@noop {} {\emph {\bibinfo {booktitle} {Turbulence and
  predictability in geophysical fluid dynamics and climate dynamics (1983
  International School of Physics ``Enrico Fermi'', Varenna)}}},\ \bibinfo
  {editor} {edited by\ \bibinfo {editor} {\bibfnamefont {M.}~\bibnamefont
  {Ghil}}, \bibinfo {editor} {\bibfnamefont {R.}~\bibnamefont {Benzi}},\ and\
  \bibinfo {editor} {\bibfnamefont {G.}~\bibnamefont {Parisi}}}\ (\bibinfo
  {publisher} {North-Holland},\ \bibinfo {address} {Amsterdam},\ \bibinfo
  {year} {1985})\BibitemShut {NoStop}%
\bibitem [{\citenamefont {Castellani}\ and\ \citenamefont
  {Peliti}(1986)}]{castellani:86}%
  \BibitemOpen
  \bibfield  {author} {\bibinfo {author} {\bibfnamefont {C.}~\bibnamefont
  {Castellani}}\ and\ \bibinfo {author} {\bibfnamefont {L.}~\bibnamefont
  {Peliti}},\ }\href {https://doi.org/10.1088/0305-4470/19/8/004} {\bibfield
  {journal} {\bibinfo  {journal} {Journal of physics A: mathematical and
  general}\ }\textbf {\bibinfo {volume} {19}},\ \bibinfo {pages} {L429}
  (\bibinfo {year} {1986})}\BibitemShut {NoStop}%
\bibitem [{\citenamefont {Halsey}\ \emph
  {et~al.}(1986{\natexlab{a}})\citenamefont {Halsey}, \citenamefont {Jensen},
  \citenamefont {Kadanoff}, \citenamefont {Procaccia},\ and\ \citenamefont
  {Shraiman}}]{halsey:86}%
  \BibitemOpen
  \bibfield  {author} {\bibinfo {author} {\bibfnamefont {T.~C.}\ \bibnamefont
  {Halsey}}, \bibinfo {author} {\bibfnamefont {M.~H.}\ \bibnamefont {Jensen}},
  \bibinfo {author} {\bibfnamefont {L.~P.}\ \bibnamefont {Kadanoff}}, \bibinfo
  {author} {\bibfnamefont {I.}~\bibnamefont {Procaccia}},\ and\ \bibinfo
  {author} {\bibfnamefont {B.~I.}\ \bibnamefont {Shraiman}},\ }\href
  {https://doi.org/10.1103/PhysRevA.33.1141} {\bibfield  {journal} {\bibinfo
  {journal} {Phys. Rev. A}\ }\textbf {\bibinfo {volume} {33}},\ \bibinfo
  {pages} {1141} (\bibinfo {year} {1986}{\natexlab{a}})}\BibitemShut {NoStop}%
\bibitem [{\citenamefont {Halsey}\ \emph
  {et~al.}(1986{\natexlab{b}})\citenamefont {Halsey}, \citenamefont {Jensen},
  \citenamefont {Kadanoff}, \citenamefont {Procaccia},\ and\ \citenamefont
  {Shraiman}}]{halsey:86b}%
  \BibitemOpen
  \bibfield  {author} {\bibinfo {author} {\bibfnamefont {T.~C.}\ \bibnamefont
  {Halsey}}, \bibinfo {author} {\bibfnamefont {M.~H.}\ \bibnamefont {Jensen}},
  \bibinfo {author} {\bibfnamefont {L.~P.}\ \bibnamefont {Kadanoff}}, \bibinfo
  {author} {\bibfnamefont {I.}~\bibnamefont {Procaccia}},\ and\ \bibinfo
  {author} {\bibfnamefont {B.~I.}\ \bibnamefont {Shraiman}},\ }\href
  {https://doi.org/10.1103/PhysRevA.34.1601} {\bibfield  {journal} {\bibinfo
  {journal} {Phys. Rev. A}\ }\textbf {\bibinfo {volume} {34}},\ \bibinfo
  {pages} {1601} (\bibinfo {year} {1986}{\natexlab{b}})}\BibitemShut {NoStop}%
\bibitem [{\citenamefont {Harte}(2001)}]{harte:01}%
  \BibitemOpen
  \bibfield  {author} {\bibinfo {author} {\bibfnamefont {D.}~\bibnamefont
  {Harte}},\ }\href {https://doi.org/10.1201/9781420036008} {\emph {\bibinfo
  {title} {Multifractals. Theory and applications}}},\ \bibinfo {edition}
  {1st}\ ed.\ (\bibinfo  {publisher} {Chapman and Hall/CRC},\ \bibinfo
  {address} {New York},\ \bibinfo {year} {2001})\BibitemShut {NoStop}%
\bibitem [{\citenamefont {Edwards}\ and\ \citenamefont
  {Anderson}(1975)}]{edwards:75}%
  \BibitemOpen
  \bibfield  {author} {\bibinfo {author} {\bibfnamefont {S.~F.}\ \bibnamefont
  {Edwards}}\ and\ \bibinfo {author} {\bibfnamefont {P.~W.}\ \bibnamefont
  {Anderson}},\ }\href {https://doi.org/10.1088/0305-4608/5/5/017} {\bibfield
  {journal} {\bibinfo  {journal} {J. Phys. F}\ }\textbf {\bibinfo {volume}
  {5}},\ \bibinfo {pages} {965} (\bibinfo {year} {1975})}\BibitemShut {NoStop}%
\bibitem [{\citenamefont {Edwards}\ and\ \citenamefont
  {Anderson}(1976)}]{edwards:76}%
  \BibitemOpen
  \bibfield  {author} {\bibinfo {author} {\bibfnamefont {S.~F.}\ \bibnamefont
  {Edwards}}\ and\ \bibinfo {author} {\bibfnamefont {P.~W.}\ \bibnamefont
  {Anderson}},\ }\href {https://doi.org/10.1088/0305-4608/6/10/022} {\bibfield
  {journal} {\bibinfo  {journal} {J. Phys. F}\ }\textbf {\bibinfo {volume}
  {6}},\ \bibinfo {pages} {1927} (\bibinfo {year} {1976})}\BibitemShut
  {NoStop}%
\bibitem [{\citenamefont {Baity-Jesi}\ \emph {et~al.}(2024)\citenamefont
  {Baity-Jesi}, \citenamefont {Calore}, \citenamefont {Cruz}, \citenamefont
  {Fernandez}, \citenamefont {Gil-Narvi{\'o}n}, \citenamefont
  {Gonz{\'a}lez-Adalid~Pemart{\'\i}n}, \citenamefont {Gordillo-Guerrero},
  \citenamefont {{\'I}{\~n}iguez}, \citenamefont {Maiorano}, \citenamefont
  {Marinari}, \citenamefont {Martin-Mayor}, \citenamefont {Moreno-Gordo},
  \citenamefont {Muñoz~Sudupe}, \citenamefont {Navarro}, \citenamefont {Paga},
  \citenamefont {Parisi}, \citenamefont {Perez-Gaviro}, \citenamefont
  {Ricci-Tersenghi}, \citenamefont {Ruiz-Lorenzo}, \citenamefont {Schifano},
  \citenamefont {Seoane}, \citenamefont {Tarancon},\ and\ \citenamefont
  {Yllanes}}]{janus:24}%
  \BibitemOpen
  \bibfield  {author} {\bibinfo {author} {\bibfnamefont {M.}~\bibnamefont
  {Baity-Jesi}}, \bibinfo {author} {\bibfnamefont {E.}~\bibnamefont {Calore}},
  \bibinfo {author} {\bibfnamefont {A.}~\bibnamefont {Cruz}}, \bibinfo {author}
  {\bibfnamefont {L.~A.}\ \bibnamefont {Fernandez}}, \bibinfo {author}
  {\bibfnamefont {J.~M.}\ \bibnamefont {Gil-Narvi{\'o}n}}, \bibinfo {author}
  {\bibfnamefont {I.}~\bibnamefont {Gonz{\'a}lez-Adalid~Pemart{\'\i}n}},
  \bibinfo {author} {\bibfnamefont {A.}~\bibnamefont {Gordillo-Guerrero}},
  \bibinfo {author} {\bibfnamefont {D.}~\bibnamefont {{\'I}{\~n}iguez}},
  \bibinfo {author} {\bibfnamefont {A.}~\bibnamefont {Maiorano}}, \bibinfo
  {author} {\bibfnamefont {E.}~\bibnamefont {Marinari}}, \bibinfo {author}
  {\bibfnamefont {V.}~\bibnamefont {Martin-Mayor}}, \bibinfo {author}
  {\bibfnamefont {J.}~\bibnamefont {Moreno-Gordo}}, \bibinfo {author}
  {\bibfnamefont {A.}~\bibnamefont {Muñoz~Sudupe}}, \bibinfo {author}
  {\bibfnamefont {D.}~\bibnamefont {Navarro}}, \bibinfo {author} {\bibfnamefont
  {I.}~\bibnamefont {Paga}}, \bibinfo {author} {\bibfnamefont {G.}~\bibnamefont
  {Parisi}}, \bibinfo {author} {\bibfnamefont {S.}~\bibnamefont
  {Perez-Gaviro}}, \bibinfo {author} {\bibfnamefont {F.}~\bibnamefont
  {Ricci-Tersenghi}}, \bibinfo {author} {\bibfnamefont {J.~J.}\ \bibnamefont
  {Ruiz-Lorenzo}}, \bibinfo {author} {\bibfnamefont {S.~F.}\ \bibnamefont
  {Schifano}}, \bibinfo {author} {\bibfnamefont {D.}~\bibnamefont {Seoane}},
  \bibinfo {author} {\bibfnamefont {A.}~\bibnamefont {Tarancon}},\ and\
  \bibinfo {author} {\bibfnamefont {D.}~\bibnamefont {Yllanes}},\ }\href
  {https://doi.org/10.1073/pnas.2312880120} {\bibfield  {journal} {\bibinfo
  {journal} {Proc. Natl. Acad. Sci. USA}\ }\textbf {\bibinfo {volume} {121}},\
  \bibinfo {pages} {e2312880120} (\bibinfo {year} {2024})}\BibitemShut
  {NoStop}%
\bibitem [{Note3()}]{Note3}%
  \BibitemOpen
  \bibinfo {note} {However, the multifractal spectrum that we compute here
  cannot be directly compared with the results of~\cite {janus:24}, because of
  the different geometry. The importance of geometry was emphasized in~\cite
  {marinari:24}.}\BibitemShut {Stop}%
\bibitem [{\citenamefont {Billoire}\ \emph {et~al.}(2018)\citenamefont
  {Billoire}, \citenamefont {Fernandez}, \citenamefont {Maiorano},
  \citenamefont {Marinari}, \citenamefont {Martin-Mayor}, \citenamefont
  {Moreno-Gordo}, \citenamefont {Parisi}, \citenamefont {Ricci-Tersenghi},\
  and\ \citenamefont {Ruiz-Lorenzo}}]{billoire:18}%
  \BibitemOpen
  \bibfield  {author} {\bibinfo {author} {\bibfnamefont {A.}~\bibnamefont
  {Billoire}}, \bibinfo {author} {\bibfnamefont {L.~A.}\ \bibnamefont
  {Fernandez}}, \bibinfo {author} {\bibfnamefont {A.}~\bibnamefont {Maiorano}},
  \bibinfo {author} {\bibfnamefont {E.}~\bibnamefont {Marinari}}, \bibinfo
  {author} {\bibfnamefont {V.}~\bibnamefont {Martin-Mayor}}, \bibinfo {author}
  {\bibfnamefont {J.}~\bibnamefont {Moreno-Gordo}}, \bibinfo {author}
  {\bibfnamefont {G.}~\bibnamefont {Parisi}}, \bibinfo {author} {\bibfnamefont
  {F.}~\bibnamefont {Ricci-Tersenghi}},\ and\ \bibinfo {author} {\bibfnamefont
  {J.~J.}\ \bibnamefont {Ruiz-Lorenzo}},\ }\href
  {https://doi.org/10.1088/1742-5468/aaa387} {\bibfield  {journal} {\bibinfo
  {journal} {J. Stat. Mech.}\ }\textbf {\bibinfo {volume} {2018}},\ \bibinfo
  {pages} {033302} (\bibinfo {year} {2018})}\BibitemShut {NoStop}%
\bibitem [{\citenamefont {Alvarez~Ba{\~n}os}\ \emph
  {et~al.}(2010{\natexlab{a}})\citenamefont {Alvarez~Ba{\~n}os}, \citenamefont
  {Cruz}, \citenamefont {Fernandez}, \citenamefont {Gil-Narvion}, \citenamefont
  {Gordillo-Guerrero}, \citenamefont {Guidetti}, \citenamefont {Maiorano},
  \citenamefont {Mantovani}, \citenamefont {Marinari}, \citenamefont
  {Mart\'{i}n-Mayor}, \citenamefont {Monforte-Garcia}, \citenamefont
  {Mu{\~n}oz~Sudupe}, \citenamefont {Navarro}, \citenamefont {Parisi},
  \citenamefont {Perez-Gaviro}, \citenamefont {Ruiz-Lorenzo}, \citenamefont
  {Schifano}, \citenamefont {Seoane}, \citenamefont {Tarancon}, \citenamefont
  {Tripiccione},\ and\ \citenamefont {Yllanes}}]{janus:10}%
  \BibitemOpen
  \bibfield  {author} {\bibinfo {author} {\bibfnamefont {R.}~\bibnamefont
  {Alvarez~Ba{\~n}os}}, \bibinfo {author} {\bibfnamefont {A.}~\bibnamefont
  {Cruz}}, \bibinfo {author} {\bibfnamefont {L.~A.}\ \bibnamefont {Fernandez}},
  \bibinfo {author} {\bibfnamefont {J.~M.}\ \bibnamefont {Gil-Narvion}},
  \bibinfo {author} {\bibfnamefont {A.}~\bibnamefont {Gordillo-Guerrero}},
  \bibinfo {author} {\bibfnamefont {M.}~\bibnamefont {Guidetti}}, \bibinfo
  {author} {\bibfnamefont {A.}~\bibnamefont {Maiorano}}, \bibinfo {author}
  {\bibfnamefont {F.}~\bibnamefont {Mantovani}}, \bibinfo {author}
  {\bibfnamefont {E.}~\bibnamefont {Marinari}}, \bibinfo {author}
  {\bibfnamefont {V.}~\bibnamefont {Mart\'{i}n-Mayor}}, \bibinfo {author}
  {\bibfnamefont {J.}~\bibnamefont {Monforte-Garcia}}, \bibinfo {author}
  {\bibfnamefont {A.}~\bibnamefont {Mu{\~n}oz~Sudupe}}, \bibinfo {author}
  {\bibfnamefont {D.}~\bibnamefont {Navarro}}, \bibinfo {author} {\bibfnamefont
  {G.}~\bibnamefont {Parisi}}, \bibinfo {author} {\bibfnamefont
  {S.}~\bibnamefont {Perez-Gaviro}}, \bibinfo {author} {\bibfnamefont {J.~J.}\
  \bibnamefont {Ruiz-Lorenzo}}, \bibinfo {author} {\bibfnamefont {S.~F.}\
  \bibnamefont {Schifano}}, \bibinfo {author} {\bibfnamefont {B.}~\bibnamefont
  {Seoane}}, \bibinfo {author} {\bibfnamefont {A.}~\bibnamefont {Tarancon}},
  \bibinfo {author} {\bibfnamefont {R.}~\bibnamefont {Tripiccione}},\ and\
  \bibinfo {author} {\bibfnamefont {D.}~\bibnamefont {Yllanes}} (\bibinfo
  {collaboration} {Janus Collaboration}),\ }\href
  {https://doi.org/10.1088/1742-5468/2010/06/P06026} {\bibfield  {journal}
  {\bibinfo  {journal} {J. Stat. Mech.}\ }\textbf {\bibinfo {volume} {2010}},\
  \bibinfo {pages} {P06026} (\bibinfo {year} {2010}{\natexlab{a}})},\ \Eprint
  {https://arxiv.org/abs/arXiv:1003.2569} {arXiv:1003.2569} \BibitemShut
  {NoStop}%
\bibitem [{\citenamefont {Baity-Jesi}\ \emph {et~al.}(2014)\citenamefont
  {Baity-Jesi}, \citenamefont {Ba\~{n}os}, \citenamefont {Cruz}, \citenamefont
  {Fernandez}, \citenamefont {Gil-Narvion}, \citenamefont {Gordillo-Guerrero},
  \citenamefont {Iniguez}, \citenamefont {Maiorano}, \citenamefont {Mantovani},
  \citenamefont {Marinari}, \citenamefont {Mart\'{i}n-Mayor}, \citenamefont
  {Monforte-Garcia}, \citenamefont {Mu{\~n}oz~Sudupe}, \citenamefont {Navarro},
  \citenamefont {Parisi}, \citenamefont {Perez-Gaviro}, \citenamefont
  {Pivanti}, \citenamefont {Ricci-Tersenghi}, \citenamefont {Ruiz-Lorenzo},
  \citenamefont {Schifano}, \citenamefont {Seoane}, \citenamefont {Tarancon},
  \citenamefont {Tripiccione},\ and\ \citenamefont {Yllanes}}]{janus:14}%
  \BibitemOpen
  \bibfield  {author} {\bibinfo {author} {\bibfnamefont {M.}~\bibnamefont
  {Baity-Jesi}}, \bibinfo {author} {\bibfnamefont {R.~A.}\ \bibnamefont
  {Ba\~{n}os}}, \bibinfo {author} {\bibfnamefont {A.}~\bibnamefont {Cruz}},
  \bibinfo {author} {\bibfnamefont {L.~A.}\ \bibnamefont {Fernandez}}, \bibinfo
  {author} {\bibfnamefont {J.~M.}\ \bibnamefont {Gil-Narvion}}, \bibinfo
  {author} {\bibfnamefont {A.}~\bibnamefont {Gordillo-Guerrero}}, \bibinfo
  {author} {\bibfnamefont {D.}~\bibnamefont {Iniguez}}, \bibinfo {author}
  {\bibfnamefont {A.}~\bibnamefont {Maiorano}}, \bibinfo {author}
  {\bibfnamefont {F.}~\bibnamefont {Mantovani}}, \bibinfo {author}
  {\bibfnamefont {E.}~\bibnamefont {Marinari}}, \bibinfo {author}
  {\bibfnamefont {V.}~\bibnamefont {Mart\'{i}n-Mayor}}, \bibinfo {author}
  {\bibfnamefont {J.}~\bibnamefont {Monforte-Garcia}}, \bibinfo {author}
  {\bibfnamefont {A.}~\bibnamefont {Mu{\~n}oz~Sudupe}}, \bibinfo {author}
  {\bibfnamefont {D.}~\bibnamefont {Navarro}}, \bibinfo {author} {\bibfnamefont
  {G.}~\bibnamefont {Parisi}}, \bibinfo {author} {\bibfnamefont
  {S.}~\bibnamefont {Perez-Gaviro}}, \bibinfo {author} {\bibfnamefont
  {M.}~\bibnamefont {Pivanti}}, \bibinfo {author} {\bibfnamefont
  {F.}~\bibnamefont {Ricci-Tersenghi}}, \bibinfo {author} {\bibfnamefont
  {J.~J.}\ \bibnamefont {Ruiz-Lorenzo}}, \bibinfo {author} {\bibfnamefont
  {S.~F.}\ \bibnamefont {Schifano}}, \bibinfo {author} {\bibfnamefont
  {B.}~\bibnamefont {Seoane}}, \bibinfo {author} {\bibfnamefont
  {A.}~\bibnamefont {Tarancon}}, \bibinfo {author} {\bibfnamefont
  {R.}~\bibnamefont {Tripiccione}},\ and\ \bibinfo {author} {\bibfnamefont
  {D.}~\bibnamefont {Yllanes}} (\bibinfo {collaboration} {Janus
  Collaboration}),\ }\href {https://doi.org/10.1016/j.cpc.2013.10.019}
  {\bibfield  {journal} {\bibinfo  {journal} {Comp. Phys. Comm}\ }\textbf
  {\bibinfo {volume} {185}},\ \bibinfo {pages} {550} (\bibinfo {year}
  {2014})},\ \Eprint {https://arxiv.org/abs/arXiv:1310.1032} {arXiv:1310.1032}
  \BibitemShut {NoStop}%
\bibitem [{\citenamefont {Hukushima}\ and\ \citenamefont
  {Nemoto}(1996)}]{hukushima:96}%
  \BibitemOpen
  \bibfield  {author} {\bibinfo {author} {\bibfnamefont {K.}~\bibnamefont
  {Hukushima}}\ and\ \bibinfo {author} {\bibfnamefont {K.}~\bibnamefont
  {Nemoto}},\ }\href {https://doi.org/10.1143/JPSJ.65.1604} {\bibfield
  {journal} {\bibinfo  {journal} {J. Phys. Soc. Japan}\ }\textbf {\bibinfo
  {volume} {65}},\ \bibinfo {pages} {1604} (\bibinfo {year} {1996})},\ \Eprint
  {https://arxiv.org/abs/arXiv:cond-mat/9512035} {arXiv:cond-mat/9512035}
  \BibitemShut {NoStop}%
\bibitem [{\citenamefont {Houdayer}(2001)}]{houdayer:01}%
  \BibitemOpen
  \bibfield  {author} {\bibinfo {author} {\bibfnamefont {J.}~\bibnamefont
  {Houdayer}},\ }\href {https://doi.org/10.1007/PL00011151} {\bibfield
  {journal} {\bibinfo  {journal} {Eur. Phys. J. B}\ }\textbf {\bibinfo {volume}
  {22}},\ \bibinfo {pages} {479} (\bibinfo {year} {2001})}\BibitemShut
  {NoStop}%
\bibitem [{\citenamefont {Fernandez}\ \emph {et~al.}(2016)\citenamefont
  {Fernandez}, \citenamefont {Marinari}, \citenamefont {Martin-Mayor},
  \citenamefont {Parisi},\ and\ \citenamefont {Ruiz-Lorenzo}}]{fernandez:16b}%
  \BibitemOpen
  \bibfield  {author} {\bibinfo {author} {\bibfnamefont {L.~A.}\ \bibnamefont
  {Fernandez}}, \bibinfo {author} {\bibfnamefont {E.}~\bibnamefont {Marinari}},
  \bibinfo {author} {\bibfnamefont {V.}~\bibnamefont {Martin-Mayor}}, \bibinfo
  {author} {\bibfnamefont {G.}~\bibnamefont {Parisi}},\ and\ \bibinfo {author}
  {\bibfnamefont {J.~J.}\ \bibnamefont {Ruiz-Lorenzo}},\ }\href
  {https://doi.org/10.1103/PhysRevB.94.024402} {\bibfield  {journal} {\bibinfo
  {journal} {Phys. Rev. B}\ }\textbf {\bibinfo {volume} {94}},\ \bibinfo
  {pages} {024402} (\bibinfo {year} {2016})}\BibitemShut {NoStop}%
\bibitem [{\citenamefont {Zhu}\ \emph {et~al.}(2015)\citenamefont {Zhu},
  \citenamefont {Ochoa},\ and\ \citenamefont {Katzgraber}}]{zhu:15}%
  \BibitemOpen
  \bibfield  {author} {\bibinfo {author} {\bibfnamefont {Z.}~\bibnamefont
  {Zhu}}, \bibinfo {author} {\bibfnamefont {A.~J.}\ \bibnamefont {Ochoa}},\
  and\ \bibinfo {author} {\bibfnamefont {H.~G.}\ \bibnamefont {Katzgraber}},\
  }\href {https://doi.org/10.1103/PhysRevLett.115.077201} {\bibfield  {journal}
  {\bibinfo  {journal} {Phys. Rev. Lett.}\ }\textbf {\bibinfo {volume} {115}},\
  \bibinfo {pages} {077201} (\bibinfo {year} {2015})}\BibitemShut {NoStop}%
\bibitem [{\citenamefont {Chilin}\ \emph {et~al.}(2026)\citenamefont {Chilin},
  \citenamefont {Marinari}, \citenamefont {Martín-Mayor}, \citenamefont
  {Parisi}, \citenamefont {Ruiz-Lorenzo},\ and\ \citenamefont
  {Yllanes}}]{chilin:26b}%
  \BibitemOpen
  \bibfield  {author} {\bibinfo {author} {\bibfnamefont {C.}~\bibnamefont
  {Chilin}}, \bibinfo {author} {\bibfnamefont {E.}~\bibnamefont {Marinari}},
  \bibinfo {author} {\bibfnamefont {V.}~\bibnamefont {Martín-Mayor}}, \bibinfo
  {author} {\bibfnamefont {G.}~\bibnamefont {Parisi}}, \bibinfo {author}
  {\bibfnamefont {J.~J.}\ \bibnamefont {Ruiz-Lorenzo}},\ and\ \bibinfo {author}
  {\bibfnamefont {D.}~\bibnamefont {Yllanes}},\ }\Eprint
  {https://arxiv.org/abs/2605.25872} {arXiv:2605.25872 [cond-mat.dis-nn]}
  (\bibinfo {year} {2026})\BibitemShut {NoStop}%
\bibitem [{\citenamefont {Baity-Jesi}\ \emph {et~al.}(2013)\citenamefont
  {Baity-Jesi}, \citenamefont {Ba\~{n}os}, \citenamefont {Cruz}, \citenamefont
  {Fernandez}, \citenamefont {Gil-Narvion}, \citenamefont {Gordillo-Guerrero},
  \citenamefont {Iniguez}, \citenamefont {Maiorano}, \citenamefont {Mantovani},
  \citenamefont {Marinari}, \citenamefont {Mart\'{i}n-Mayor}, \citenamefont
  {Monforte-Garcia}, \citenamefont {Mu{\~n}oz~Sudupe}, \citenamefont {Navarro},
  \citenamefont {Parisi}, \citenamefont {Perez-Gaviro}, \citenamefont
  {Pivanti}, \citenamefont {Ricci-Tersenghi}, \citenamefont {Ruiz-Lorenzo},
  \citenamefont {Schifano}, \citenamefont {Seoane}, \citenamefont {Tarancon},
  \citenamefont {Tripiccione},\ and\ \citenamefont {Yllanes}}]{janus:13}%
  \BibitemOpen
  \bibfield  {author} {\bibinfo {author} {\bibfnamefont {M.}~\bibnamefont
  {Baity-Jesi}}, \bibinfo {author} {\bibfnamefont {R.~A.}\ \bibnamefont
  {Ba\~{n}os}}, \bibinfo {author} {\bibfnamefont {A.}~\bibnamefont {Cruz}},
  \bibinfo {author} {\bibfnamefont {L.~A.}\ \bibnamefont {Fernandez}}, \bibinfo
  {author} {\bibfnamefont {J.~M.}\ \bibnamefont {Gil-Narvion}}, \bibinfo
  {author} {\bibfnamefont {A.}~\bibnamefont {Gordillo-Guerrero}}, \bibinfo
  {author} {\bibfnamefont {D.}~\bibnamefont {Iniguez}}, \bibinfo {author}
  {\bibfnamefont {A.}~\bibnamefont {Maiorano}}, \bibinfo {author}
  {\bibfnamefont {F.}~\bibnamefont {Mantovani}}, \bibinfo {author}
  {\bibfnamefont {E.}~\bibnamefont {Marinari}}, \bibinfo {author}
  {\bibfnamefont {V.}~\bibnamefont {Mart\'{i}n-Mayor}}, \bibinfo {author}
  {\bibfnamefont {J.}~\bibnamefont {Monforte-Garcia}}, \bibinfo {author}
  {\bibfnamefont {A.}~\bibnamefont {Mu{\~n}oz~Sudupe}}, \bibinfo {author}
  {\bibfnamefont {D.}~\bibnamefont {Navarro}}, \bibinfo {author} {\bibfnamefont
  {G.}~\bibnamefont {Parisi}}, \bibinfo {author} {\bibfnamefont
  {S.}~\bibnamefont {Perez-Gaviro}}, \bibinfo {author} {\bibfnamefont
  {M.}~\bibnamefont {Pivanti}}, \bibinfo {author} {\bibfnamefont
  {F.}~\bibnamefont {Ricci-Tersenghi}}, \bibinfo {author} {\bibfnamefont
  {J.~J.}\ \bibnamefont {Ruiz-Lorenzo}}, \bibinfo {author} {\bibfnamefont
  {S.~F.}\ \bibnamefont {Schifano}}, \bibinfo {author} {\bibfnamefont
  {B.}~\bibnamefont {Seoane}}, \bibinfo {author} {\bibfnamefont
  {A.}~\bibnamefont {Tarancon}}, \bibinfo {author} {\bibfnamefont
  {R.}~\bibnamefont {Tripiccione}},\ and\ \bibinfo {author} {\bibfnamefont
  {D.}~\bibnamefont {Yllanes}} (\bibinfo {collaboration} {Janus
  Collaboration}),\ }\href {https://doi.org/10.1103/PhysRevB.88.224416}
  {\bibfield  {journal} {\bibinfo  {journal} {Phys. Rev. B}\ }\textbf {\bibinfo
  {volume} {88}},\ \bibinfo {pages} {224416} (\bibinfo {year} {{2013}})},\
  \Eprint {https://arxiv.org/abs/arXiv:1310.2910} {arXiv:1310.2910}
  \BibitemShut {NoStop}%
\bibitem [{\citenamefont {{Br\'ezin, E.}}(1982)}]{brezin:82}%
  \BibitemOpen
  \bibfield  {author} {\bibinfo {author} {\bibnamefont {{Br\'ezin, E.}}},\
  }\href {https://doi.org/10.1051/jphys:0198200430101500} {\bibfield  {journal}
  {\bibinfo  {journal} {J. Phys. France}\ }\textbf {\bibinfo {volume} {43}},\
  \bibinfo {pages} {15} (\bibinfo {year} {1982})}\BibitemShut {NoStop}%
\bibitem [{Note4()}]{Note4}%
  \BibitemOpen
  \bibinfo {note} {The subtraction $\xi (T,L)-\xi (T_\protect \text {c},L)$
  highlights the differences from critical scaling. Since $a_3>1$, see \protect
  \eqref {eq:scaling-2}, $\xi (T_\protect \text {c},L)$ is subdominant compared
  with $\xi (T,L)$ at any fixed $T<T_\protect \text {c}$. An additional benefit
  is that the subtraction largely suppresses scaling corrections.}\BibitemShut
  {Stop}%
\bibitem [{\citenamefont {Barber}(1983)}]{barber:83}%
  \BibitemOpen
  \bibfield  {author} {\bibinfo {author} {\bibfnamefont {M.~N.}\ \bibnamefont
  {Barber}}\ }(\bibinfo  {publisher} {Academic Press},\ \bibinfo {year}
  {1983})\BibitemShut {NoStop}%
\bibitem [{\citenamefont {Wallace}\ and\ \citenamefont
  {Zia}(1979)}]{wallace:79}%
  \BibitemOpen
  \bibfield  {author} {\bibinfo {author} {\bibfnamefont {D.~J.}\ \bibnamefont
  {Wallace}}\ and\ \bibinfo {author} {\bibfnamefont {R.~K.~P.}\ \bibnamefont
  {Zia}},\ }\href {https://doi.org/10.1103/PhysRevLett.43.808} {\bibfield
  {journal} {\bibinfo  {journal} {Phys. Rev. Lett.}\ }\textbf {\bibinfo
  {volume} {43}},\ \bibinfo {pages} {808} (\bibinfo {year} {1979})}\BibitemShut
  {NoStop}%
\bibitem [{\citenamefont {Br\'ezin}\ and\ \citenamefont
  {Zinn-Justin}(1976)}]{brezin:76}%
  \BibitemOpen
  \bibfield  {author} {\bibinfo {author} {\bibfnamefont {E.}~\bibnamefont
  {Br\'ezin}}\ and\ \bibinfo {author} {\bibfnamefont {J.}~\bibnamefont
  {Zinn-Justin}},\ }\href {https://doi.org/10.1103/PhysRevLett.36.691}
  {\bibfield  {journal} {\bibinfo  {journal} {Phys. Rev. Lett.}\ }\textbf
  {\bibinfo {volume} {36}},\ \bibinfo {pages} {691} (\bibinfo {year}
  {1976})}\BibitemShut {NoStop}%
\bibitem [{\citenamefont {Ba\~nos}\ \emph {et~al.}(2012)\citenamefont
  {Ba\~nos}, \citenamefont {Fernandez}, \citenamefont {Martin-Mayor},\ and\
  \citenamefont {Young}}]{banos:12}%
  \BibitemOpen
  \bibfield  {author} {\bibinfo {author} {\bibfnamefont {R.~A.}\ \bibnamefont
  {Ba\~nos}}, \bibinfo {author} {\bibfnamefont {L.~A.}\ \bibnamefont
  {Fernandez}}, \bibinfo {author} {\bibfnamefont {V.}~\bibnamefont
  {Martin-Mayor}},\ and\ \bibinfo {author} {\bibfnamefont {A.~P.}\ \bibnamefont
  {Young}},\ }\href {https://doi.org/10.1103/PhysRevB.86.134416} {\bibfield
  {journal} {\bibinfo  {journal} {Phys. Rev. B}\ }\textbf {\bibinfo {volume}
  {86}},\ \bibinfo {pages} {134416} (\bibinfo {year} {2012})},\ \Eprint
  {https://arxiv.org/abs/arXiv:1207.7014} {arXiv:1207.7014} \BibitemShut
  {NoStop}%
\bibitem [{\citenamefont {Aguilar-Janita}\ \emph {et~al.}(2025)\citenamefont
  {Aguilar-Janita}, \citenamefont {Martín-Mayor}, \citenamefont
  {Moreno-Gordo},\ and\ \citenamefont {Ruiz-Lorenzo}}]{aguilar-janita:25}%
  \BibitemOpen
  \bibfield  {author} {\bibinfo {author} {\bibfnamefont {M.}~\bibnamefont
  {Aguilar-Janita}}, \bibinfo {author} {\bibfnamefont {V.}~\bibnamefont
  {Martín-Mayor}}, \bibinfo {author} {\bibfnamefont {J.}~\bibnamefont
  {Moreno-Gordo}},\ and\ \bibinfo {author} {\bibfnamefont {J.~J.}\ \bibnamefont
  {Ruiz-Lorenzo}},\ }\href {https://doi.org/10.1088/1742-5468/ae1a4a}
  {\bibfield  {journal} {\bibinfo  {journal} {J. Stat. Mech.}\ }\textbf
  {\bibinfo {volume} {2025}},\ \bibinfo {pages} {113301} (\bibinfo {year}
  {2025})}\BibitemShut {NoStop}%
\bibitem [{\citenamefont {Cardy}(1984)}]{cardy:84b}%
  \BibitemOpen
  \bibfield  {author} {\bibinfo {author} {\bibfnamefont {J.~L.}\ \bibnamefont
  {Cardy}},\ }\href {https://doi.org/10.1088/0305-4470/17/7/003} {\bibfield
  {journal} {\bibinfo  {journal} {Journal of Physics A: Mathematical and
  General}\ }\textbf {\bibinfo {volume} {17}},\ \bibinfo {pages} {L385}
  (\bibinfo {year} {1984})}\BibitemShut {NoStop}%
\bibitem [{\citenamefont {Burkhardt}\ and\ \citenamefont
  {Derrida}(1985)}]{burkhardt:85}%
  \BibitemOpen
  \bibfield  {author} {\bibinfo {author} {\bibfnamefont {T.~W.}\ \bibnamefont
  {Burkhardt}}\ and\ \bibinfo {author} {\bibfnamefont {B.}~\bibnamefont
  {Derrida}},\ }\href {https://doi.org/10.1103/PhysRevB.32.7273} {\bibfield
  {journal} {\bibinfo  {journal} {Phys. Rev. B}\ }\textbf {\bibinfo {volume}
  {32}},\ \bibinfo {pages} {7273} (\bibinfo {year} {1985})}\BibitemShut
  {NoStop}%
\bibitem [{Note5()}]{Note5}%
  \BibitemOpen
  \bibinfo {note} {The p-values from the $\chi ^2$ tests with two degrees of
  freedom are $0.19\protect \,(T\protect \!=\protect \!0.7)$, $0.18\protect
  \,(T\protect \!=\protect \!0.8),$ $0.05\protect \, (T\protect \!=\protect
  \!0.9),$ and $0.11\protect \,(T\protect \!=\protect \!1)$. These four $\chi
  ^2$ tests are not statistically independent, because data at different $T$
  come from the same simulation. Although we have not computed $u(T)$ close
  enough to $T_\protect \text {c}$ to perform a stringent test of \protect
  \eqref {eq:matching}, these $u(T\geq 0.8)$ can be fitted to $u(T)=(T_\protect
  \text {c}-T)^{p_3}[h_1+h_2(T_\protect \text {c}-T)]$ with $p_3=0.854$ ($\chi
  ^2/\protect \text {dof}=0.95/1$, p-value=0.33, the fit parameters being $h_1$
  and $h_2$).}\BibitemShut {Stop}%
\bibitem [{\citenamefont {Gunnarsson}\ \emph {et~al.}(1991)\citenamefont
  {Gunnarsson}, \citenamefont {Svedlindh}, \citenamefont {Nordblad},
  \citenamefont {Lundgren}, \citenamefont {Aruga},\ and\ \citenamefont
  {Ito}}]{gunnarsson:91}%
  \BibitemOpen
  \bibfield  {author} {\bibinfo {author} {\bibfnamefont {K.}~\bibnamefont
  {Gunnarsson}}, \bibinfo {author} {\bibfnamefont {P.}~\bibnamefont
  {Svedlindh}}, \bibinfo {author} {\bibfnamefont {P.}~\bibnamefont {Nordblad}},
  \bibinfo {author} {\bibfnamefont {L.}~\bibnamefont {Lundgren}}, \bibinfo
  {author} {\bibfnamefont {H.}~\bibnamefont {Aruga}},\ and\ \bibinfo {author}
  {\bibfnamefont {A.}~\bibnamefont {Ito}},\ }\href
  {https://doi.org/10.1103/PhysRevB.43.8199} {\bibfield  {journal} {\bibinfo
  {journal} {Phys. Rev. B}\ }\textbf {\bibinfo {volume} {43}},\ \bibinfo
  {pages} {8199} (\bibinfo {year} {1991})}\BibitemShut {NoStop}%
\bibitem [{\citenamefont {Palassini}\ and\ \citenamefont
  {Caracciolo}(1999)}]{palassini:99}%
  \BibitemOpen
  \bibfield  {author} {\bibinfo {author} {\bibfnamefont {M.}~\bibnamefont
  {Palassini}}\ and\ \bibinfo {author} {\bibfnamefont {S.}~\bibnamefont
  {Caracciolo}},\ }\href {https://doi.org/10.1103/PhysRevLett.82.5128}
  {\bibfield  {journal} {\bibinfo  {journal} {Phys. Rev. Lett.}\ }\textbf
  {\bibinfo {volume} {82}},\ \bibinfo {pages} {5128} (\bibinfo {year}
  {1999})},\ \Eprint {https://arxiv.org/abs/arXiv:cond-mat/9904246}
  {arXiv:cond-mat/9904246} \BibitemShut {NoStop}%
\bibitem [{\citenamefont {Ballesteros}\ \emph {et~al.}(2000)\citenamefont
  {Ballesteros}, \citenamefont {Cruz}, \citenamefont {Fernandez}, \citenamefont
  {Mart{\'i}n-Mayor}, \citenamefont {Pech}, \citenamefont {Ruiz-Lorenzo},
  \citenamefont {Tarancon}, \citenamefont {Tellez}, \citenamefont {Ullod},\
  and\ \citenamefont {Ungil}}]{ballesteros:00}%
  \BibitemOpen
  \bibfield  {author} {\bibinfo {author} {\bibfnamefont {H.~G.}\ \bibnamefont
  {Ballesteros}}, \bibinfo {author} {\bibfnamefont {A.}~\bibnamefont {Cruz}},
  \bibinfo {author} {\bibfnamefont {L.~A.}\ \bibnamefont {Fernandez}}, \bibinfo
  {author} {\bibfnamefont {V.}~\bibnamefont {Mart{\'i}n-Mayor}}, \bibinfo
  {author} {\bibfnamefont {J.}~\bibnamefont {Pech}}, \bibinfo {author}
  {\bibfnamefont {J.~J.}\ \bibnamefont {Ruiz-Lorenzo}}, \bibinfo {author}
  {\bibfnamefont {A.}~\bibnamefont {Tarancon}}, \bibinfo {author}
  {\bibfnamefont {P.}~\bibnamefont {Tellez}}, \bibinfo {author} {\bibfnamefont
  {C.~L.}\ \bibnamefont {Ullod}},\ and\ \bibinfo {author} {\bibfnamefont
  {C.}~\bibnamefont {Ungil}},\ }\href
  {https://doi.org/10.1103/PhysRevB.62.14237} {\bibfield  {journal} {\bibinfo
  {journal} {Phys. Rev. B}\ }\textbf {\bibinfo {volume} {62}},\ \bibinfo
  {pages} {14237} (\bibinfo {year} {2000})},\ \Eprint
  {https://arxiv.org/abs/arXiv:cond-mat/0006211} {arXiv:cond-mat/0006211}
  \BibitemShut {NoStop}%
\bibitem [{\citenamefont {Guchhait}\ and\ \citenamefont
  {Orbach}(2014)}]{guchhait:14}%
  \BibitemOpen
  \bibfield  {author} {\bibinfo {author} {\bibfnamefont {S.}~\bibnamefont
  {Guchhait}}\ and\ \bibinfo {author} {\bibfnamefont {R.}~\bibnamefont
  {Orbach}},\ }\href {https://doi.org/10.1103/PhysRevLett.112.126401}
  {\bibfield  {journal} {\bibinfo  {journal} {Phys. Rev. Lett.}\ }\textbf
  {\bibinfo {volume} {112}},\ \bibinfo {pages} {126401} (\bibinfo {year}
  {2014})}\BibitemShut {NoStop}%
\bibitem [{\citenamefont {Fernandez}\ \emph {et~al.}(2019)\citenamefont
  {Fernandez}, \citenamefont {Marinari}, \citenamefont {Martin-Mayor},
  \citenamefont {Paga},\ and\ \citenamefont {Ruiz-Lorenzo}}]{fernandez:19b}%
  \BibitemOpen
  \bibfield  {author} {\bibinfo {author} {\bibfnamefont {L.~A.}\ \bibnamefont
  {Fernandez}}, \bibinfo {author} {\bibfnamefont {E.}~\bibnamefont {Marinari}},
  \bibinfo {author} {\bibfnamefont {V.}~\bibnamefont {Martin-Mayor}}, \bibinfo
  {author} {\bibfnamefont {I.}~\bibnamefont {Paga}},\ and\ \bibinfo {author}
  {\bibfnamefont {J.~J.}\ \bibnamefont {Ruiz-Lorenzo}},\ }\href
  {https://doi.org/10.1103/PhysRevB.100.184412} {\bibfield  {journal} {\bibinfo
   {journal} {Phys. Rev. B}\ }\textbf {\bibinfo {volume} {100}},\ \bibinfo
  {pages} {184412} (\bibinfo {year} {2019})}\BibitemShut {NoStop}%
\bibitem [{\citenamefont {Boettcher}(2005)}]{boettcher:05}%
  \BibitemOpen
  \bibfield  {author} {\bibinfo {author} {\bibfnamefont {S.}~\bibnamefont
  {Boettcher}},\ }\href {https://doi.org/10.1103/PhysRevLett.95.197205}
  {\bibfield  {journal} {\bibinfo  {journal} {Phys. Rev. Lett.}\ }\textbf
  {\bibinfo {volume} {95}},\ \bibinfo {pages} {197205} (\bibinfo {year}
  {2005})},\ \Eprint {https://arxiv.org/abs/arXiv:cond-mat/0508061}
  {arXiv:cond-mat/0508061} \BibitemShut {NoStop}%
\bibitem [{\citenamefont {Alvarez~Ba{\~n}os}\ \emph
  {et~al.}(2010{\natexlab{b}})\citenamefont {Alvarez~Ba{\~n}os}, \citenamefont
  {Cruz}, \citenamefont {Fernandez}, \citenamefont {Gil-Narvion}, \citenamefont
  {Gordillo-Guerrero}, \citenamefont {Guidetti}, \citenamefont {Maiorano},
  \citenamefont {Mantovani}, \citenamefont {Marinari}, \citenamefont
  {Mart\'{i}n-Mayor}, \citenamefont {Monforte-Garcia}, \citenamefont
  {Mu{\~n}oz~Sudupe}, \citenamefont {Navarro}, \citenamefont {Parisi},
  \citenamefont {Perez-Gaviro}, \citenamefont {Ruiz-Lorenzo}, \citenamefont
  {Schifano}, \citenamefont {Seoane}, \citenamefont {Tarancon}, \citenamefont
  {Tripiccione},\ and\ \citenamefont {Yllanes}}]{janus:10b}%
  \BibitemOpen
  \bibfield  {author} {\bibinfo {author} {\bibfnamefont {R.}~\bibnamefont
  {Alvarez~Ba{\~n}os}}, \bibinfo {author} {\bibfnamefont {A.}~\bibnamefont
  {Cruz}}, \bibinfo {author} {\bibfnamefont {L.~A.}\ \bibnamefont {Fernandez}},
  \bibinfo {author} {\bibfnamefont {J.~M.}\ \bibnamefont {Gil-Narvion}},
  \bibinfo {author} {\bibfnamefont {A.}~\bibnamefont {Gordillo-Guerrero}},
  \bibinfo {author} {\bibfnamefont {M.}~\bibnamefont {Guidetti}}, \bibinfo
  {author} {\bibfnamefont {A.}~\bibnamefont {Maiorano}}, \bibinfo {author}
  {\bibfnamefont {F.}~\bibnamefont {Mantovani}}, \bibinfo {author}
  {\bibfnamefont {E.}~\bibnamefont {Marinari}}, \bibinfo {author}
  {\bibfnamefont {V.}~\bibnamefont {Mart\'{i}n-Mayor}}, \bibinfo {author}
  {\bibfnamefont {J.}~\bibnamefont {Monforte-Garcia}}, \bibinfo {author}
  {\bibfnamefont {A.}~\bibnamefont {Mu{\~n}oz~Sudupe}}, \bibinfo {author}
  {\bibfnamefont {D.}~\bibnamefont {Navarro}}, \bibinfo {author} {\bibfnamefont
  {G.}~\bibnamefont {Parisi}}, \bibinfo {author} {\bibfnamefont
  {S.}~\bibnamefont {Perez-Gaviro}}, \bibinfo {author} {\bibfnamefont {J.~J.}\
  \bibnamefont {Ruiz-Lorenzo}}, \bibinfo {author} {\bibfnamefont {S.~F.}\
  \bibnamefont {Schifano}}, \bibinfo {author} {\bibfnamefont {B.}~\bibnamefont
  {Seoane}}, \bibinfo {author} {\bibfnamefont {A.}~\bibnamefont {Tarancon}},
  \bibinfo {author} {\bibfnamefont {R.}~\bibnamefont {Tripiccione}},\ and\
  \bibinfo {author} {\bibfnamefont {D.}~\bibnamefont {Yllanes}} (\bibinfo
  {collaboration} {Janus Collaboration}),\ }\href
  {https://doi.org/10.1103/PhysRevLett.105.177202} {\bibfield  {journal}
  {\bibinfo  {journal} {Phys. Rev. Lett.}\ }\textbf {\bibinfo {volume} {105}},\
  \bibinfo {pages} {177202} (\bibinfo {year} {2010}{\natexlab{b}})},\ \Eprint
  {https://arxiv.org/abs/arXiv:1003.2943} {arXiv:1003.2943} \BibitemShut
  {NoStop}%
\bibitem [{\citenamefont {Kawamura}(1992)}]{kawamura:92}%
  \BibitemOpen
  \bibfield  {author} {\bibinfo {author} {\bibfnamefont {H.}~\bibnamefont
  {Kawamura}},\ }\href {https://doi.org/10.1103/PhysRevLett.68.3785} {\bibfield
   {journal} {\bibinfo  {journal} {Phys. Rev. Lett.}\ }\textbf {\bibinfo
  {volume} {68}},\ \bibinfo {pages} {3785} (\bibinfo {year}
  {1992})}\BibitemShut {NoStop}%
\bibitem [{\citenamefont {Hukushima}\ and\ \citenamefont
  {Kawamura}(2000)}]{hukushima:00}%
  \BibitemOpen
  \bibfield  {author} {\bibinfo {author} {\bibfnamefont {K.}~\bibnamefont
  {Hukushima}}\ and\ \bibinfo {author} {\bibfnamefont {H.}~\bibnamefont
  {Kawamura}},\ }\href {https://doi.org/10.1103/PhysRevE.61.R1008} {\bibfield
  {journal} {\bibinfo  {journal} {Phys. Rev. E}\ }\textbf {\bibinfo {volume}
  {61}},\ \bibinfo {pages} {R1008(R)} (\bibinfo {year} {2000})}\BibitemShut
  {NoStop}%
\bibitem [{\citenamefont {Kawamura}\ and\ \citenamefont
  {Taniguchi}(2015)}]{kawamura:15}%
  \BibitemOpen
  \bibfield  {author} {\bibinfo {author} {\bibfnamefont {H.}~\bibnamefont
  {Kawamura}}\ and\ \bibinfo {author} {\bibfnamefont {T.}~\bibnamefont
  {Taniguchi}},\ }in\ \href@noop {} {\emph {\bibinfo {booktitle} {Handbook of
  Magnetic Materials}}},\ Vol.~\bibinfo {volume} {24},\ \bibinfo {editor}
  {edited by\ \bibinfo {editor} {\bibfnamefont {K.~H.~J.}\ \bibnamefont
  {Buschow}}}\ (\bibinfo  {publisher} {Elsevier},\ \bibinfo {year}
  {2015})\BibitemShut {NoStop}%
\bibitem [{\citenamefont {Ogawa}\ \emph {et~al.}(2020)\citenamefont {Ogawa},
  \citenamefont {Uematsu},\ and\ \citenamefont {Kawamura}}]{ogawa:20}%
  \BibitemOpen
  \bibfield  {author} {\bibinfo {author} {\bibfnamefont {T.}~\bibnamefont
  {Ogawa}}, \bibinfo {author} {\bibfnamefont {K.}~\bibnamefont {Uematsu}},\
  and\ \bibinfo {author} {\bibfnamefont {H.}~\bibnamefont {Kawamura}},\ }\href
  {https://doi.org/10.1103/PhysRevB.101.014434} {\bibfield  {journal} {\bibinfo
   {journal} {Phys. Rev. B}\ }\textbf {\bibinfo {volume} {101}},\ \bibinfo
  {pages} {014434} (\bibinfo {year} {2020})}\BibitemShut {NoStop}%
\bibitem [{\citenamefont {Lee}\ and\ \citenamefont {Young}(2003)}]{lee:03}%
  \BibitemOpen
  \bibfield  {author} {\bibinfo {author} {\bibfnamefont {L.~W.}\ \bibnamefont
  {Lee}}\ and\ \bibinfo {author} {\bibfnamefont {A.~P.}\ \bibnamefont
  {Young}},\ }\href {https://doi.org/10.1103/PhysRevLett.90.227203} {\bibfield
  {journal} {\bibinfo  {journal} {Phys. Rev. Lett.}\ }\textbf {\bibinfo
  {volume} {90}},\ \bibinfo {pages} {227203} (\bibinfo {year}
  {2003})}\BibitemShut {NoStop}%
\bibitem [{\citenamefont {Fernandez}\ \emph {et~al.}(2009)\citenamefont
  {Fernandez}, \citenamefont {Mart{\'i}n-Mayor}, \citenamefont {Perez-Gaviro},
  \citenamefont {Tarancon},\ and\ \citenamefont {Young}}]{fernandez:09b}%
  \BibitemOpen
  \bibfield  {author} {\bibinfo {author} {\bibfnamefont {L.~A.}\ \bibnamefont
  {Fernandez}}, \bibinfo {author} {\bibfnamefont {V.}~\bibnamefont
  {Mart{\'i}n-Mayor}}, \bibinfo {author} {\bibfnamefont {S.}~\bibnamefont
  {Perez-Gaviro}}, \bibinfo {author} {\bibfnamefont {A.}~\bibnamefont
  {Tarancon}},\ and\ \bibinfo {author} {\bibfnamefont {A.~P.}\ \bibnamefont
  {Young}},\ }\href {https://doi.org/10.1103/PhysRevB.80.024422} {\bibfield
  {journal} {\bibinfo  {journal} {Phys. Rev. B}\ }\textbf {\bibinfo {volume}
  {80}},\ \bibinfo {pages} {024422} (\bibinfo {year} {2009})}\BibitemShut
  {NoStop}%
\bibitem [{\citenamefont {Nakamura}(2019)}]{nakamura:19}%
  \BibitemOpen
  \bibfield  {author} {\bibinfo {author} {\bibfnamefont {T.}~\bibnamefont
  {Nakamura}},\ }\href {https://doi.org/10.1103/PhysRevE.99.023301} {\bibfield
  {journal} {\bibinfo  {journal} {Phys. Rev. E}\ }\textbf {\bibinfo {volume}
  {99}},\ \bibinfo {pages} {023301} (\bibinfo {year} {2019})}\BibitemShut
  {NoStop}%
\bibitem [{\citenamefont {Nattermann}(1998)}]{nattermann:98}%
  \BibitemOpen
  \bibfield  {author} {\bibinfo {author} {\bibfnamefont {T.}~\bibnamefont
  {Nattermann}},\ }in\ \href@noop {} {\emph {\bibinfo {booktitle} {Spin glasses
  and random fields}}},\ \bibinfo {editor} {edited by\ \bibinfo {editor}
  {\bibfnamefont {A.~P.}\ \bibnamefont {Young}}}\ (\bibinfo  {publisher} {World
  Scientific},\ \bibinfo {address} {Singapore},\ \bibinfo {year}
  {1998})\BibitemShut {NoStop}%
\bibitem [{\citenamefont {Kardar}\ \emph {et~al.}(1986)\citenamefont {Kardar},
  \citenamefont {Parisi},\ and\ \citenamefont {Zhang}}]{kardar:86}%
  \BibitemOpen
  \bibfield  {author} {\bibinfo {author} {\bibfnamefont {M.}~\bibnamefont
  {Kardar}}, \bibinfo {author} {\bibfnamefont {G.}~\bibnamefont {Parisi}},\
  and\ \bibinfo {author} {\bibfnamefont {Y.-C.}\ \bibnamefont {Zhang}},\ }\href
  {https://doi.org/10.1103/PhysRevLett.56.889} {\bibfield  {journal} {\bibinfo
  {journal} {Phys. Rev. Lett.}\ }\textbf {\bibinfo {volume} {56}},\ \bibinfo
  {pages} {889} (\bibinfo {year} {1986})}\BibitemShut {NoStop}%
\bibitem [{\citenamefont {Wilson}(1975)}]{wilson:75}%
  \BibitemOpen
  \bibfield  {author} {\bibinfo {author} {\bibfnamefont {K.~G.}\ \bibnamefont
  {Wilson}},\ }\href {https://doi.org/10.1103/RevModPhys.47.773} {\bibfield
  {journal} {\bibinfo  {journal} {Rev. Mod. Phys.}\ }\textbf {\bibinfo {volume}
  {47}},\ \bibinfo {pages} {773} (\bibinfo {year} {1975})}\BibitemShut
  {NoStop}%
\bibitem [{\citenamefont {M{\'e}zard}\ \emph {et~al.}(1987)\citenamefont
  {M{\'e}zard}, \citenamefont {Parisi},\ and\ \citenamefont
  {Virasoro}}]{mezard:87}%
  \BibitemOpen
  \bibfield  {author} {\bibinfo {author} {\bibfnamefont {M.}~\bibnamefont
  {M{\'e}zard}}, \bibinfo {author} {\bibfnamefont {G.}~\bibnamefont {Parisi}},\
  and\ \bibinfo {author} {\bibfnamefont {M.}~\bibnamefont {Virasoro}},\ }\href
  {https://doi.org/10.1142/0271} {\emph {\bibinfo {title} {Spin-Glass Theory
  and Beyond}}}\ (\bibinfo  {publisher} {World Scientific},\ \bibinfo {address}
  {Singapore},\ \bibinfo {year} {1987})\BibitemShut {NoStop}%
\bibitem [{\citenamefont {Kogut}(1979)}]{kogut:79}%
  \BibitemOpen
  \bibfield  {author} {\bibinfo {author} {\bibfnamefont {J.~B.}\ \bibnamefont
  {Kogut}},\ }\href {https://doi.org/10.1103/RevModPhys.51.659} {\bibfield
  {journal} {\bibinfo  {journal} {Rev. Mod. Phys.}\ }\textbf {\bibinfo {volume}
  {51}},\ \bibinfo {pages} {659} (\bibinfo {year} {1979})}\BibitemShut
  {NoStop}%
\bibitem [{\citenamefont {Lucibello}\ \emph {et~al.}(2014)\citenamefont
  {Lucibello}, \citenamefont {Morone},\ and\ \citenamefont
  {Rizzo}}]{lucibello:14}%
  \BibitemOpen
  \bibfield  {author} {\bibinfo {author} {\bibfnamefont {C.}~\bibnamefont
  {Lucibello}}, \bibinfo {author} {\bibfnamefont {F.}~\bibnamefont {Morone}},\
  and\ \bibinfo {author} {\bibfnamefont {T.}~\bibnamefont {Rizzo}},\ }\href
  {https://doi.org/10.1103/PhysRevE.90.012140} {\bibfield  {journal} {\bibinfo
  {journal} {Phys. Rev. E}\ }\textbf {\bibinfo {volume} {90}},\ \bibinfo
  {pages} {012140} (\bibinfo {year} {2014})}\BibitemShut {NoStop}%
\bibitem [{\citenamefont {Marinari}\ \emph {et~al.}(2024)\citenamefont
  {Marinari}, \citenamefont {Martin-Mayor}, \citenamefont {Parisi},
  \citenamefont {Ricci-Tersenghi},\ and\ \citenamefont
  {Ruiz-Lorenzo}}]{marinari:24}%
  \BibitemOpen
  \bibfield  {author} {\bibinfo {author} {\bibfnamefont {E.}~\bibnamefont
  {Marinari}}, \bibinfo {author} {\bibfnamefont {V.}~\bibnamefont
  {Martin-Mayor}}, \bibinfo {author} {\bibfnamefont {G.}~\bibnamefont
  {Parisi}}, \bibinfo {author} {\bibfnamefont {F.}~\bibnamefont
  {Ricci-Tersenghi}},\ and\ \bibinfo {author} {\bibfnamefont {J.~J.}\
  \bibnamefont {Ruiz-Lorenzo}},\ }\href
  {https://doi.org/10.1088/1742-5468/ad13fe} {\bibfield  {journal} {\bibinfo
  {journal} {J. Stat. Mech.}\ }\textbf {\bibinfo {volume} {2024}},\ \bibinfo
  {pages} {013301} (\bibinfo {year} {2024})}\BibitemShut {NoStop}%
\end{thebibliography}%

\end{document}

\title{Low energy excitations in a long prism geometry:\\ testing the lower critical dimension of the Ising spin glass\\[2mm]\normalsize Supporting Information}

\author{M.~Bernaschi, L.A.~Fern\'{a}ndez, I.~Gonz\'{a}lez-Adalid Pemart\'{i}n, V.~Mart\'{i}n-Mayor, G.~Parisi and F.~Ricci-Tersenghi}
\thanks{Corresponding author: I.~Gonz\'{a}lez-Adalid Pemart\'{i}n. E-mail: i.gonzalez@iac.cnr.it}

\maketitle

\renewcommand{\thesection}{S\arabic{section}}
\renewcommand{\theequation}{S\arabic{equation}}
\renewcommand{\thefigure}{S\arabic{figure}}
\renewcommand{\thetable}{S\arabic{table}}

In this Supporting Information, we will give additional details about our work.
\begin{itemize}
\setlength{\itemsep}{0pt}
  \setlength{\parsep}{0pt}
\item In Sect.~\ref{SMsect:simulations} we describe our simulations in $L^2\times M$ lattices and the algorithms that we employed. Since we have used both Periodic Boundary Conditions (PBC) and Open Boundary Conditions (OBC), we give in Sect.~\ref{SMsect:OBC} specific details about our OBC simulations.

\item  In Sect.~\ref{SMsec:equiltests} we discuss the performance of the algorithm used in the simulation and present some thermalization tests.

\item In Sect.~\ref{SMSect:CorrOBC}, we explain how we have analyzed the correlation functions computed with Open Boundary Conditions (OBC) along the Z dimension. Indeed, as explained in the main text (see also Sect.~\ref{SMsect:1D-IEA-model}), OBCs have proven very advantageous. However, this choice of boundary conditions breaks translational invariance, and some care is needed in the analysis.

\item Sect.~\ref{SMsect:IntegralEstimators} explains our use of integral estimators to estimate statistical errors for the correlation length reliably.

\item The quantitative computation of the sub-leading contributions to the correlation function is addressed in Sect.~\ref{SMsect:determinanti}.

\item In Sect.~\ref{SMsect:1D-IEA-model}, we derive some (not widely known) results for the one-dimensional Ising-Edwards-Anderson model that have turned out to be of great importance for our study.

\item At this point, we shall be ready to discuss in Sect.~\ref{SMsect:finite-M} how large $M$ should be (given our statistical accuracy) if our data are to be representative of the large-$M$ limit.

\item A different approach, which employs very short prisms with OBC, will be found in Sect.~\ref{SMsect:finite-M-OBC}.

\item Finally, in Sect.~\ref{SMsect:surface-tension}, we explain that for models without Goldstone bosons, the correlation length for our rectangular right prisms scales very differently from what was discussed (and found) in the main text.
\end{itemize}

\section{Description of our simulations}\label{SMsect:simulations}
Hereafter, we describe our CUDA implementation for simulating the three-dimensional tubular Ising spin glass on the largest lattices (for values of $L<16$, a 128-sample multispin code is employed on the CPU instead). The code implements two levels of parallelism: we use multispin coding to pack 64 spins into 64-bit words, and then use CUDA threads to concurrently update the {\em even} or {\em odd} spins according to the classic checkerboard decomposition (Ref.~\cite{barma:77} provides an early example).

Multiple {\em replicas} and {\em samples} can be simulated in a single execution; the actual number of samples depends on the resources (mainly memory) available on the GPU. We started with the classic Metropolis algorithm, using random numbers generated on the GPU following the procedure described in Ref.~\cite{bernaschi:24}. The code has been enhanced by adding a Parallel Tempering (PT) procedure that simulates multiple copies of the system at different temperatures concurrently.

We recall that in the PT, copies at neighboring temperatures swap their temperatures with probability $p=\text{min}[1,{\mathrm e}^{(\beta-\beta')(E-E')}]$ where $\beta$ and $E$ are the inverse of the temperature and the energy of one copy; $\beta'$ and $E'$ have the same meaning for the other copy. Copy exchange moves permit copies to diffuse in the temperature space. Although the PT provides a speedup, the dynamics of the spin glass remain slow, especially at low temperatures. In other systems, such as ferromagnets, a widely adopted approach to achieving significant acceleration is to leverage cluster algorithms \cite{swendsen:87, wolff:89} that update multiple spins simultaneously.

\begin{figure}
\centering\includegraphics[width=\columnwidth]{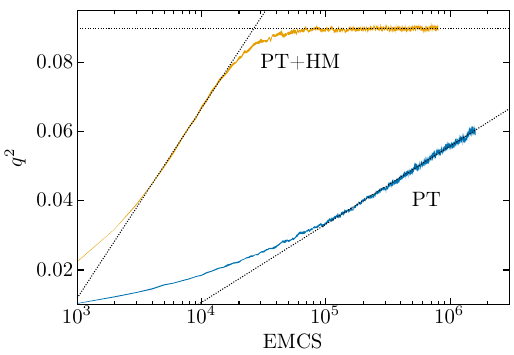}
    \caption{Overlap squared evolution with and without Houdayer moves in a $12^2\times 320$ lattice at $T=0.7$. We simulate 8 identical sets of 128 samples with both methods. The displayed errors are estimated by comparing independent runs and therefore do not include sample fluctuations. Dashed lines serve as a guide to the eye.}
    \label{fig:TAUSOBS}
\end{figure}

Houdayer introduced a clustering algorithm \cite{houdayer:01} for a spin glass that applies cluster moves between replicas at the same temperature while conserving total energy and maintaining detailed balance. Houdayer’s algorithm requires an underlying geometry with a percolation threshold (i.e., the critical value of the probability of occupation at which a large-scale connected cluster first emerges in the system) above 50\% to provide actual acceleration, which is not the case for a three-dimensional Ising spin glass in a cubic lattice~\cite{munster:24}. As a matter of fact, our case is somewhat borderline since we have a 3D lattice but with a very elongated rectangular shape, so we decided also to add the Houdayer move to the simulation of our system.

This has required significant code extensions. In particular, we had to include new CUDA {\em kernels} to compute the overlap between replicas, unpack the results of the overlap (which are computed in the multispin coding format), create the clusters, and apply an algorithm {\it a la} Swendsen-Wang using the approach proposed in \cite{komura:15}. Finally, the results of the cluster update are packed in the multispin coding format and applied to the spins. The cluster algorithm \cite{komura:15} has been modified to account for the possibility that, along the Z direction, the boundary conditions may be periodic or open (see Section \ref{SMsect:OBC}).

Moreover, we made a few other changes to improve the clustering algorithm's performance, such as using a one-byte data type (\verb|char|) instead of the original 4-byte data type (\verb|int|).  The Houdayer's move is carried out every 16 Metropolis iterations. The replicas are divided into two equal-size groups, and from each group, two replicas are randomly selected and involved in the move. The two groups remain unchanged throughout the simulation. The total number of replicas is 8 for each sample.

Our tests confirmed that the Houdayer's move dramatically increases the thermalization of the system below the critical temperature, as confirmed by Figure \ref{fig:TAUSOBS}, which compares the evolution of the squared overlap $q^2$ for the case $12^2 \times 320$ obtained using only the PT and combining the PT and Houdayer's move. On the other hand, since the GPU-based clustering algorithm remains computationally expensive, we limit its application to temperatures below the critical one, as the potential advantage is negligible (if any) at high temperatures and thus does not justify the computational burden. Computing the cluster only for a subset of temperatures and using the combination of PT and cluster moves to speed up the dynamics and reduce computational cost is a natural choice that has been explored previously~\cite{zhu:15}. The total amount of GPU-hours required by the simulations described in the present paper exceeds 400000.

More details about our code can be obtained by direct inspection of the source code available from~\cite{github}.

\begin{table}[htb]
    \centering
        \caption{\boldmath Simulation details for the different system sizes $L^2\times M$ and Boundary Conditions along the Z-axis (BC-Z). We simulated NB groups of NS samples using a Multi-Sample code. Samples in the same group share the random numbers. We consider NR replicas grouped in two sets and construct clusters using one replica of each set. For the PT, we use NT copies of the system in the temperature range $[0.7,2]$. For each PT step, we perform NMET Metropolis sweeps and one HM, defining one Monte Carlo step, EMCS.}
    \label{tab:SimDetails}
    \begin{tabular}{ccccccccc}
        $L$ & $M$ & BC-Z & NB & NS & NR & NT & NMET & EMCS \\
        \midrule
         $4$ & $192$ & PBC & 256 & 128 & 4 & 35 & 10 & 300000 \\
         $6$ & $192$ & PBC & 256 & 128 & 4 & 35 & 10 & 300000\\
         $8$ & $192$ & PBC & 256 & 128 & 8 & 35 & 16 & 300000 \\
         $12$ & $320$ & PBC & 256 & 128 & 4 & 60 & 10 & 1024000 \\
         $16$ & $512$ & PBC & 64 & 16 & 8 & 114 & 16 & 2457600 \\
         $16$ & $48$ & OBC & 1024 & 16 & 8 & 40 & 16 & 1310720 \\
         $24$ & $88$ & OBC & 128 & 16 & 8 & 82 & 16 & 12255232 \\
         \bottomrule
   \end{tabular}
\end{table}

\subsection{Open boundary conditions}\label{SMsect:OBC}
\begin{figure}[htb]
\centering\includegraphics[width=\columnwidth]{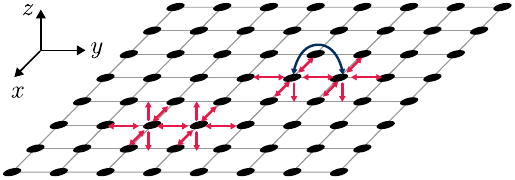}
    \caption{The open boundary condition on the last $z$ plane. The link to $z+1$ is replaced by an additional link along the $y$ direction.}
    \label{fig:OBC}
\end{figure}

Open boundary conditions require a change on the first and the last plane along the Z direction to maintain the number of interacting spins equal to the other planes. There are several possible alternatives to fulfill this requirement. We choose to add one more link along the $y$ direction. A sketch of this solution is shown in Figure \ref{fig:OBC}. Its main advantage is that code changes are minimal.

In the beginning, we tested three variants of this scheme that differ for the value of the coupling along the auxiliary link: i) \textit{weak}, the value of the coupling is the opposite of the value of the coupling along $y$, so that they cancel each other; ii) \textit{strong}, the value of the coupling is the same as the coupling along $y$ so that they reinforce each other; and iii) \textit{independent}, the value of the coupling is independent of the value of the coupling along $y$.

The tests revealed no differences in the physical observables, although the value of $z_{\text{discard}}$ (see \ref{SMSect:CorrOBC}) is different in the various cases, so we decided to use a single variant (the third one, $independent$) for the production phase.

\subsection{On our equilibration test}\label{SMsec:equiltests}

A significant challenge for spin glass simulations is ensuring that thermal equilibrium has been reached. In this section, we first describe some specific tests performed on a $12^2\times 320$ lattice using a CPU-based code. This code simulates 128 different samples in parallel but uses basically the same algorithm as in GPU: a Metropolis update followed by a Houdayer's move on a single cluster chosen at random, within a PT procedure.

In Fig~\ref{fig:TAUSOBS}, we plot the squared overlap $q^2$ with and without cluster update. Although we cannot measure the equilibrium time from the data without clusters, we observe that the inflection point shifts by a factor of 200, while the straight lines tangent to the inflection points cross the horizontal limit line at points corresponding to a factor of 1200 increase.

\begin{figure}[htb]
\centering\includegraphics[width=\columnwidth]{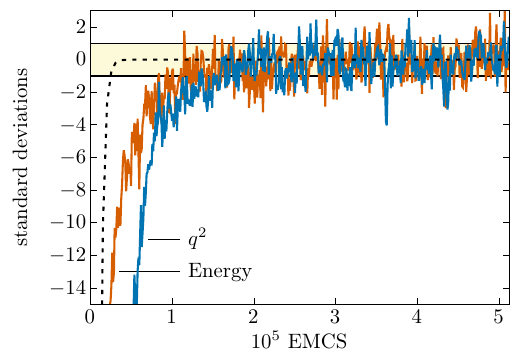}
    \caption{Number of standard deviations from equilibrium (computed using the second half of data) for the energy and the squared overlap versus the number of Elementary Monte Carlo Steps for the $12^2\times 320$. Points correspond to averages over 100 measurements. For comparison, we also plot a pure exponential with an amplitude equal to the number of deviations of $q^2$ in the first block of 100 measurements, with a decay rate equal to the Temperature Random Walk autocorrelation time.}
    \label{fig:SUBSTRACTED}
\end{figure}

We have also studied the autocorrelation time of the Temperature Random Walk, $\tau_\text{TRW}$, generated by the Parallel Tempering, which has proved extremely useful in a local Parallel Tempering simulation to monitor thermalization~\cite{billoire:11}. As we cannot reach equilibrium without using Houdayer moves in the $12^2\times 320$ lattice, we first compare $\tau_\text{TRW}$ in an $8^2\times 64$ lattice at equilibrium, with and without cluster moves. While without clusters, the autocorrelation time spans from 1000 to 20,000  EMCS, the use of clusters reduces $\tau_\text{TRW}$ to the range 300 to 400 EMCS.

Unfortunately, unlike in a local algorithm simulation, we have observed that the TRW autocorrelation time is not coupled to the equilibration time of observables. In Fig~\ref{fig:SUBSTRACTED}, we present the evolution of the energy and squared overlap, $q^2$, in the $12^2\times 320$ lattice at $T=0.7$ in the first half of the simulation, subtracting the mean values of the second half. In this way, we cancel out the sample-to-sample fluctuations. To better compare the two quantities, we plot the deviation from zero in units of their respective standard deviations. The points are computed by averaging over 100 contiguous measurements (corresponding to 1000 EMCS). The peach colored band corresponds to the $\pm 1~\sigma$ band. We first remark that the equilibration times for both quantities are similar, despite the energy being a local quantity and $q^2$ is not. However, both times are much greater than the TRW autocorrelation time (about 4000--5000 EMCS for this system). For a graphical comparison, we plot a plain exponential with this decay rate, and an amplitude similar to that of $q^2$ in the first block of data.

We remark that in this work, we have used a rather strict thermalization criterion: we double-check that there are no statistically significant differences between the averages of the observables computed using large blocks of contiguous data.

As the two quantities considered above behave very similarly, we found that it suffices for the energy to meet the requirement.

In the $12^2\times 320$ lattice, we have also tried to characterize the evolution of the observables at large times using a plain exponential, but this is not possible, indicating that the signal is good enough to perform a fit. We can obtain a more successful fit by trying a stretched exponential: $\sim\exp[-(x/a)^\alpha]$. Although the data accuracy does not allow a safe determination of the parameters, the fits point to a value of $\alpha \simeq 0.4$, clearly different from 1. This suggests a strong dispersion of the equilibration time of different samples (although the $\tau_\text{TRW}$ are similar). We have tried to measure the exponential decay time of $q^2$ from a set of $128$ samples on the $12^2\times 320$ lattice referred to above. We observe a large dispersion in decay times, with a few cases where the equilibration time is 6 times the TRW autocorrelation time, suggesting a long tail in the probability distribution, consistent with the behavior of the sample mean.

\section{Correlation function with Open Boundary Conditions}\label{SMSect:CorrOBC}

Periodic boundary conditions (PBC) are the most commonly used boundary conditions in simulations, because
they ensure translational invariance: $C^{(n)}(z_1,z_2)$ is a function of the plane-to-plane distance
$|z_1-z_2|$ for any prism length $M$. Furthermore, with PBC, the infinite-$M$ limit is approached exponentially fast in $M/\xi$. In fact, all our previous experience with non-disordered systems (wrongly) suggested that finite-$M$ effects would be very strongly suppressed with PBC. We were expecting that extremely accurate data would be necessary to resolve any residual $M$-dependence  in prisms with $M> 6 \xi$. However, the real data, see Sect.~\ref{SMsect:finite-M} below, turned out to differ significantly from our expectations. Fortunately, the analysis of our 1D toy model in Sect.~\ref{SMsect:1D-IEA-model} not only explained the difficulties with PBC but also suggested a radical cure: changing to open boundary conditions (OBC). Yet,  the analysis of correlation functions for systems with open-boundary conditions was somewhat unfamiliar to us. We briefly describe our analysis for OBC here.

\begin{figure}[htb]
    \centering
    \includegraphics[width=\columnwidth]{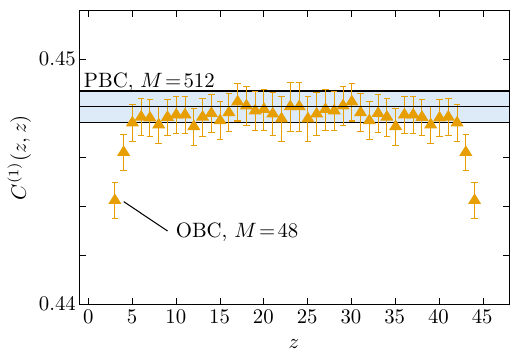}
    \caption{Diagonal elements of the correlation matrix $C^{(1)}(z_1,z_2)$ as a function of $z$ for a system $16\times 16\times 48$ with OBC. The horizontal band corresponds to the value of the correlation function at zero distance for a system of size $16\times 16\times 512$ with PBC. Both data sets, OBC and PBC, were obtained at the same temperature $T=0.7$.}
    \label{fig:C(z,z)_OBC}
\end{figure}

By definition, the correlation function is a symmetric matrix, the plane labels $z_1$ and $z_2$ being the matrix indices: $C^{\text{OBC},(n)}(z_1,z_2)=C^{\text{OBC},(n)}(z_2,z_1)=C^{\text{OBC},(n)}(M-1-z_2,M-1-z_1)$, where the last equality follows from the inversion symmetry with respect to the center of the prism. We have enforced these symmetries in our statistical analysis by averaging our numerical estimates over the two equivalent arrangements for the matrix indices $(z1,z2)$ and $(M-1-z_2,M-1-z_1)$.

Nevertheless, the OBC correlation matrix is more complex than its PBC counterpart, which can be treated as a vector depending on the single index $r=z_2-z_1$. However, quite amazingly, for the 1D toy-model, the OBC correlation matrix is vector-like, regardless of $M$, $\left.C^{\text{1D},(n)}\right|_{\text{OBC}}=\text{e}^{-|z_2-z_1|/\xi_{n}}$, see \eqref{SMeq:C2n-1D}.

The beautiful simplicity of the 1D OBC correlation matrix gave us hope that translational invariance could be recovered in the prism with OBC as well. To asses
how this happens, we show in Fig.~\ref{fig:C(z,z)_OBC} the diagonal terms of $C^{\text{OBC},(1)}$. As soon as one gets away from the prism borders, at $z=0$ and
$z=M-1$, the diagonal terms of the PBC correlation matrix become not only independent of $z$ (within an error of $2\, 10^{-5}$), but also compatible with
$C^{\text{PBC},(1)}(r=0)$ from a prism with $M\approx 11.63\xi$ (and hence representative of the limit $M\to\infty$, see Sect.~\ref{SMsect:finite-M}). This encouraging result for the diagonal terms suggests considering the off-diagonal terms as well, see Fig.~\ref{fig:C(z,z+r)_OBC}. It turns out that we can choose a safe distance for the borders, $z_{\text{discard}}$, such that if we choose
the matrix indices $z_1<z_2$ with $z_1>z_\text{discard}$ and $z_2<M-1-z_\text{discard}$, the OBC correlation matrix becomes a function of $|z_2-z_1|$ that (within our statistical precision) equals the PBC correlation function in a long prism. Hence, to obtain a vector-like correlation $C^{(n)}(r)$ from our OBC simulations, we have chosen $z_\text{discard}=5$ (for $L=16$) or $z_\text{discard}=10$ (for $L=24$), and we have averaged over all pairs $(z_1,z_2=z_1+r)$ that verify the restriction imposed by $z_\text{discard}$.

\begin{figure}[htb]
    \centering
    \includegraphics[width=\columnwidth]{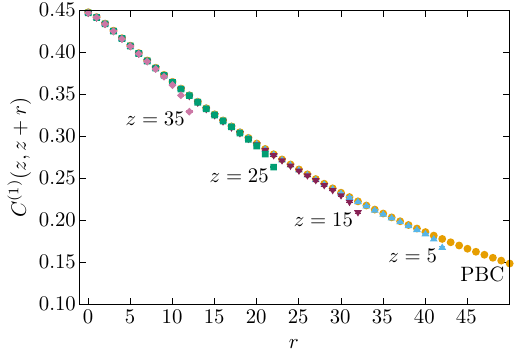}
    \caption{Rows of the matrix $C^{(1)}(z_1=z,z_2=z+r)$ as a function of the distance $r=z_2-z_1$ for several values of $z=5,15,25$, and $35$. Data for a system $16\times 16\times 48$ with OBC. Data for a system of size $16\times 16\times 512$ with PBC are also included to compare the two boundary conditions. Both data are at temperature $T=0.7$.}
    \label{fig:C(z,z+r)_OBC}
\end{figure}

\section{Integral estimators for the correlation length}\label{SMsect:IntegralEstimators}

When one attempts to extract the correlation length from a fit to an exponential function, an unexpected problem appears. We are referring to large statistical
correlations between the numerical determination of $C^{(n)}(r)$ at different distances $r$. In principle, this problem may be addressed by obtaining the covariance matrix for $C^{(n)}(r)$ at different $r$ and minimizing the fit's figure of merit $\chi^2$. The problem is that $\chi^2$ should be computed using the inverse of the covariance matrix. Unfortunately, obtaining the covariance matrix with sufficient accuracy to ensure safe matrix inversion can be daunting for a large prism length $M$. Using only the diagonal elements of the covariance matrix could look like a reasonable alternative. However, it often leads to fits with $\chi^2$ significantly smaller than the number of degrees of freedom. This not only complicates the computation of statistical uncertainties for the fit parameters but also makes it difficult to assess the reliability of the fit. Integral estimators of the correlation length were introduced as a simple and effective remedy for this situation~\cite{cooper:82,edwards:89}.

\begin{figure}[htb]
    \centering
    \includegraphics[width=\columnwidth]{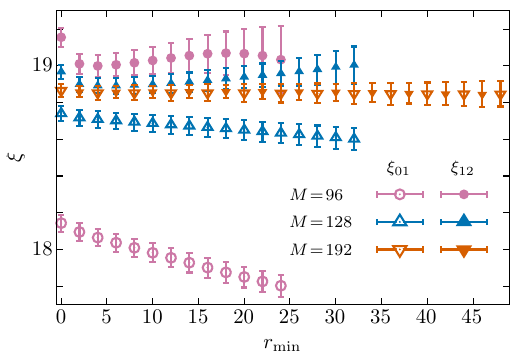}
    \caption{Integral estimator of the correlation length for a system of size $8\times 8 \times M$ with PBC at temperature $T=0.7$. Open symbols correspond to $\xi_{0,1}$, while closed ones are $\xi_{1,2}$. Different symbols (and colors) are used to represent different values of $M$. For every $M$ value, there are two curves (same color but open or filled symbols). For clarity, we plot only one point every two and, for $M=192$, every four alternating estimators.}
    \label{fig:xi_vs_r_min_different_M}
\end{figure}

We shall explain our integral estimators for PBC and OBC separately, as they differ. The estimators should meet two requirements regardless of the choice of boundary conditions: (i) should the correlation function be \emph{exactly} of exponential form $C(r)=constant\times\text{e}^{-r/\xi}$, the estimator should output $\xi$, and (ii) the estimator should make it possible to discard distances $r<r_{\text{min}}$, so that short-distance deviations from the purely exponential form can be assessed. Note that in the case of PBC, the distances $r>M-r_\text{min}$ should also be discarded. Finally, note that the statistical errors for these estimators can be easily computed using a bootstrap method.

\subsection{Estimators for PBC}

For these boundary conditions, we simply modify the standard choice~\cite{edwards:89}  in such a way that allows us to include the minimal distance $r_\text{min}$. Let us assume a purely exponential decay, as adapted to PBC
\begin{equation}\label{SMeq:pure-exponential-PBC}
    C(r)=A \,[\,\text{e}^{-r/\xi}\ +\ \text{e}^{-(M-r)/\xi}\,]\,,
\end{equation}
(here, $C$ is a shorthand for whatever correlation function we are analyzing, and $A$ is an amplitude). The crucial observation is that, if we consider only the range
of $r_\text{min}\leq r < M-r_\text{min}$, we can
rewrite \eqref{SMeq:pure-exponential-PBC} as
\begin{equation}\label{SMeq:pure-exponential-PBC-rewrite}
    C(r_{\text{min}}\!+\!r)=A \,\text{e}^{-r_{\text{min}}/\xi}\,[\,\text{e}^{-r/\xi} +\text{e}^{-(M-2 r_{\text{min}}-r)/\xi}\,]\,,
\end{equation}
with $r=0,1,\ldots,M-2r_{\text{min}}-1$. We see that introducing the short-distance cutoff $r_{\text{min}}$ merely amounts to a redefinition of the constant amplitude and a shorter effective length of the system. Hence, we can directly use the standard formulae that we recall next.

Let us define
\begin{equation}
M^*=M-2r_{\text{min}}\,,\quad k_{\text{min}}=\frac{2\pi}{M^*}\,,\label{SMeq:kmin-def}
\end{equation}
then we obtain $G_n$, the Fourier transform  of $C(r)$, for $n=0,1,2$:
\begin{equation}
G_n=\sum_{r=0}^{M^*-1}\,C(r+r_{\text{min}})\cos (n k_\text{min}r)\,.
\end{equation}
The next step is to extract the so-called pole-mass term $m_\text{p}$ by assuming that $G_n$  has the form of a free-field propagator $G_n\propto 1/[m_\text{p}^2+4\sin^2 ( n\,k_\text{min}/2)]$, at least for small $n$.  In order to compute $m_\text{p}$, let $R_{0,1}=G_0/G_1$ and $R_{1,2}=G_1/G_2$. Then the two estimators of the pole mass are
\begin{eqnarray}
m_{\text{p};0,1}&=& 2\sin(k_\text{min}/2)/\sqrt{R_{0,1}-1}\,,\\
m_{\text{p};1.2}&=& 2\sin(k_\text{min}) \sqrt{\frac{1-R_{1,2}/\big(4\,\cos^2(k_\text{min}/2)\big)}{R_{1,2}-1}},
\end{eqnarray}
where $k_\text{min}$ was defined in \eqref{SMeq:kmin-def}. Finally, we use the
general relationship between the pole mass and the
correlation length
\begin{equation}
\xi=\frac{1}{2 \,\text{arcsinh}\,(m_{\text{p}}/2)}\,,
\end{equation}
to obtain the two integral estimators $\xi_{0,1}$ and $\xi_{1,2}$.

The statistical compatibility of $\xi_{0,1}$ and $\xi_{1,2}$ is an important test of consistency. Nevertheless, of the two estimators only $\xi_{1,2}$
remains unchanged if a constant background is added to \eqref{SMeq:pure-exponential-PBC}.
This insensitivity of $\xi_{1,2}$ makes it more convenient in the presence of the constant background predicted by \eqref{SMeq:three-terms}, below.

\subsection{Estimators for OBC}
We generalize here the PBC integral estimators to the OBC case, where a pure exponential decay goes as $C(r)=A\text{e}^{-r/\xi}$ [recall that the image term in \eqref{SMeq:pure-exponential-PBC}, namely $\text{e}^{-(M-r)/\xi}$, is absent with these boundary conditions].

The basic quantities for this choice of boundary conditions are
defined in terms of $M_{\text{\tiny OBC}}=M-r_\text{min}$ and $k_\text{\tiny OBC}=2\pi/M_{\text{\tiny OBC}}$
\begin{eqnarray}
N&=&\sum_{r=0}^{M_\text{\tiny OBC}-1}\,C(r+r_{\text{min}})\, \big[1-\cos (k_{\text{\tiny OBC}}\,r)\big]\,,\\
D&=&\sum_{r=0}^{M_\text{\tiny OBC}-1}\,C(r+r_{\text{min}})\, \sin(k_{\text{\tiny OBC}}\,r)\,.\\
\end{eqnarray}
Notice that the kernels for $N$ and $D$
were chosen to suppress the contributions of $C(r\approx r_{\text{min}})$ and $C(r\approx M)$,
in order to avoid any residual problem with translational invariance in
the computation of $C(r)$.  Next, we compute the intermediate quantity
\begin{equation}
    \mathcal{W}=\frac{N}{D}\,\frac{\sin k_{\text{\tiny OBC}}}{\big(1-\cos k_\text{\tiny OBC}\big)}\,,
\end{equation}
and from it we obtain our sought integral estimator
\begin{equation}
    \xi=1\Big/\log\left(\frac{\mathcal{W}+1}{\mathcal{W}-1}\right)\,.
\end{equation}

\section{Subdominant correlation lengths}\label{SMsect:determinanti}

We provide here a quantitative answer to the problem outlined in the End Matter Section, namely, the computation of subdominant corrections to the long-distance behavior of the correlation function $C^{(1)}(r)=a_1\text{e}^{-r/\xi}+ a_2 \text{e}^{-r/\xi_{n=1;k=2}}+ \ldots$. Our goal here is to compute $\xi_{n=1;k=2}$.
To that purpose, we shall adopt a trick that \emph{eliminates} the leading term  $\text{e}^{-r/\xi}$.

Specifically, let us consider three functions of the plane overlap, $A_1(z)=Q(z), A_2(z)=Q^3(z)$, and $A_3(z)=\text{sign}[Q(z)]$. All three change sign under the global spin reversal of either of the two replicas used to compute the spin overlap.
Therefore, the three functions $A_{i=1,2,3}$ are represented by odd-parity operators $\hat{A}_{i=1,2,3}$ in the transfer-matrix formalism~\cite{parisi:94}. From the $A_i(z)$, we form the matrix-propagator $\mathcal{C}(r)$ with matrix elements
\begin{equation}\label{SMeq:Cmaxtrix-elements}
    C_{i,j}(r)=\overline{\langle A_i(z_1) A_j(z_2=z_1+r)\rangle}\,.
\end{equation}
In particular, $C_{1,1}(r)$ is the correlation function $C^{(1)}(r)$ that has been considered for  most of this work.
We shall consider either a single $3\times 3$  matrix or the three $2\times 2$ matrices that can be formed with our three operators.  The transfer matrix prediction for the large-$M$ limit is
\begin{equation}\label{SMeq:C-expansion}
\mathcal{C}= \sum_{k=1}^{\infty} \mathcal{M}_k\, \text{e}^{-E^{\text{odd}}_k\,r}\,,\quad E^{\text{odd}}_k=\frac{1}{\xi_{n=1;k}}.
\end{equation}
where we have used the same notation as the End-Matter Section, and the $\mathcal{M}_k$ are rank-\emph{one} matrices with matrix elements
\begin{equation}
    [\mathcal{M}_k]_{i,j} = \langle 0|\hat{A}_i|k\rangle \langle k|\hat{A}_j|0\rangle\,.
\end{equation}
The superscript in $E_k^{\text{odd}}$ emphasizes that only states with \emph{odd}-parity appear in the expansion~\eqref{SMeq:C-expansion}. Note that $E_1^{\text{odd}}$ is the inverse of the correlation length $\xi$, while the $E_{k>1}^{\text{odd}}$ are the inverses of the sub-leading correlation lengths.

\begin{figure}[htb]
    \centering
    \includegraphics[width=\columnwidth]{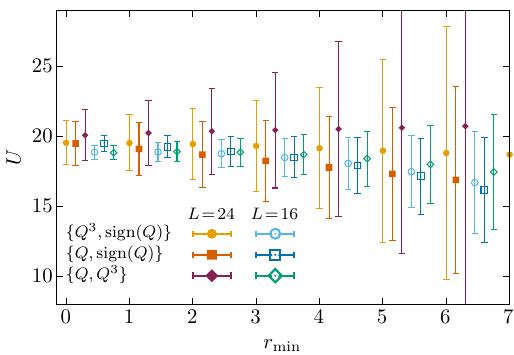}
    \caption{Estimation of the ratio of correlation length $U$, see \eqref{SMeq:Udef}, as a function of $r_\text{min}$ for OBC (data for $T=0.7$ and $L=16,24$). Different symbols (and colors) are used for each of the subsets of observables under consideration. Closed symbols (and hot colors) are used to represent $L=24$, while open symbols (and cold colors) are used for $L=16$. For clarity, data points are slightly shifted horizontally.}
    \label{fig:U_vs_r_min}
\end{figure}

Let us now show that, if we compute the determinant of $\mathcal{C}(r)$, a major simplification occurs. The simplification is because the $\mathcal{M}$ matrices are of rank one (i.e., every column in the matrix is proportional to the first column).
Let $\text{Col}_j(\mathcal{M}_k)$ be the $j$-th column of $\mathcal{M}_k$, the determinant, which is a multi-linear function of the matrix columns, is expanded as
\begin{eqnarray}
    \det{\mathcal{C}(r)}&=&\sum_{k_1,k_2,k_3}\,\text{e}^{-(E^{\text{odd}}_{k_1}+E^{\text{odd}}_{k_2}+E^{\text{odd}}_{k_3})\,r}\,\label{SMeq:det-expansion}\\
    &\times& \det{\Big[\text{Col}_1 (\mathcal{M}_{k_1}),\text{Col}_2(\mathcal{M}_{k_2}),\text{Col}_3(\mathcal{M}_{k_3})\Big]}\,.\nonumber
\end{eqnarray}
Notice, now, that all terms in the expansion~\eqref{SMeq:det-expansion} with two coincident indices are zero, because the corresponding matrix columns are proportional. Therefore, the leading behavior is
\begin{equation}
 \det{\mathcal{C}(r)}=A\, \text{e}^{-(E^{\text{odd}}_{1}+E^{\text{odd}}_{2}+E^{\text{odd}}_{3})\,r}\,+\ldots
\end{equation}
where $A$ is an amplitude. An analogous argument gives us the leading behavior of the $2\times 2$ minors of $\det{\mathcal{C}}(r)$
\begin{equation}
 \det{\mathcal{C}^{(i,j)}(r)}=A^{(i,j)} \text{e}^{-(E^{\text{odd}}_{1}+E^{\text{odd}}_{2})\,r}\,+\ldots
\end{equation}

This strategy is put to work in Fig.~\ref{fig:U_vs_r_min}, where we use the OBC integral estimator to obtain the correlation length of the minors and compute from it the dimensionless quotient
\begin{equation}\label{SMeq:Udef}
U=\frac{E^{\text{odd}}_1+E^{\text{odd}}_2}{E^{\text{odd}}_1}=1+\frac{\xi}{\xi_{n=1;k=2}}\,.
\end{equation}
The estimate of $\xi$ in Fig.~\ref{fig:U_vs_r_min} is obtained from $C^{(1)}(r)$.  $U$ turns out to be remarkably independent of (i) the minor choice, (ii) $r_\text{min}$, and (iii) the transverse size of the prism $L$. The ratio $(E^{\text{odd}}_{1}+E^{\text{odd}}_{2}+E^{\text{odd}}_{3})/E^{\text{odd}}_{1}$ turns out to be much larger ($\approx 60$), which indicates that it is safe to neglect all correlation lengths $\xi_{n=1;k>2}$.
Two major conclusions follow from this analysis:
\begin{enumerate}
\item  $\xi_{n=1;k=2}/\xi$ tends to a positive constant when $L$ grows, and
\item $\xi_{n=1;k=2}\approx \xi/17$. This large ratio explains \emph{a posteriori} the large accuracy, of a few percent, that we reached in the computation of $\xi$. As soon as $r_\text{min}$ becomes larger than $\xi_{n=1;k=2}$ (and this is easy, given the smallness of $\xi_{n=1;k=2}$), the contribution of this correction term becomes negligible as compared to the main contribution.
\end{enumerate}

\section{The 1D Ising Edwards-Anderson model}\label{SMsect:1D-IEA-model}

The Droplet Model (DM) regards the 1D Ising-Edwards-Anderson model as the actual effective model for the prism, once length scales are renormalized by a factor $L$. In order to connect the DM prediction with the 1D-chain, we shall need to recall the following scaling relation from the main text:
\begin{equation}\label{SMeq:scaling-2}
\xi(L) \propto L^{a_D}\,.
\end{equation}
Instead, from the point of view of the Replica Symmetry Breaking theory, the 1D-chain merely provides a useful toy model.

Specifically, we introduce an Edwards-Anderson Ising spin chain of length $ M$, the same as the prism length. In this toy model, every plane $x_3=z$ in $D=3$ is represented by a spin $\sigma_z=\pm 1$, in the hope that $C^{(n)} (z_1,z_2)\approx  C^{\text{1D},(n)} (z_1,z_2) =\overline{\langle \sigma_{z_1} \sigma_{z_2}\rangle^{2n}}$ for large $|z_1-z_2|$. The 1D spins have their own Hamiltonian, $H^\text{1D}=-\sum_{z} \mathcal{J}_{z,z+1}\sigma_z\sigma_{z+1}$, with the corresponding coupling distribution and effective temperature:
\begin{equation}\label{eq:H-1D}
P^\text{1D}(\mathcal{J})\underset{\mathcal{J}\ll 1}{\sim} {\cal J}^\lambda\,,\quad T_{\text{1D},L}\sim L^{-\rho}\,.
\end{equation}
Even if the 3D couplings are $J=\pm 1$, the 1D couplings ${\cal J}$ are continuously distributed.
In fact, the analysis of the 1D model, see below, reveals that the only relevant feature of $P^\text{1D}$ is its scaling at ${\cal J}=0$. Three probably not widely known results for the 1D model turn out to be of great relevance to our study:
\begin{enumerate}
\item In the $M\to\infty$ limit, $C^{\text{1D},(n)} (z_1,z_2)=\text{e}^{-|z_1-z_2|/\xi_n}$, $\xi_n$ diverges when  $T_\text{1D,L}\to 0$ as $\xi_n\sim 1/T^{1+\lambda}_{\text{1D},L}$. So, comparing with \eqref{SMeq:scaling-2},
$\rho=a_3/(1+\lambda)$. and predicts $\lambda=0$ and $\rho=1+y_D$~\cite{fisher:88b,carter:02}.
\item We have
$C^{\text{1D},(n)}\sim [C^{\text{1D,(1)}}]^{\tau_n}$ with $\tau_n\!<\!n$,
hence a multifractal with a singularity spectrum
\begin{equation}\label{SMeq:singularity-spectrum}
    \tau_n=\frac{I_{2n}}{I_2}\,,\quad I_k=\int_0^\infty \mathrm{d}u\, u^\lambda (1-\text{tanh}^{k} u)\,.
\end{equation}
Plugging the DM's prediction $\lambda=0$ in \eqref{SMeq:singularity-spectrum} one gets for the multifractal spectrum $\tau_n=\sum_{k=0}^{n-1}1/(2k+1)$.
\item With open boundary conditions (OBC), the exponential decay, $C^{\text{1D},(n)}(z_1,z_2)=\text{e}^{-|z_1-z_2|/\xi_n}$, holds true even for finite values of $M$. Instead, for periodic boundary conditions (PBC), it holds only in the $M\to\infty$ limit, and $C^{\text{1D,PBC,}(n)}(z_1,z_2)$ is plagued by finite-$M$ corrections.
This probably explains the very large $M$ values required for the PBC prism simulation.
\end{enumerate}

To ease the discussion, we have divided this paragraph into three sections. Sect.~\ref{SMsubsect:detailed-def} provides some basic definitions. The behavior of the 1D-chain in the thermodynamic limit (i.e. $M\to\infty$) is addressed in Sect.~\ref{SMsubsect:scaling-limit}. How the $M\to\infty$ limit is approached is studied in Sect.~\ref{SMsubsect:PBC}.

\subsection{Detailed definitions}\label{SMsubsect:detailed-def}

In this section, we shall consider the Hamiltonian for a spin chain with Open Boundary conditions (OBC)
\begin{equation}\label{SMeq:OBC-H}
H_{\text{\tiny OBC}}^\text{1D}=-\sum_{z=0}^{M-2} {\cal J}_{z,z+1}\sigma_z\sigma_{z+1}\,,\quad \sigma_z=\pm 1\,.
\end{equation}
or for periodic boundary conditions (PBC)
\begin{equation}
H^\text{1D}=-\sum_{z=0}^{M-1} {\cal J}_{z,z+1}\sigma_z\sigma_{z+1}\,,\quad \sigma_M\equiv \sigma_0\,.
\end{equation}
With either set of boundary conditions, we consider quenched disorder. The couplings are independent and identically distributed random variables. The coupling distribution is an even function of ${\cal J}$
\begin{equation}\label{SMeq:even}
P^\text{1D}({\cal J})= P^\text{1D}(-{\cal J})\,.
\end{equation}
We shall be concerned with the scaling limit (i.e., the infinite-correlation-length limit). For a one-dimensional system, a scaling limit may appear only when the inverse temperature $\beta_{\text{1D}}$ goes to infinity.

The correlation functions can be readily computed using the high-temperature expansion (see, e.g., Ref.~\cite{parisi:94}). Consider the two spins at positions $z$ and $z+r$, $r>0$,  and the  OBC Hamiltonian in \eqref{SMeq:OBC-H}, one readily obtains
\begin{equation}\label{SMeq:C-OBC}
    \langle \sigma_{z}\sigma_{z+r}\rangle_{\text{\tiny OBC}} = \prod_{i=1}^r \tanh{(\beta_{\text{1D} {\cal J}_{z+i-1,z+i}})}\,.
\end{equation}
The average over the random couplings further simplifies the result  because the different factors in \eqref{SMeq:C-OBC}, $\tanh{(\beta_{\text{1D} {\cal J}_{z+i-1,z+i}})}$, are statistically independent:
\begin{equation}\label{SMeq:C2n-1D}
    \left.C^{\text{1D},(n)}(r)\right|_{\text{\tiny OBC}}=\left.\overline{\langle \sigma_{z}\sigma_{z+r}\rangle^{2n}}\right|_{\text{\tiny OBC}}= t_{2n}^r\,,
\end{equation}
where we have defined
\begin{equation}\label{SMeq:t2n-def}
    t_{2n}=\int_{-\infty}^\infty\mathrm{d}{\cal J}\, P^{\text{1D}}({\cal J})\, \tanh^{2n}{(\beta_{\text{1D}}{\cal J})}\,.
\end{equation}
Remarkably, \eqref{SMeq:C2n-1D} is independent of $M$. Therefore, \eqref{SMeq:C2n-1D} tells us that the spin-glass correlation functions in the $M\to\infty$ limit display a simple exponential decay
\begin{equation}\label{SMeq:C2n-1D-final}
    \left.C^\text{1D,(n)}(r)\right|_{M=\infty}=\text{e}^{-m_n\,r}\,,\quad  m_n=\frac{1}{\xi_n}= - \log {t_{2n}}\,.
\end{equation}
[Mind that $m_{p+1}>m_p$ because $t_{2p+2}<t_{2p}$; $t_0=1$ follows from the normalization condition for the probability density $P^{\text{1D}}({\cal J})$]. We shall analyze the scaling limit for these correlation functions in Sect.~\ref{SMsubsect:scaling-limit}. Of course, the same limit for the correlation functions is eventually reached for PBC. How \eqref{SMeq:C2n-1D} is recovered with PBC is the subject of Sect.~\ref{SMsubsect:PBC}.

\subsection{The 1D-scaling limit for an infinite chain}\label{SMsubsect:scaling-limit}

Let us now consider how the correlation lengths $\xi_n$ diverge as the inverse temperature $\beta_{\text{1D}}$  goes to $\infty$ (or, equivalently, $T_{\text{1D}}\to 0$ for the temperature). We shall rather work with the masses $m_n=1/\xi_n$ and rewrite \eqref{SMeq:C2n-1D-final} in a form more convenient to discuss the scaling limit:
\begin{equation}
m_n = -\log{[1-A_n(\beta_\text{1D})}]\,,
\end{equation}
where we have defined
\begin{equation}
A_n(\beta_\text{1D})= 2\int_0^{\infty} \mathrm{d}{\cal J}\,P^{\text{1D}}({\cal J})\,[1-\tanh^{2n}{(\beta_{\text{1D}}{\cal J})}]\,.\label{SMeq:scaling1D-1}
\end{equation}
Next, we need to make some further assumptions about $P^{\text{1D}}$. In particular, let us assume that
\begin{equation}\label{SMeq:g-def}
    P^{\text{1D}}({\cal J})=|\mathcal{J}|^\lambda g(\mathcal{J})\,,\quad g(\mathcal{J}=0)>0.
\end{equation}
In particular, note that the Droplet's model prediction $P^{\text{1D}}({\cal J}=0)\sim 1$ implies $\lambda=0$.
Hence, the change of variable $u=\beta_{\text{1D}}{\cal J}$ allows one to rewrite $A_n$ in \eqref{SMeq:scaling1D-1} as
\begin{eqnarray}
    A_n(\beta_\text{1D})&=&\frac{B_n(\beta_\text{1D})}{\beta_\text{1D}^{1+\lambda}}\,,\\[1mm]
    B_n(\beta_\text{1D})&=&2\int_0^{\infty} \mathrm{d}u\, u^\lambda g(u/\beta_\text{1D})\,[1-\tanh^{2n}{u}].
\end{eqnarray}
Now, we need to add two additional (and fairly mild) hypotheses about the function $g$ introduced in \eqref{SMeq:g-def}. First, we assume that $g$ is continuous at $\mathcal{J}=0$. Second, we impose that the product $|\mathcal{J}|^\lambda g(\mathcal{J})=P^{\text{1D}}({\cal J})$ is bounded.
The combination of the two assumptions is sufficient to show that
\begin{equation}
\lim_{\beta_\text{1D}\to\infty} B_n(\beta_\text{1D})=2 g(0) \int_0^{\infty} \mathrm{d}u\, u^\lambda \,[1-\tanh^{2n}{u}].
\end{equation}
At this point, it is straightforward to show that
\begin{equation}
\lim_{\beta_\text{1D}\to\infty} \beta_\text{1D}^{1+\lambda} m_n =
2 g(0)\int_0^{\infty} \mathrm{d}u\, u^\lambda\,[1-\tanh^{2n}{u}]
\,.
\end{equation}
Hence, the correlation lengths $\xi_{n}$ diverge as $1/T_{\text{1D}}^{1+\lambda}$ in the scaling limit at $T_{\text{1D}}=0$.

In order to compute the multifractal spectrum $\tau_n$ we start from the observation that
$\text{e}^{-m_n\,r}=(\text{e}^{-m_1\,r})^{m_n/m_1}$. Hence,
\begin{equation}
\tau_n=\lim_{\beta_{\text{1D}\to\infty}}\frac{m_n}{m_{n=1}}=\frac{I_{2n}}{I_2}\,,
\end{equation}
where
\begin{equation}
I_k=\int_0^\infty \mathrm{d}u\, u^\lambda (1-\text{tanh}^{k} u)\,.
\end{equation}
In the particular case $\lambda=0$, one may explicitly compute the
multifractal spectrum using the change of variable $y=\tanh {u}$
\begin{eqnarray}
I_2^{\lambda=0}&=&\int_0^1 \frac{\mathrm{d}\,y}{1-y^2}(1-y^2)=1\,,\\
\tau_n^{\lambda=0}&=& \int_0^1 \frac{\mathrm{d}\,y}{1-y^2}(1-y^{2n})\,,\\
&=&\sum_{k=0}^{n-1}\,\int_0^1 \mathrm{d}\,y\,y^{2k}\,,\\
&=&\sum_{k=0}^{n-1}\,\frac{1}{2k+1}\,.
\end{eqnarray}

\subsection{The difficult PBC-journey towards infinite chain length}\label{SMsubsect:PBC}

To simplify the expressions as much as possible, let us consider the spins at $z=0$ and $z=r$ (there is no loss of generality, because the disorder average induces translation invariance in the PBC spin chain). Hence, the analog of \eqref{SMeq:C-OBC} for PBC is nicely expressed in terms of the following two statistically independent random variables:
\begin{eqnarray}
    k&=&\Pi_{z=1}^r \tanh{(\beta_{\text{1D}} \mathcal{J}_{z-1,z})}\,,\\
    \tilde k&=&\Pi_{z=r}^{M-1} \tanh(\beta_{\text{1D}} \mathcal{J}_{z,z+1})\,,\\
    \langle \sigma_0 \sigma_d\rangle_{\text{PBC}}&=&\frac{k+\tilde{k}}{1+k\tilde{k}}\,.
\end{eqnarray}

Let us now consider the simplest spin-glass correlation function $C^\text{1D,(1)}(r)=\overline{\langle \sigma_0 \sigma_r\rangle^2}$
\begin{equation}
\left.C^\text{1D,(1)}(r)\right|_{\text{PBC}}\ =\
\overline{\frac{k^2+\tilde{k}^2+2 k\tilde{k}}{(1+k\tilde{k})^2}}\,.
\end{equation}
We shall find that it is the sum of three different contributions,
\begin{equation}\label{SMeq:three-terms}
    \left.C^{\text{1D},(1)}(r)\right|_{\text{PBC}}=\mathcal{C}_\text{direct}(r)+\mathcal{C}_\text{image}(r)+\mathcal{C}_{\text{const}}\,.
\end{equation}
The comparison of \eqref{SMeq:compare-1} with \eqref{SMeq:compare-2} will tell us that the image term is simply $\mathcal{C}_\text{image}(r)= \mathcal{C}_\text{direct}(M-r)$. In fact, the image term is completely standard when working with PBC. Instead, the constant term $\mathcal{C}_{\text{const}}$ lacks an analog in our experience with non-disordered systems.

The computation of the disorder average starts with the Taylor expansion
\begin{equation}
    \frac{1}{(1+k\tilde{k})^2} =\sum_{n=0}^\infty\, (-1)^n (n+1) (k\tilde{k})^n\,,
\end{equation}
which is safe because $ |k\tilde{k}|<1$. When computing the average over the couplings, a simple rule will be helpful. Let $n$ and $s$ be zero or positive integers
\begin{equation}
\overline{ k^{2n} \tilde{k}^{2s}} = t_{2n}^r t_{2s}^{M-r},\,
\end{equation}
where $t_{2n}$ was defined in \eqref{SMeq:t2n-def}. Instead, if any of the two exponents turns out to be an odd integer, the disorder average vanishes:
\begin{equation}
\overline{ k^{2n+1} \tilde{k}^{s}} = \overline{ k^{n} \tilde{k}^{2s+1}}=0\,.
\end{equation}

Using the notational conventions of \eqref{SMeq:C2n-1D-final}, we find that the direct term is
\begin{eqnarray}
\mathcal{C}_\text{direct}(r) &=&
    \overline{\frac{k^2}{(1+k\tilde{k})^2}}\\
    &=&\sum_{n=0}^\infty\, \overline{(-1)^n\, (n+1)\, k^{n+2} \tilde{k}^n}\,,\\
    &=&\sum_{p=0}^\infty\, (2p+1)\,t_{2p}^M\, \left(\frac{t_{2p+2}}{t_{2p}}\right)^r\,,\label{SMeq:compare-1}\\
    &=& \text{e}^{-m_1 r}\,+\, \mathcal{L}(r;M)\,.\label{SMeq:bad-corrections}
\end{eqnarray}
So, we have the final asymptotic term in~\eqref{SMeq:C2n-1D-final} plus a finite-$M$ correction $\mathcal{L}(r;M)$:
\begin{equation}\label{SMeq:very-bad-corrections}
\mathcal{L}(r)=\sum_{p=1}^\infty\, (2p+1)\,\text{e}^{-m_p M}\,\text{e}^{-(m_{p+1}-m_{p})r}\,.
\end{equation}
Although every term in $\mathcal{L}(r;M)$  is exponentially suppressed by the factor $\text{e}^{-m_p M}$, the exponential decay $\text{e}^{-(m_{p+1}-m_{p})r}$ can be pretty slow because the mass differences $m_{p+1}-m_p$ might be small (due to the multifractal scaling).

The image term follows from
\begin{eqnarray}
\mathcal{C}_\text{image}(r) &=&
    \overline{\frac{\tilde{k}^2}{(1+k\tilde{k})^2}}\\
    &=&\sum_{n=0}^\infty\, \overline{(-1)^n\, (n+1)\, k^{n} \tilde{k}^{n+2}}\,,  \\
    &=&\sum_{p=0}^\infty\, (2p+1)\,t_{2p}^r\,t_{2p+2}^{M-r}\,.\\
    &=&\sum_{p=0}^\infty\, (2p+1)\,t_{2p}^M\, \left(\frac{t_{2p+2}}{t_{2p}}\right)^{M-r}\,.\label{SMeq:compare-2}
\end{eqnarray}

The constant term is
\begin{eqnarray}
\mathcal{C}_\text{const} &=&
    \overline{\frac{2 k \tilde{k}}{(1+k\tilde{k})^2}}\\
    &=&2\sum_{n=0}^\infty\, \overline{(-1)^n\, (n+1)\, (k\tilde{k})^{n+1}}\,,\\
    &=&-2 \sum_{p=0}^\infty\, (2p+2)\, t_{2p+2}^M\,,\\
    &=&-4\sum_{p=1}^\infty\,p\, \text{e}^{-m_p M}\,.
\end{eqnarray}
As we see, the constant term is negative and decays exponentially in $M$ (the dominant contribution is $-4\,\text{e}^{-m_1 M}=-4\,\text{e}^{-M/\xi}$).

A similar analysis is possible for higher-order correlation functions
$C^{\text{1D},(n>1)}(r)$, but it adds little new information and we
shall omit it.

\section{How large \boldmath$M$ needs to be?}\label{SMsect:finite-M}

Both the analysis of the 1D toy-model in Sect.~\ref{SMsect:1D-IEA-model} and general transfer-matrix considerations agree in that a quantitative answer to the question in the title of this section should be given in terms of the dimensionless ratio $M/\xi$.

Indeed, let us compare in Fig.~\ref{fig:xi_vs_r_min_different_M} the estimators $\xi_{0,1}$ and $\xi_{1,2}$ for $\xi$, as obtained on prisms of sizes $8\times 8\times M$ and PBC. The first striking observation is that the $M$-dependence of $\xi_{1,2}$ is significantly milder than that of $\xi_{0,1}$.
This difference among the two estimators could have been anticipated from the constant term in \eqref{SMeq:three-terms}. Indeed, although the constant term is exponentially suppressed in $M/\xi$, $\xi_{1,2}$ (i.e., the estimator that involves non-zero wavenumbers) is totally blind to this constant. However, even the $\xi_{1,2}$ estimators have some $M$- and $r_\text{min}$-dependence, which indicates that the constant term does not suffice to explain all relevant  finite-$M$ effects.
Indeed, the analysis of the 1D-toy model suggests that the small mass differences in \eqref{SMeq:very-bad-corrections} will cause undesirable drifts as $r_\text{min}$ is varied. Only for $M=192$, namely $M/\xi\approx 10.3$ we find mutual consistency among the different $r_\text{min}$ and the two integral estimators $\xi_{0,1}$ and $\xi_{1,2}$. We also observe from Fig.~5 in the main text that our data from prism's size $16\times 16\times 512$ and PBC are $r_\text{min}$ independent ($M/\xi\approx  11.5$ in that case). We conclude that,  given our accuracy level, data representative of the large $M$-limit will be obtained in a PBC prism only if $M/\xi>10$.

As for OBC systems, from Fig.~5 in the main text, we note that data representative of the large-$M$ limit are obtained from prism's size $16\times 16\times 48$ (i.e., $M/\xi\approx 1.08$ and $z_{\text{discard}}=5$. Both lengths $M$
and $z_{\text{discard}}$ were conservatively scaled for our $L=24$ simulations
---  $M=88$ and $z_{\text{discard}}=10$ while $\xi_{n=1,L=24}/\xi_{n=1,L=16}\approx 1.65$--- despite the larger errors due to the smaller number of samples for $L=24$.

\section{Comparing data for OBC prisms of different lengths}\label{SMsect:finite-M-OBC}

The main results of this paper have been obtained for a lattice long enough to extract the correlation functions in the infinite-length limit with a reasonably small bias.
At this end, we had to use different tricks, such as defining an unusable region near the two ends, which we discarded. In this way, it is enough to take a sufficiently large $M$ measure of the correlation functions in this geometry at points sufficiently distant from the boundaries.
In this paragraph, we describe a different approach: computing the correlations for the overlap at the two ends of the prism. As we shall show, this approach offers a number of practical advantages (in particular, when equilibrating long prisms is unfeasible due to algorithmic or computational difficulties).

Let us consider a prism of length $M$, where $z=0,M-1$. We shall be computing
\begin{eqnarray}
    R(M)&=&\frac{\overline{\langle Q(z_1=0) Q(z_2=M-1)\rangle}}{\overline{\langle [Q(z_1=0)]^2\rangle}}\,,\\[1mm]
    &=&\frac{C^{(1)}(0,M-1)}{C^{(1)}(0,0)}\,.\label{SMeq:R-def}
\end{eqnarray}
The transfer-matrix analysis of a finite-$M$ correlation function is more involved than its $M\to\infty$ limit, because it has contributions from both even-parity and odd-parity states (see, e.g., Ref.~\cite{bernaschi:24b} for an example worked out in detail). The reader will find in Sect.~\ref{SMsubsect:Transfer}, below, the sketch of the derivation of the finite-$M$ expansion for $R(M)$:
\begin{equation}\label{SMeq:R-expansion-1}
    R(M)=\frac{a_1 \text{e}^{-M E_1^{\text{odd}}}+a_2 \text{e}^{-M E_2^{\text{odd}}}+\ldots}{1+b_1 \text{e}^{-M E_1^{\text{even}}}+\ldots}\,.
\end{equation}
We have already encountered the odd-eigenvalues of the transfer matrix $\text{e}^{E_k^{\text{odd}}}$ with $E_1^\text{odd}=1/\xi$, $E_{k>1}^\text{odd}=1/\xi_{n=1;k>1}$. This is the first time that we encounter an excited even-state for the transfer matrix $\text{e}^{E_1^{\text{even}}}$. Also, $E_1^{\text{even}}$ can
be interpreted as the inverse of a correlation length. The dots in \eqref{SMeq:R-expansion-1} stand for the contributions of higher excited states. The coefficients $a_1,a_2$ and $b_1$ are expected to be of order one. Hence, under the hypothesis that $M E_1^{\text{even}}$ and $M E_2^{\text{odd}}$ are large-enough (larger than two would be enough in practice), it is meaningful to truncate the expansions to the order shown in  \eqref{SMeq:R-expansion-1} and Taylor-expand in the supposedly small quantity $b_1 \text{e}^{-M E_1^{\text{even}}}$, obtaining
\begin{eqnarray}\label{SMeq:R-expansion-2}
R(M)&=& a_1 \text{e}^{-M E_1^{\text{odd}}}+a_2 \text{e}^{-M E_2^{\text{odd}}}\nonumber\\
&-& a_1b_1\text{e}^{-M \Big(E_1^{\text{odd}}+E_1^{\text{even}}\Big)}+\ldots\,.
\end{eqnarray}
\eqref{SMeq:R-expansion-2} is finally ready for the data analysis that we explain next.

Take a set of $R(M)$, obtained in prisms of varying length $M$. Because each $R(M)$ comes from a different simulation, they will be statistically independent. In practice, we have found it convenient to use the \emph{strong} version of the OBC explained in Sect.~\ref{SMsect:OBC}, as they seem to have smaller coefficients $a_2$ and $b_1$. Then, the $R(M)$ are fitted to the ansatz
\begin{equation}\label{SMeq:R-ansatz}
    R(M) = a\,\text{e}^{-M/\xi}+b\,\text{e}^{-M/\xi'}\,,
\end{equation}
where the fit parameter $\xi$ stands for the dominant correlation length that we aim to compute, namely $\xi$, while the term $b\exp(-M/\xi')$ tries to mimic the effect of the sub-leading correction terms in \eqref{SMeq:R-expansion-2}. The correlation length $\xi'$ is not exactly equal to $1/E_2^{\text{odd}}$ or $1/(E_1^{\text{odd}}+E_1^{\text{even}})$.

Before presenting the results, let us describe the advantages and disadvantages of this approach.
On the positive side, we have:
\begin{itemize}
\item There is no region to be discarded in the correlation functions. This is compensated by the large-distance behavior, which is more complex and includes corrections from the even excited states of the transfer matrix that are absent in the standard approach.
\item We can use data for relatively small values  $M$. Systems with small $M$  are easier to equilibrate.
\item The data at different $M$ are independent of each other, so it is possible to perform the fits with a standard interpretation of the $\chi^2$ test.
\end{itemize}
Among the disadvantages, we have:
\begin{itemize}
\item If $M$ is not sufficiently large, we are not in the fully asymptotic regime, so the value of the correlation length strongly depends on the corrections to the one-length scaling. In other words, there are possible systematic errors that should be carefully addressed.
\item The data $M\simeq L$ are challenging to obtain because the Houdayer trick is not efficient in this nearly-cubic region. Lattices with $M\gg L$ thermalize more easily than those with $M\simeq L$. In the pre-Houdayer era, this approach would be invaluable; in the post-Houdayer era, it may not be so valuable for spin glasses at zero magnetic field. However, it can be handy in systems where the Houdayer trick does not work, for example, ferromagnetic systems in a random magnetic field.
\item The approach provides independent measurement of the correlation length, with systematic errors that differ from those in the main text, where we do control the systematic errors.
\end{itemize}

We now present the results obtained with this method for $L=16$ and $L=24$.

\begin{figure}[htb]
    \centering
    \includegraphics[width=0.9\columnwidth]{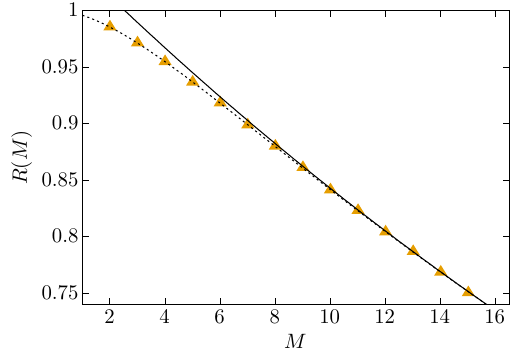}
    \caption{Correlation $R(M)$, see \eqref{SMeq:R-def}, versus the prism length $M$, as computed in prism of transverse dimension $L=16$, with the \emph{strong} version of the OBC. The dotted line is a fit to \eqref{SMeq:R-ansatz}. The solid line is the asymptotically dominant term in the fit $a\,\text{e}^{-M/\xi}$.}
    \label{fig:short16}
\end{figure}

For $L=16$ we have done simulations up to $M=15$. The figure of merit for the fit shown in Fig.~\ref{fig:short16} is quite satisfactory, that is
$\chi^2/\text{DoF}=10.11/10$. The fitted coefficients are  $\xi=43.60 (24)$ and $\xi'=2.52 (11)$. The value of $\xi$ is pretty close to the estimates for $\xi$ in the main text, which are more reliable. Regarding the subdominant term, $(1+\xi/\xi')=18.30\ldots$ is intriguingly close to the dimensionless ratio $U$ that is shown in the more controlled computation of Fig.~\ref{fig:U_vs_r_min}.

\begin{figure}[htb]
    \centering
    \includegraphics[width=0.9\columnwidth]{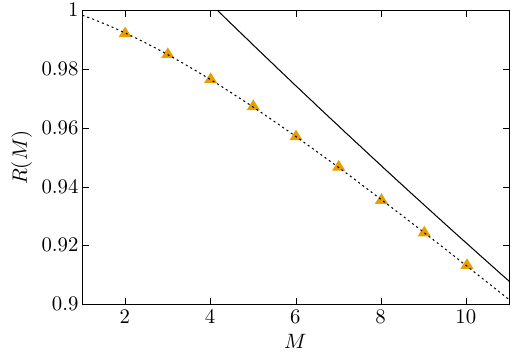}
    \caption{Correlation $R(M)$, see \eqref{SMeq:R-def}, versus the prism length $M$, as computed in prism of transverse dimension $L=16$, with the \emph{strong} version of the OBC. The dotted line is a fit to \eqref{SMeq:R-ansatz}, the data point with $M=2$ was excluded from the fit. The solid line is the asymptotically dominant term in the fit $a\,\text{e}^{-M/\xi}$.}
    \label{fig:short24}
\end{figure}

For $L=24$, we performed simulations with $M=2,3,\ldots, 10$. The fit in
Fig.~\ref{fig:short24} is fair, $\chi^2/\text{DoF}=4.01/4$. The fitted coefficients are  $\xi=71(7)$ and $\xi'=5.0 (1.6)$. The value of $\xi$ is, again, compatible with the estimates of $\xi$ in the main text. As for $\xi'$, the ratio $1+\xi/\xi'=15.2$ is a bit too low, but not too much given the size of the errors (in this case, the lattice is much shorter and we do not reach the region of $M$ where the pre-asymptotic terms are small).

Overall, it is refreshing to note that, for lattices 9 times shorter than those in the main text, we obtain compatible results.

\subsection{Transfer matrix analysis of the short prism}\label{SMsubsect:Transfer}

To analyze a prism of finite length $M$, we need to address the fact that the transfer matrix has eigenvalues of even and odd parity. This means we need a more precise notation than we have been using so far.  The transfer matrix has a ground state $|0\rangle$ of even parity,
\begin{equation}
\mathcal{T}|0\rangle = |0\rangle\,.
\end{equation}
The excited eigenvectors with even or odd parity will be denoted $|k^{\text{even}}\rangle$ or $|k^{\text{odd}}\rangle$, respectively.
For consistency, we should have written the Ground State as $|0^{\text{even}}\rangle$,
but we prefer to name it $|0\rangle$ to emphasize its uniqueness. We ordered the spectrum of
$\mathcal{T}$ as
\begin{eqnarray}
    \mathcal{T}|k^\text{even}\rangle&=& \text{e}^{-E_k^{\text{even}}}| k^\text{even}\rangle\,,\\
    \mathcal{T}|k^\text{odd}\rangle &=& \text{e}^{-E_k^{\text{odd}}}| k^\text{odd}\rangle\,,\ \\
    E_0^{\text{even}}=0&<& E_1^{\text{even}} \leq E_2^{\text{even}} \leq E_3^{\text{even}}\leq\ldots\\
    E_0^{\text{even}}=0&<& E_1^{\text{odd}} \leq E_2^{\text{odd}} \leq E_3^{\text{odd}}\leq\ldots
\end{eqnarray}
Odd operators such as $\hat Q$ have non-vanishing matrix elements $\langle A|\hat Q|B\rangle$ for parity eigenvectors only if the parity of $|A\rangle$ and $|B\rangle$ differ. Instead, even operators such as $\hat Q^2$ have non-vanishing matrix elements for parity eigenvectors only if the parity of the two states is the same.

The partition function of the replicated system is
\begin{equation}
    \mathcal{Z}=\langle B|\mathcal{T}^M|B\rangle\,,
\end{equation}
where $|B\rangle$ is the boundary state (it depends on the choice of the boundary
conditions, see Section~\ref{SMsect:OBC}). $|B\rangle$ is an even state. The reduced partition function
$\hat{\mathcal{Z}}=\mathcal{Z}/|\langle B|0\rangle|^2$ can be formally expanded using the eigenstates of the transfer matrix as
\begin{eqnarray}
     \hat{\mathcal{Z}}&=& 1 + \sum_{k>1} \frac{|\langle B|k^{\text{even}}\rangle|^2}{|\langle B|0\rangle|^2} \,\text{e}^{-M E_k^\text{even}}\,.
\end{eqnarray}
The contribution of the odd states is missing because $0=\langle B|k^{\text{odd}}\rangle$, due to the different parity.
Then, we have for the numerator of \eqref{SMeq:R-def}
\begin{equation}\label{SMeq:numerador}
    \hat{\mathcal{Z}}\,\overline{\langle Q(0) Q(M-1)\rangle}=\frac{\langle B|\hat{Q}\mathcal{T}^M\hat Q|B\rangle}{\langle B|0\rangle^2}\,.
\end{equation}
For the denominator of \eqref{SMeq:R-def} we have instead
\begin{equation}\label{SMeq:denominador}
    \hat{\mathcal{Z}}\,\overline{\langle [Q(0)]^2\rangle}=\frac{\langle B|\hat{Q}^2 \mathcal{T}^M|B\rangle}{\langle B|0\rangle^2}\,.
\end{equation}
The quotient of the l.h.s. of Eqs.~\eqref{SMeq:numerador} and~\eqref{SMeq:denominador} gives us $R(M)$, while the quotient of the r.h.s. of both equations can be expanded using the eigenvectors of $\mathcal{T}$.
Let us define the normalized expansion coefficients
\begin{equation}
A_k=\frac{|\langle B|\hat{Q}|k^\text{odd}\rangle|^2}{|\langle B| 0\rangle|^2}\,,\quad
C_k=\frac{\langle B| \hat{Q}^2 |k^\text{even}\rangle\langle k^\text{even}|B\rangle}{|\langle B|  0\rangle|^2},
\end{equation}
which are of order one due to the normalization with $|\langle B|  0\rangle|^2$. Then, we have for $R(M)$
\begin{equation}
    R(M)=\frac{ \sum_{k>0} A_k\,\text{e}^{-M E_k^{\text{odd}}}}{C_0+\sum_{k>0} C_k\,\text{e}^{-M E_k^{\text{even}}}}\,.
\end{equation}
At this point, we need just to divide by $C_0$ the numerator and the denominator of the above expression, and define $a_k=A_k/C_0$ and $b_k=C_k/C_0$ to recover \eqref{SMeq:R-expansion-1}.

\section{The ferromagnetic Ising model: what happens in the absence of soft excitations?}\label{SMsect:surface-tension}

In the absence of soft excitations, we expect a positive surface tension $\Sigma(T)$ for all $T<T_{\text{c}}$. Near zero temperature, the surface tension diverges as $\Sigma(T)\propto 1/T$, while near the critical point it vanishes as $\Sigma(T)\sim( T_{\text{c}}- T)^\mu$, where the exponent $\mu$ relates to the correlation-length exponent $\nu$ through the hyperscaling relation $\mu=\nu(D-1)$~\cite{widom:65}. As we shall show in the following, the surface tension deeply modifies the results found in the main text, in the sense that the correlations along the longitudinal dimension of the prism have a correlation length that grows exponentially with
$L^{D-1}$ (recall that $L$ is the prism's transverse size), rather than with a power law in $L$.

Let us start considering prisms of equal transverse and longitudinal dimensions, $L=M$ and compare the partition function $Z_{++}^D$ (i.e., spins locked to point to the North-pole at both prism's ends, $x_D=0$ and $x_{D}=L-1$), and $Z_{+-}^D$ (spins aligned with the North-pole at $x_D=0$ and with the South-pole at $x_D=L-1)$. The opposite boundary conditions for $Z_{+-}^D$ enforce the formation of an interface that separates the (mostly) North-pole-aligned spins from the (mainly) South-pole-aligned region.
The surface tension follows from the ratio of partition functions:
\begin{equation}\label{SMeq:Y-def}
    \mathcal{Y}=\frac{Z_{+-}^D}{Z_{++}^D}\,,\quad \varSigma(T)=-\lim_{L\to\infty}\frac{\log{\mathcal{Y}}}{L^{D-1}}\,.
\end{equation}
Hence, $\mathcal{Y}$ is exponentially small in $\varSigma(T) L^{D-1}$. However, while this is the leading scaling, there might be sub-leading corrections. For example, because at low temperatures the interface is flat, one may expect $\mathcal{Y} \sim L\, \text{e}^{-L^{D-1} \Sigma(T)}$
because there are $\sim L$ equivalent positions for the interface.

To make further progress, we consider an effective 1D ferromagnetic Ising model, with a partition function
\begin{equation}
Z^{\text{1D}}=\sum_{\{\sigma_z=\pm 1\}}\ \text{e}^{\kappa_\text{FM,1D}\,\sum_z \sigma_z\sigma_{z+1}}\,,\quad \sigma_z=\pm 1\,.
\end{equation}
This spin chain can be thought of as the result of a Renormalization Group transformation that replaces a plane $x_D=z$ in the $D$-dimensional system with a single spin in the chain. Hence, the correlation length in the prism (along the $D$-direction), is expected to coincide with the correlation length in the effective spin-chain. Our next task is to determine the effective coupling $\kappa_\text{FM,1D}$. To do so, we start by  computing the ratio of partition functions for the spin-chain of length $L$, using the same boundary conditions used in \eqref{SMeq:Y-def} (the Transfer matrix, see e.g.~\cite{parisi:94}, makes the computation straightforward):
\begin{equation}
    \frac{Z_{+-}^{\text{1D}}}{Z_{+-}^{\text{1D}}}=\frac{1-\tanh^{L-1}{\kappa_\text{FM,1D}}}
    {1+\tanh^{L-1}{\kappa_\text{FM,1D}}}\,.
\end{equation}
Our matching condition for  $\kappa_{\text{FM,1D}}$ simply states that
\begin{equation}
    \mathcal{Y}=\frac{Z_{+-}^{\text{1D}}}{Z_{++}^{\text{1D}}}\quad\text{ or }\quad \tanh{\kappa_\text{FM,1D}}=\left(\frac{1 - \mathcal{Y}}{1 + \mathcal{Y}}\right)^{\frac{1}{L-1}}\,.
\end{equation}
We are finally ready to go to the infinitely long prism $M\to\infty$ for which it is well known that $\xi_{\text{FM,1D}}=1/|\log{\tanh{\kappa_\text{FM,1D}}}|$:
\begin{equation}
    \xi_{\text{FM,1D}}(L)=\frac{1}{\frac{1}{L-1}\log{\frac{1+\mathcal{Y}}{1-\mathcal{Y}}}}=\frac{L-1}{2\mathcal{Y}}\ +\ \mathcal{O}(L)\,,
\end{equation}
which is exponentially large in $\Sigma(T) L^{D-1}$, as anticipated above and in the main text.

\FloatBarrier
\bibliographystyle{apsrev4-2}
\bibliography{biblio.bib}